\documentclass[11pt]{imsart}
\RequirePackage[OT1]{fontenc}
\RequirePackage{amsthm,amsmath}
\RequirePackage[colorlinks,citecolor=blue,urlcolor=blue]{hyperref}
\RequirePackage{hypernat}
\usepackage{graphicx}
\usepackage{enumerate}
\usepackage{latexsym}
\usepackage{amssymb}

\numberwithin{equation}{section}

\startlocaldefs \theoremstyle{plain}
\newtheorem{Theorem}{Theorem}[section]

\newtheorem{Proposition}[Theorem]{Proposition}
\newtheorem{Assumption}{Assumption}
\renewcommand{\theAssumption}{A.\arabic{Assumption}}
\newtheorem{lemma}{\indent \bf Lemma}

\theoremstyle{definition}
\newtheorem{Example}{Example}[section]
\makeatletter
\@addtoreset{equation}{section}\makeatother

\endlocaldefs

\begin{document}

\begin{frontmatter}

\title{High-Dimensional Bayesian Inference in Nonparametric Additive Models}
\runtitle{Bayesian High-Dimensional Inference}
\author{Zuofeng Shang and Ping Li}
\affiliation{Cornell University}

\address{Department of Statistical Science\\
Cornell University\\
Ithaca, NY 14853\\
Email: shang9@purdue.edu\\
pingli@stat.rutgers.edu}

\runauthor{Z. Shang and P. Li}
\begin{abstract}
A fully Bayesian approach is proposed for ultrahigh-dimensional nonparametric
additive models in which the number of additive components may be larger than the
sample size, though ideally the true model is believed to include only a small number of components. Bayesian approaches can conduct stochastic model search
and fulfill flexible parameter estimation by stochastic draws. The theory shows that the proposed model selection method has satisfactory properties.
For instance, when the hyperparameter associated with the model prior is correctly specified, the true model has posterior probability approaching one as the sample size
goes to infinity; when this hyperparameter is incorrectly specified, the selected model is still acceptable since asymptotically it is shown to be nested in the true model.
To enhance model flexibility, two new $g$-priors are proposed and their theoretical performance is investigated.
We also propose an efficient reversible jump
MCMC algorithm
to handle the computational issues. Several simulation examples are provided to demonstrate the advantages of our method.
\end{abstract}


\end{frontmatter}

\section{Introduction}\label{sec:intro}
Suppose the data $\{Y_i,X_{1i},\ldots,X_{pi}\}_{i=1}^n$ are \textit{iid} copies of $Y,X_1,\ldots,X_p$ generated from the following model
\begin{equation}\label{true:model}
Y_i=\sum\limits_{j=1}^p f_j(X_{ji})+\epsilon_i,\,\,i=1,\ldots,n,
\end{equation}
where $\epsilon_i$'s denote the zero-mean random errors,
and for each $j=1,\ldots,p$, $X_j$ is a random variable taking values in $[0,1]$,
$f_j$ is a function of $X_j$ satisfying $E\{f_j(X_j)\}=0$.
The zero-expectation constraint is assumed for identifiability issue.
Model (\ref{true:model}) is called the additive component model; see \cite{S85,HT90} for an excellent introduction.
Suppose model (\ref{true:model}) contains $s_n$ significant covariates,
and the remaining $p-s_n$ covariates are insignificant.
Here we assume $p/n\rightarrow\infty$ as $n\rightarrow\infty$, denoted as $p\gg n$ or equivalently $n\ll p$,
but ideally restrict $s_n=o(n)$,
i.e., the true model is sparse. Our goal is to explore an automatic fully Bayesian procedure for
selecting and estimating the significant (nonvanishing) $f_j$'s in model (\ref{true:model}).

When each $f_j$ is linear in $X_j$, (\ref{true:model}) reduces to a linear model.
There has been a considerable amount of frequentist approaches exploring issues on model selection
in ultrahigh-dimensional situations, i.e., $\log{p}=O(n^k)$ for some $k>0$.
The representative ones include regularization-based approaches such as
\cite{ZY06,MB06,VD08,ZH08,HHM08,MY09,LF09,SPZ12,YZ13}, and correlation-based approaches such as \cite{FL08,FS10,XZ11}.
An insightful review is given by \cite{FL10}.

Model selection on the basis of a Bayesian framework is conceptually different.
Specifically, Bayesian approaches conduct stochastic search of the models and evaluate
each model by its posterior probability.
Three major advantages of Bayesian selection methods are worth mentioning:
(1) Bayesian approaches can perform model selection, parameter estimation and inference
in a unified manner through posterior samples, no additional procedures such as prescreening, thresholding or data splitting
are needed; (2) the choice of the hyperparameters is flexible by fulfilling stochastic draws;
and (3) Bayesian methods allow the practitioners to
incorporate prior information in the process of model search. The last feature might be attractive in small sample
problems where prior information may be useful to address data insufficiency.
There has been an amount of literature
on Bayesian model selection in linear models. For example, when $p$ is fixed,
\cite{FLS01,BB04,GMCM10,LPMCB08,CGMM09} show that, under certain regularity conditions, the posterior probability of the true model
converges to one as $n$ increases, in other words, \emph{posterior model consistency} holds.
This means that the proposed Bayesian selection method is asymptotically valid.
Later on, these results were generalized by \cite{SC11,JR12} to the growing $p$ situation with $p=O(n)$.
In ultrahigh-dimensional situations,
\cite{SL13} considered a fully Bayesian hierarchical model with a prior controlling the model size
and obtained posterior model consistency. A straightforward MCMC algorithm was developed for model search.
Based on an extended Bayesian information criteria, \cite{LSY13}
established posterior model consistency in generalized linear models.

However, in many practical applications there might be little evidence confirming
linearity of the $f_j$'s, for which a nonparametric assumption on the $f_j$'s will largely enhance model flexibility,
leading to the so-called nonparametric additive models.
Surprisingly, theoretical studies relating to model selection in nonparametric additive models
are almost all in frequentist settings. For instance,
\cite{LZ06,KY08,RLLW09} explored issues relating to component selection with smoothing constraints assumed on the nonparametric functions.
\cite{MSB09,HHW10} proposed penalty-based approaches and studied their asymptotic properties.
\cite{FFS11} proposed a learning approach based on independent correlation and proved selection consistency.
To the best of our knowledge, theoretical studies in Bayesian settings are nonexistent, especially when $p\gg n$.
In terms of empirical evaluation, \cite{SHK11} proposed an objective Bayesian approach using penalized splines,
\cite{CGM10} proposed a Bayesian framework based on adaptive regression trees,
and \cite{SFK12} proposed a Bayesian framework based on a spike-and-slab prior
induced from normal-mixture-inverse-gamma distributions.
However, theoretical validity of these methods in ultrahigh-dimensional scenarios
has not been justified.

In this paper, we propose a fully Bayesian hierarchical model
which involves a new spike-and-slab prior on the function coefficients
and a novel prior controlling the model size, namely, the \textit{size-control prior}.
The spike-and-slab prior has two important features: first, it either removes or includes the entire block of function coefficients,
which is useful for model selection purpose; second,
within each block, suitable decay rates are assumed on the function coefficients via their prior variances
to produce smooth estimate of the nonparametric function.
The size-control prior, which involves a \emph{size-control parameter}, effectively restricts the scope of the target models,
and facilitates both theoretical and computing issues.
Based on the proposed Bayesian framework, we show that
when the size-control parameter is correctly specified,
posterior model consistency uniformly holds when the hyperparameters
are confined by suitable ranges; when the size-control parameter is incorrectly specified,
the selection results are still acceptable in the sense that
the selected model is asymptotically nested in the true model,
in other words, the number of false positives asymptotically vanishes.
Interestingly, the asymptotic results are shown to be true even in the hyper-$g$ prior settings.
Furthermore, a novel and nontrivial blockwise MCMC procedure is proposed for computation.
Our MCMC procedure allows stochastic search of
all critical hyperparameters including the blocks of the function coefficients,
the indicator variables representing inclusion/exclusion of the variables,
the size-control parameter, and even the number of basis functions used for model fitting.
The most challenging part in computation is the so-called trans-dimensional problem,
which is successfully resolved by a novel and nontrivial variation of the ``dimension-matching"
technique proposed by \cite{P95} in the reversible jump MCMC approach.
Simulation results demonstrate satisfactory selection and estimation accuracy of the proposed method.
Performance under different basis structures is also examined.
To the best of our knowledge, our work is the first one
establishing a both theoretically and empirically effective fully Bayesian procedure for function component selection
in ultrahigh-dimensional settings.

The rest of the paper is organized as follows.
In Section \ref{sec:model}, we carefully describe our fully Bayesian model and the prior distributions
on the model parameters.
In Section \ref{sec:main:results}, asymptotic results are provided for both well specified and misspecified
model spaces. In the meantime, two new types of $g$-priors are constructed and their theoretical properties
are carefully studied. Section \ref{sec:MCMC} contains the details of the MCMC algorithm.
Section \ref{numer:study} includes the simulation examples showing the satisfactory performance of the proposed method.
Section \ref{sec:conc} summarizes the conclusions.
Technical arguments are provided in appendix.

\section{A Bayesian Nonparametric Size-Control Model}\label{sec:model}

Before describing our model, we introduce some notation and assumptions that are used frequently in this paper.
We associate each $f_j$, $j=1,\ldots,p$, a $0\backslash1$ variable $\gamma_j$ indicating the exclusion$\backslash$inclusion of $f_j$
in the model (\ref{true:model}). Specifically, when $\gamma_j=0$, $f_j=0$ implies that $f_j$ is not included in model (\ref{true:model});
when $\gamma_j=1$, $f_j\neq0$ implies that $f_j$ is included in model (\ref{true:model}).
Define $\boldsymbol{\gamma}=(\gamma_1,\ldots,\gamma_p)^T$. For simplicity, we denote $j\in\boldsymbol{\gamma}$ to mean that $\gamma_j=1$,
and denote $j\in-\boldsymbol{\gamma}$ to mean $\gamma_j=0$.
Throughout we use $|\boldsymbol{\gamma}|$
to denote the number of ones in $\boldsymbol{\gamma}$, which is called the size of $\boldsymbol{\gamma}$.
It is clear that there are totally $2^p$ possible $\boldsymbol{\gamma}$'s representing $2^p$ different models,
each of which determines a subset of $\{f_j,j=1,\ldots,p\}$ that are included in model (\ref{true:model}).
In other words, under $\boldsymbol{\gamma}$, model (\ref{true:model}) is equivalent to
$Y_i=\sum_{j\in\boldsymbol{\gamma}} f_j(X_{ji})+\epsilon_i$, $i=1,\ldots,n$.
For any $\boldsymbol{\gamma}=(\gamma_1,\ldots,\gamma_p)^T$ and $\boldsymbol{\gamma}'=(\gamma_1',\ldots,\gamma_p')^T$, let
$(\boldsymbol{\gamma}\backslash\boldsymbol{\gamma}')_j=I(\gamma_j=1,\gamma'_j=0)$, and $(\boldsymbol{\gamma}\cap\boldsymbol{\gamma}')_j=I(\gamma_j=1,\gamma'_j=1)$.
Thus, $\boldsymbol{\gamma}\backslash\boldsymbol{\gamma}'$ is the $0\backslash1$ vector indicating the functional components
present in model $\boldsymbol{\gamma}$ but absent in model $\boldsymbol{\gamma}'$,
and $\boldsymbol{\gamma}\cap\boldsymbol{\gamma}'$ is the $0\backslash1$ vector indicating the functional components
present in both models $\boldsymbol{\gamma}$ and $\boldsymbol{\gamma}'$. We say that $\boldsymbol{\gamma}$ is
nested in $\boldsymbol{\gamma}'$ (denoted by $\boldsymbol{\gamma}\subset\boldsymbol{\gamma}'$) if
$\boldsymbol{\gamma}\backslash\boldsymbol{\gamma}'$ is zero. We further assume
$\{f_j^0, j=1,\ldots,p\}$ to be the true functional components, and denote
$\boldsymbol{\gamma}^0=(\gamma_1^0,\ldots,\gamma_p^0)^T$ with $\gamma_j^0=I(f_j^0\neq0)$. That is, the data
$\{Y_i,X_{1i},\ldots,X_{pi}\}_{i=1}^n$ are truly sampled from model
$Y_i=\sum_{j\in\boldsymbol{\gamma}^0} f_j^0(X_{ji})+\epsilon_i$, $i=1,\ldots,n$.
Thus, $\boldsymbol{\gamma}^0$ represents the true model where data are generated, and
$s_n=|\boldsymbol{\gamma}^0|$ denotes the size of the true model, i.e., the number of components $f_j$'s
included in the true model.

For $j=1,\ldots,p$, define an inner product $\langle f_j,\tilde{f}_j\rangle_j=E\{f_j(X_{j})\tilde{f}_j(X_{j})\}$
for any $f_j, \tilde{f}_j\in\mathcal{H}_j$, where $\mathcal{H}_j$ is the class of functions on $[0,1]$ satisfying
$E\{|f_j(X_{j})|^2\}<\infty$ and $E\{f_j(X_{j})\}=0$.
This inner product induces a norm denoted by $\|\cdot\|_j$, that is, $\|f_j\|_j=\sqrt{E\{|f_j(X_{j})|^2\}}$.
Suppose the density function $d_j(x_j)$ of $X_j$ satisfies
$0<K_1\le d_j(x_j)\le K_2<\infty$ for any $x_j\in[0,1]$ and $j=1,\ldots,p$,
where $K_1,K_2$ are constants.
Clearly, under $\langle\cdot,\cdot\rangle_j$,
$\mathcal{H}_j$ is a well-defined Hilbert space. Let $\{\varphi_{jl}, l=1,2,\ldots\}\subset\mathcal{H}_j$ be the
orthonormal basis functions for $\mathcal{H}_j$ under $\langle\cdot,\cdot\rangle_j$.
Any function $f_j\in\mathcal{H}_j$ thus admits the Fourier series $f_j=\sum_{l=1}^\infty \beta_{jl}\varphi_{jl}$,
with $\beta_{jl}=\langle f_j,\varphi_{jl}\rangle_j$ being the Fourier coefficients.
It can be shown that $f_j=0$ if and only if all the Fourier coefficients $\beta_{jl}$'s are zero.
Therefore, to detect whether $f_j$ vanishes or not, it is sufficient to detect whether the $\beta_{jl}$'s are zero.
In general, $f_j$ might correspond to infinitely many Fourier coefficients.
Handling all the Fourier coefficients is computationally infeasible. Furthermore,
it is commonly believed that only a finite subset of the Fourier coefficients
capture most of the information possessed by $f_j$.
Thus, we consider the \emph{partial Fourier series} $f_j\approx\sum_{l=1}^m \beta_{jl}\varphi_{jl}$ with truncation parameter $m$,
where $m=m_n$ is a sequence increasing with $n$. General theory on Fourier analysis
leads to that $\|f_j-\sum_{l=1}^m \beta_{jl}\varphi_{jl}\|_j$ approaches zero as $m\rightarrow\infty$,
showing the validity of such approximation.
We introduce some additional matrix notation to simplify the expression of our model. For $j=1,\ldots,p$
and $l=1,\ldots,m$,
define $\boldsymbol{\beta}_j=(\beta_{j1},\ldots,\beta_{jm})^T$,
$\boldsymbol{\Phi}_{jl}=(\varphi_{jl}(X_{j1}),\ldots,\varphi_{jl}(X_{jn}))^T$,
and $\textbf{Z}_j=(\boldsymbol{\Phi}_{j1},\ldots,\boldsymbol{\Phi}_{jm})$. Thus, each $\textbf{Z}_j$
is $n$ by $m$.
For and $\boldsymbol{\gamma}$, define $\textbf{Z}_{\boldsymbol{\gamma}}=(\textbf{Z}_j, j\in\boldsymbol{\gamma})$,
the $n$ by $m|\boldsymbol{\gamma}|$ matrix formed by $\textbf{Z}_j$'s with $j\in\boldsymbol{\gamma}$,
and define $\boldsymbol{\beta}_{\boldsymbol{\gamma}}$ to be the $m|\boldsymbol{\gamma}|$-vector
of Fourier coefficients formed by $\boldsymbol{\beta}_j$'s with $j\in\boldsymbol{\gamma}$.
Define $\textbf{Y}=(Y_1,\ldots,Y_n)^T$ to be the response vector.

We assume that the model errors $\epsilon_i$'s are \textit{iid} zero-mean Gaussian with variance $\sigma^2$, therefore,
model (\ref{true:model}), given $\boldsymbol{\gamma}$,
$f_j$'s and $\sigma^2$, becomes
\begin{equation}\label{fbm:0}
Y_i\sim N(\sum_{j\in\boldsymbol{\gamma}} f_j(X_{ji}),\sigma^2),\,\,i=1,\ldots,n.
\end{equation}
Since each $f_j$ can be well approximated by $\sum_{l=1}^m \beta_{jl}\varphi_{jl}$ for some sufficiently large $m$,
the mean of $Y_i$ is approximately $\sum_{j\in\boldsymbol{\gamma}} \sum_{l=1}^m \beta_{jl}\varphi_{jl}(X_{ji})$.
Thus, (\ref{fbm:0}) is approximately
$Y_i\sim N(\sum_{j\in\boldsymbol{\gamma}} \sum_{l=1}^m \beta_{jl}\varphi_{jl}(X_{ji}),\sigma^2)$.
In matrix form, this becomes
\begin{equation}\label{fbm:1}
\textbf{Y}|\boldsymbol{\gamma},
\boldsymbol{\beta}_{\boldsymbol{\gamma}},\sigma^2\sim N(\textbf{Z}_{\boldsymbol{\gamma}}\boldsymbol{\beta}_{\boldsymbol{\gamma}},
\sigma^2 \textbf{I}_n).
\end{equation}
When $\gamma_j=0$, $f_j=0$ implies that all the Fourier coefficients $\beta_{jl}$'s are zero.
When $\gamma_j=1$, $f_j\neq0$, we place normal prior distributions over its Fourier coefficients.
Explicitly, for $j=1,\ldots,p$, we adopt the spike-and-slab prior for $\beta_{jl}$'s, i.e,
\begin{equation}\label{fbm:2}
\beta_{jl}|\gamma_j,\sigma^2\sim (1-\gamma_j)\delta_0+\gamma_j N(0,c_j\sigma^2\tau_l^2),\,\,l=1,\ldots,m,
\end{equation}
where $\delta_0(\cdot)$ is the point mass measure concentrating on zero, $\{\tau_l^2,l\ge1\}$ is a fixed nonincreasing sequence,
and $c_j$'s are temporarily assumed to be fixed. Note that the $c_j$'s
are used to control the variance of the nonzero coefficients, and therefore can be viewed as the \emph{variance-control parameters}.
In many applications we may choose $\tau_l^2=l^{-(2\omega+1)}$ for $l\ge1$, where $\omega>1/2$ is a fixed constant characterizing the
degree of smoothness;
see, e.g., \cite{BG03}. The prior (\ref{fbm:2}) can be viewed as a direct multivariate extension of the conventional
spike-and-slab prior on scalar coefficients considered by \cite{CPV98}. Note that $\gamma_j=0$ or 1 will
exclude or include the entire block of the coefficients $\beta_{jl}$'s,
and within the nonvanishing block, the coefficients follow the zero-mean Gaussian priors
with variances decaying at the rates $\tau_l^2$'s, which may be useful to produce smooth estimates of the functions.
In \cite{SFK12}, a different type of spike-and-slab prior was considered.
Specifically, each coefficient block is represented as the product of a scalar
with normal-mixture-inverse-gamma prior and a vector whose entries follow the bivariate mixture normal priors
with a constant variance.

A variety of priors can be assumed on $\sigma^2$. For convenience,
we consider the inverse $\chi^2$ prior, i.e.,
\begin{equation}\label{fmb:3}
1/\sigma^2\sim \chi_\nu^2,
\end{equation}
where $\nu$ is a fixed hyperparameter. Other priors such as the noninformative priors or the inverse Gamma priors can also be applied.
All the results developed in this paper can be extended to such situations without substantial difficulty.

In high-dimensional inference, it is commonly believed that only a small subset of covariates contribute substantially to the model.
Treating this as prior information, the models with larger sizes should be assigned with zero prior probabilities,
and only the models with smaller sizes should be assigned with positive weights. We call this a size-control prior on the model space. Namely,
we choose the prior on $\boldsymbol{\gamma}$ as
\begin{equation}\label{fbm:5}
p(\boldsymbol{\gamma})=\left\{\begin{array}{cc} \pi_{\boldsymbol{\gamma}},&\,\,\textrm{if $|\boldsymbol{\gamma}|\le t_n$,}\\
                                    0, &\textrm{otherwise,}\end{array}\right.
\end{equation}
where $\pi_{\boldsymbol{\gamma}}$ for $|\boldsymbol{\gamma}|\le t_n$ are fixed positive numbers, and $t_n\in (0,n)$ is an integer-valued hyperparameter controlling the sizes of the candidate models. We name the set of models whose sizes are not exceeding $t_n$
as the \emph{target model space}.

Denote $\textbf{D}_n=\{Y_i,X_{1i},\ldots,X_{pi}\}_{i=1}^n$ to be the full data variable. It can be shown by direct calculations that,
based on the above hierarchical model (\ref{fbm:1})-(\ref{fbm:5}),
the joint posterior distribution of $(\boldsymbol{\gamma},\boldsymbol{\beta}_{\boldsymbol{\gamma}},\sigma^2)$ is
\begin{eqnarray}\label{post:dist}
&&p(\boldsymbol{\gamma},\boldsymbol{\beta}_{\boldsymbol{\gamma}},\sigma^2|\textbf{D}_n)\nonumber\\
&\propto&p(\textbf{Y}|\boldsymbol{\gamma},\boldsymbol{\beta}_{\boldsymbol{\gamma}},\sigma^2,\textbf{X}_j\textrm{'s})
p(\boldsymbol{\beta}_{\boldsymbol{\gamma}}|\boldsymbol{\gamma},\sigma^2)p(\boldsymbol{\gamma})p(\sigma^2)\nonumber\\
&\propto&\sigma^{-(n+\nu+2)}
\exp\left(-\frac{\|\textbf{Y}-\textbf{Z}_{\boldsymbol{\gamma}}\boldsymbol{\beta}_{\boldsymbol{\gamma}}\|^2+1}{2\sigma^2}\right)
p(\boldsymbol{\gamma})
\prod\limits_{j\in\boldsymbol{\gamma}}\left(\sqrt{2\pi c_j}\sigma\right)^{-m}\det(\textbf{T}_m)^{-1/2}
\exp\left(-\frac{\boldsymbol{\beta}_j^T\textbf{T}_m^{-1}\boldsymbol{\beta}_j}{2c_j\sigma^2}\right),\nonumber\\
\end{eqnarray}
where $\|\cdot\|$ denotes the Euclidean norm of a vector, and $\textbf{T}_m=\textrm{diag}(\tau_1^2,\ldots,\tau_m^2)$.
Integrating out $\boldsymbol{\beta}_{\boldsymbol{\gamma}}$ and $\sigma^2$ in (\ref{post:dist}),
it can be checked that the marginal posterior of $\boldsymbol{\gamma}$ is
\begin{equation}\label{p:gamma}
p(\boldsymbol{\gamma}|\textbf{D}_n)\propto \det(\textbf{W}_{\boldsymbol{\gamma}})^{-1/2}p(\boldsymbol{\gamma})
\left(1+\textbf{Y}^T
(\textbf{I}_n-\textbf{Z}_{\boldsymbol{\gamma}}\textbf{U}_{\boldsymbol{\gamma}}^{-1}
\textbf{Z}_{\boldsymbol{\gamma}}^T)\textbf{Y}\right)^{-(n+\nu)/2},
\end{equation}
where $\textbf{W}_{\boldsymbol{\gamma}}=\boldsymbol{\Sigma}_{\boldsymbol{\gamma}}^{1/2}\textbf{U}_{\boldsymbol{\gamma}}
\boldsymbol{\Sigma}_{\boldsymbol{\gamma}}^{1/2}$,
$\textbf{U}_{\boldsymbol{\gamma}}=\boldsymbol{\Sigma}_{\boldsymbol{\gamma}}^{-1}
+\textbf{Z}_{\boldsymbol{\gamma}}^T\textbf{Z}_{\boldsymbol{\gamma}}$,
and $\boldsymbol{\Sigma}_{\boldsymbol{\gamma}}=\textrm{diag}(c_j \textbf{T}_m, j\in\boldsymbol{\gamma})$
is the $m|\boldsymbol{\gamma}|$ by $m|\boldsymbol{\gamma}|$ diagonal matrix
with diagonal elements $(c_j\tau_1^2,\ldots,c_j\tau_m^2)$ for $j\in\boldsymbol{\gamma}$.
We adopt the convention $\textbf{Z}_{\emptyset}=0$ and
$\boldsymbol{\Sigma}_\emptyset=\textbf{U}_\emptyset=\textbf{W}_\emptyset=1$, where $\emptyset$ means the null model,
i.e., the vector $\boldsymbol{\gamma}$ with all elements being zero. So (\ref{p:gamma}) is meaningful
for $\boldsymbol{\gamma}=\emptyset$.
The selected model $\widehat{\boldsymbol{\gamma}}$ is defined to be the one that maximizes $p(\boldsymbol{\gamma}|\textbf{D}_n)$.
Clearly, $\widehat{\boldsymbol{\gamma}}$ belongs to the target model space since any model outside the target space
has zero posterior probability.

\section{Main Results}\label{sec:main:results}

Suppose the data are truly drawn from the model $Y_i=\sum_{j\in\boldsymbol{\gamma}^0} f_j^0(X_{ji})+\epsilon_i$,
where $\epsilon_i$'s $\overset{iid}{\sim} N(0,\sigma_0^2)$ are independent of $X_{ji}$'s,
$\sigma_0^2$ is a fixed (unknown) positive number,
and $f_j^0\in\mathcal{H}_j$ for $j\in\boldsymbol{\gamma}^0$.
Recall that $\boldsymbol{\gamma}^0$ is a $p$-dimensional $0\backslash1$-vector
representing the true model, and $s_n=|\boldsymbol{\gamma}^0|$ denotes its size.
We only consider $s_n>0$, i.e., the true model is nonnull.
Any $f_j^0$ for $j\in\boldsymbol{\gamma}^0$ admits the Fourier expansion $f_j^0=\sum_{l=1}^\infty\beta_{jl}^0\varphi_{jl}$,
where $\beta_{jl}^0$'s represent the ``true" unknown Fourier coefficients of $f_j^0$.
When $m$ is sufficiently large, $f_j^0$ is approximated by its partial Fourier series,
that is, $f_j^0\approx \sum_{l=1}^m \beta_{jl}^0\varphi_{jl}$.
To insure that such partial Fourier series is a valid approximation,
we assume a uniform error rate on the tails of the Fourier series.
Specifically, we assume that there are some positive constants $a>1$ and $C_{\beta}$ such that
\begin{equation}\label{TFS:error:rate}
\max\limits_{m\ge1}\max\limits_{j\in\boldsymbol{\gamma}^0}m^a\sum\limits_{l=m+1}^\infty |\beta_{jl}^0|^2\le C_\beta.
\end{equation}
It is easy to see that (\ref{TFS:error:rate}) is equivalent to
$\max\limits_{j\in\boldsymbol{\gamma}^0}\|f_j^0-\sum\limits_{l=1}^m\beta_{jl}^0\varphi_{jl}\|_j^2=O(m^{-a})$, uniformly for $m\ge1$.
That is, the errors of the partial Fourier series of the nonzero $f_j^0$'s uniformly decrease to zero at rate $m^{-a}$.
For instance, when $f_j^0$'s uniformly belong to the Sobolev's ellipsoid of order $a/2$, i.e.,
$\max_{j\in\boldsymbol{\gamma}^0}\sum_{l=1}^\infty l^a|\beta_{jl}^0|^2<\infty$,
for some constant $a>0$, it can be checked that (\ref{TFS:error:rate}) holds.
Namely, $a$ measures the degree of smoothness of the nonzero functions. A larger $a$
implies that the nonzero functions are more smooth.

Define $l_n=\sum_{j\in\boldsymbol{\gamma}^0}\|f_j^0\|_j^2$
and $\theta_n=\min_{j\in\boldsymbol{\gamma}^0}\|f_j^0\|_j$.
Define $\textbf{P}_{\boldsymbol{\gamma}}=
\textbf{Z}_{\boldsymbol{\gamma}}(\textbf{Z}_{\boldsymbol{\gamma}}^T\textbf{Z}_{\boldsymbol{\gamma}})^{-1}\textbf{Z}_{\boldsymbol{\gamma}}^T$
to be the $n$ by $n$ projection (or smoothing) matrix corresponding to $\boldsymbol{\gamma}$.
We adopt the convention $\textbf{P}_\emptyset=0$.
Let $\lambda_{-}(\textbf{A})$ and $\lambda_{+}(\textbf{A})$ be the minimal and maximal eigenvalues of matrix $\textbf{A}$.
Suppose the truncation parameter $m$ is chosen within the range $[m_1,m_2]$, where
$m_1=m_{1n}$, $m_2=m_{2n}$ with $m_1\le m_2$ are positive sequences
approaching infinity as $n\rightarrow\infty$.
The variance-control parameters $c_j$'s are chosen within $[\underline{\phi}_n,\bar{\phi}_n]$
for some positive sequences $\underline{\phi}_n$, $\bar{\phi}_n$.

\subsection{Well Specified Target Model Space}\label{sec:wstms}
\renewcommand{\theAssumption}{A.\arabic{Assumption}}
\setcounter{Assumption}{0}
In this section we present our first theorem on posterior consistency of our model selection procedure.
We consider the situation $t_n\ge s_n$, that is, the hyperparameter $t_n$ is correctly
specified as being no less than the size of the true model.
Thus, the true model is among our target model space, for which we say that the target model space is well specified.
We will present a set of sufficient conditions and show that under these conditions,
the posterior probability of the true model converges to one in probability. Thus,
the selection procedure asymptotically yields the true model.

Define $S_1(t_n)=\{\boldsymbol{\gamma}|\boldsymbol{\gamma}^0\subset\boldsymbol{\gamma},
\boldsymbol{\gamma}\neq\boldsymbol{\gamma}^0, |\boldsymbol{\gamma}|\le t_n\}$ and
$S_2(t_n)=\{\boldsymbol{\gamma}|\boldsymbol{\gamma}^0\,\, \textrm{is not nested in}\,\,
\boldsymbol{\gamma}, |\boldsymbol{\gamma}|\le t_n\}$. It
is clear that $S_1(t_n)$ and $S_2(t_n)$ are disjoint, and $S(t_n)$ defined by $S(t_n)=S_1(t_n)\bigcup S_2(t_n)
\bigcup\{\boldsymbol{\gamma}^0\}$ is
the class of all models with size not exceeding $t_n$, i.e., the target model space.
We first list some conditions that are used to show our theorem.

\begin{Assumption}\label{A1}
There exists a positive constant $c_0$ such that, as $n\rightarrow\infty$, with probability approaching one
\[
1/c_0\le \min\limits_{m\in[m_1,m_2]}\min\limits_{\boldsymbol{\gamma}\in
S_2(t_n)}\lambda_{-}\left(\frac{1}{n}\textbf{Z}_{\boldsymbol{\gamma}^0\backslash\boldsymbol{\gamma}}^T
(\textbf{I}_n-\textbf{P}_{\boldsymbol{\gamma}})\textbf{Z}_{\boldsymbol{\gamma}^0\backslash\boldsymbol{\gamma}}\right)\le
\max\limits_{m\in[m_1,m_2]}\max\limits_{\boldsymbol{\gamma}\in S_2(t_n)}\lambda_{+}
\left(\frac{1}{n}\textbf{Z}_{\boldsymbol{\gamma}^0\backslash\boldsymbol{\gamma}}^T
\textbf{Z}_{\boldsymbol{\gamma}^0\backslash\boldsymbol{\gamma}}\right)\le c_0,
\]
and
\[
\min\limits_{m\in[m_1,m_2]}\min\limits_{\boldsymbol{\gamma}\in
S_1(t_n)}\lambda_-\left(\frac{1}{n}\textbf{Z}_{\boldsymbol{\gamma}\backslash\boldsymbol{\gamma}^0}^T
(\textbf{I}_n-\textbf{P}_{\boldsymbol{\gamma}^0})\textbf{Z}_{\boldsymbol{\gamma}\backslash\boldsymbol{\gamma}^0}\right)\ge 1/c_0.
\]
\end{Assumption}

\begin{Assumption}\label{A2}
$\sup\limits_n\max\limits_{\boldsymbol{\gamma}\in S(t_n)}\frac{p(\boldsymbol{\gamma})}{p(\boldsymbol{\gamma}^0)}<\infty$.
\end{Assumption}

\begin{Assumption}\label{A3'} There exists a positive sequence $\{h_m,m\ge1\}$
such that, as $m,m_1,m_2\rightarrow\infty$, $h_m\rightarrow\infty$,
$m^{-a}h_m$ decreasingly converges to zero, $mh_m$ increasingly converges to $\infty$,
and $\sum_{m_1\le m\le m_2}1/h_m=o(1)$. Furthermore,
the sequences $m_1,m_2,h_m,s_n,t_n,\theta_n,l_n,\underline{\phi}_n,\bar{\phi}_n$ satisfy
\begin{enumerate}[(1).]
\item $m_2h_{m_2}s_n=o(n\min\{1,\theta_n^2\})$ and
$m_1^{-a}h_{m_1}s_n^2=o(\min\{1,n^{-1}m_1\log(\underline{\phi}_n),\theta_n^2,\theta_n^4\})$;
\item $t_n\ge s_n$ and $t_n\log{p}=o(n\log(1+\min\{1,\theta_n^2\}))$;
\item $l_n=O(\underline{\phi}_n\tau_{m_2}^2)$ and $\log{p}=o(m_1\log{(n\underline{\phi}_n\tau_{m_2}^2)})$;
\item\label{A3':special} $m_2 s_n\log(1+n\bar{\phi}_n)=o(n\log(1+\min\{1,\theta_n^2\}))$.
\end{enumerate}
\end{Assumption}

In the following proposition we show that Assumption \ref{A1} holds under suitable
dependence assumption among the predictors $X_j$'s. To clearly describe this assumption,
let $\{X_j\}_{j=1}^\infty$ be a stationary sequence taking values in $[0,1]$, and define its
$\rho$-mixing coefficient to be
$\rho(|j-j'|)=\sup_{f,g}|E\{f(X_j)g(X_{j'})\}-E\{f(X_j)\}E\{g(X_{j'})\}|$,
where the supremum is taken over the measurable functions $f$ and $g$ with $E\{f(X_j)^2\}=E\{g(X_{j'})^2\}=1$.
Ideally we assume that the predictors
$X_1,\ldots,X_p$ in model (\ref{true:model}) are simply the first $p$ elements of $\{X_j\}_{j=1}^\infty$.
\begin{Proposition}\label{prop:1}
Suppose $\sum_{r=1}^\infty \rho(r)<1/2$, $t_n^2 m_2^2\log{p}=o(n)$,
and $\max_{1\le j\le p}\sup_{l\ge1}\|\varphi_{jl}\|_{\sup}<\infty$,
where $\|\cdot\|_{\sup}$ denotes the supnorm.
Then there is a constant $c_0>0$ such that
with probability approaching one
\begin{equation}\label{prop1:eq}
c_0^{-1}\le \min\limits_{m\in[m_1,m_2]}\min\limits_{0<|\boldsymbol{\gamma}|\le 2t_n}
\lambda_-\left(\frac{1}{n}\textbf{Z}_{\boldsymbol{\gamma}}^T\textbf{Z}_{\boldsymbol{\gamma}}\right)\le
\max\limits_{m\in[m_1,m_2]}\max\limits_{0<|\boldsymbol{\gamma}|\le 2t_n}
\lambda_+\left(\frac{1}{n}\textbf{Z}_{\boldsymbol{\gamma}}^T\textbf{Z}_{\boldsymbol{\gamma}}\right)\le c_0.
\end{equation}
Furthermore, (\ref{prop1:eq}) implies Assumption \ref{A1}.
\end{Proposition}

Assumption \ref{A2} holds if we choose $p(\boldsymbol{\gamma})$ to be constant for all $|\boldsymbol{\gamma}|\le t_n$,
i.e., we adopt an indifference prior over the target model space.
To see when Assumption \ref{A3'} holds, we look at a special example.
We choose $\tau_l^2=l^{-5}$ for $l\ge 1$.
Suppose $\log{p}\propto n^k$ for $0<k<1$, $s_n\propto 1$, $\psi_n\propto 1$, and the smoothness parameter $a=4$.
Choose $t_n\propto 1$, $m_1=\zeta n^{1/5}+c_{1n}$ and $m_2=\zeta n^{1/5}+c_{2n}$, where $\zeta>0$ is constant,
$c_{1n}=o((\log{n})^r)$, $c_{2n}=o((\log{n})^r)$, $c_{1n}\le c_{2n}$, and $r>0$ is a constant.
Note that such choice of $m_1$ and $m_2$ yields minimax error rate in univariate regression.
Let $h_m=(\log{m})^r$ for $m\ge1$.
Ideally we suppose that the selected $t_n$ is greater than $s_n$.
Choose $\underline{\phi}_n$ and $\bar{\phi}_n$ as $\log(\underline{\phi}_n)\propto n^{k_1}$
and $\log(\bar{\phi}_n)\propto n^{k_2}$ with
$\max\{0,k-1/5\}<k_1<k_2<4/5$. In this simple situation,
it can be directly verified that Assumption \ref{A3'} holds.
Furthermore, Proposition \ref{prop:1} says that to satisfy Assumption \ref{A1},
an additional sufficient condition is $t_n^2m_2^2\log{p}=o(n)$, which implies $k<3/5$.
Therefore, the dimension $p$ cannot exceed the order $\exp(O(n^{3/5}))$,
which coincides with the finding by \cite{RLLW09}.

\begin{Theorem}\label{main:thm1}
Under Assumptions \ref{A1} to \ref{A3'}, as $n\rightarrow\infty$,
\begin{equation}\label{pstcon}
\min\limits_{m\in[m_1,m_2]}\inf\limits_{\underline{\phi}_n\le c_1,\ldots,c_p\le \bar{\phi}_n} p(\boldsymbol{\gamma}^0|\textbf{D}_n)\rightarrow 1,\,\,\textrm{in probability}.
\end{equation}
\end{Theorem}

Theorem \ref{main:thm1} says that under mild conditions the posterior probability of the true model converges to one in probability.
This means, with probability approaching one, our Bayesian method selects the true model,
which guarantees the validity of the proposed approach. Here, convergence holds uniformly over $c_j$'s $\in[\underline{\phi}_n,\bar{\phi}_n]$
and $m\in[m_1,m_2]$. This means, the selection result is insensitive to the choice of $c_j$'s and $m$
when they belong to suitable ranges. It is well known that choosing the truncation parameter $m$ is a practically
difficult problem in nonparametrics; see \cite{RLLW09,FFS11}. Therefore, a method that is insensitive to the choice
of the truncation parameter within certain range will be highly useful. In Theorem \ref{main:thm1} we theoretically
show that the proposed Bayesian selection method is among the ones which provide insensitive selection results.
On the other hand, we also show that our method is insensitive to the choice of the variance-control parameters $c_j$'s.
This is both theoretically and practically useful since it allows us to place an additional prior, such as the $g$-priors, over the $c_j$'s
while preserving the desired posterior model consistency; see Section \ref{g:prior}.
By slightly modifying the assumptions, it is possible
to show that (\ref{pstcon}) actually holds uniformly for $t_n$
within some range, as established by \cite{SL13} in the linear model setting.
That is, posterior model consistency is also insensitive to the choice of $t_n$.
We ignore this part since in our nonparametric models with $g$-priors,
insensitivity of the truncation parameter $m$ and the variance control parameters $c_j$'s
should be paid more attention. This may also simplify the statements so that the results become more readable.
To the best of our knowledge, Theorem \ref{main:thm1} is the first theoretical result
showing the validity of the Bayesian methods in function component selection in ultrahigh-dimensional settings.

\subsection{Misspecified Target Model Space}\label{sec:mtms}

\renewcommand{\theAssumption}{B.\arabic{Assumption}}
\setcounter{Assumption}{0}

In this section, we investigate the case $0<t_n<s_n$, that is, $t_n$ is misspecified as being smaller than the size of the true model.
Therefore, the true model is outside the target model space, for which we say that the target model space is misspecified.
We conclude that in this false setting
the selected model is still not ``bad" because it can be asymptotically nested in the true model,
uniformly for the choice of $m$ and $c_j$'s.

Define $T_0(t_n)=\{\boldsymbol{\gamma}| 0\le |\boldsymbol{\gamma}|\le t_n, \boldsymbol{\gamma}\subset\boldsymbol{\gamma}^0\}$,
$T_1(t_n)=\{\boldsymbol{\gamma}| 0<|\boldsymbol{\gamma}|\le t_n,\boldsymbol{\gamma}\cap\boldsymbol{\gamma}^0\neq\emptyset,
\,\,\textrm{$\boldsymbol{\gamma}$ is not nested in $\boldsymbol{\gamma}^0$}\}$,
and $T_2(t_n)=\{\boldsymbol{\gamma}| 0<|\boldsymbol{\gamma}|\le t_n, \boldsymbol{\gamma}\cap\boldsymbol{\gamma}^0=\emptyset\}$.
It is easy to see that $T_0(t_n), T_1(t_n), T_2(t_n)$ are disjoint and
$T(t_n)=T_0(t_n)\cup T_1(t_n)\cup T_2(t_n)$ is exactly the target model space,
i.e., the class of
$\boldsymbol{\gamma}$ with $|\boldsymbol{\gamma}|\le t_n$. Throughout this section,
we make the following assumptions.

\begin{Assumption}\label{B1}
There exist a positive constant $d_0$ and a positive sequence $\rho_n$ such that, when $n\rightarrow\infty$, with probability approaching one,
\begin{equation}\label{B1:eq}
d_0^{-1}\le \min\limits_{m\in[m_1,m_2]}\min
\limits_{0<|\boldsymbol{\gamma}|\le s_n}\lambda_{-}\left(\frac{1}{n}\textbf{Z}_{\boldsymbol{\gamma}}^T
\textbf{Z}_{\boldsymbol{\gamma}}\right)\le
\max\limits_{m\in[m_1,m_2]}\max\limits_{0<|\boldsymbol{\gamma}|\le s_n}
\lambda_{+}\left(\frac{1}{n}\textbf{Z}_{\boldsymbol{\gamma}}^T \textbf{Z}_{\boldsymbol{\gamma}}\right)\le
d_0,\,\,\textrm{and}\,\,
\end{equation}
\begin{equation}\label{B1:eq2}
\max\limits_{m\in[m_1,m_2]}\max\limits_{\boldsymbol{\gamma}\in T(s_n-1)}\lambda_{+}\left(\textbf{Z}_{\boldsymbol{\gamma}^0\backslash\boldsymbol{\gamma}}^T
\textbf{P}_{\boldsymbol{\gamma}} \textbf{Z}_{\boldsymbol{\gamma}^0\backslash\boldsymbol{\gamma}}\right)\le \rho_n.
\end{equation}
\end{Assumption}

\begin{Assumption}\label{B2}
$\sup\limits_n
\max\limits_{\boldsymbol{\gamma},\boldsymbol{\gamma}'\in T(t_n)}\frac{p(\boldsymbol{\gamma})}{p(\boldsymbol{\gamma}')}<\infty$.
\end{Assumption}

\begin{Assumption}\label{B3'} There exists a positive sequence $\{h_m,m\ge1\}$
such that, as $m,m_1,m_2\rightarrow\infty$, $h_m\rightarrow\infty$,
$m^{-a}h_m$ decreasingly converges to zero, $mh_m$ increasingly converges to $\infty$,
and $\sum_{m_1\le m\le m_2}1/h_m=o(1)$. Furthermore,
the sequences $m_1,m_2,h_m,s_n,\theta_n,l_n,\underline{\phi}_n$ satisfy
\begin{enumerate}[(1).]
\item $m_2h_{m_2}s_n=o(n\min\{1,\theta_n^2\})$ and $m_1^{-a}h_{m_1}s_n^2
=o(\min\{1,n^{-1}m_1\log(\underline{\phi}_n),\theta_n^2\})$;
\item $l_n=O(\underline{\phi}_n\tau_{m_2}^2)$;
\item $\max\{\rho_n,s_n^2\log{p}\}=o(\min\{n,m_1\log(n\underline{\phi}_n\tau_{m_2}^2)\})$.
\end{enumerate}
\end{Assumption}

The following result presents a situation in which Assumption \ref{B1} holds. For technical convenience,
we require the predictors to be independent. It is conjectured that this result may hold in more general settings.

\begin{Proposition}\label{prop:2}
Suppose that the predictors $X_1,\ldots,X_p$ are $iid$ random variables taking values in $[0,1]$, $s_n^2 m_2^2 \log{p}=o(n)$,
and $\max_{1\le j\le p}\sup_{l\ge 1}\|\varphi_{jl}\|_{\sup}<\infty$.
Then Assumption \ref{B1} holds with $\rho_n\propto m_2 s_n^2\log{p}$.
\end{Proposition}

Assumption \ref{B2} holds when we place indifference prior over the models with size not exceeding $t_n$.
To examine Assumption \ref{B3'}, we again look at a special case. For simplicity, we suppose the setting of
Proposition \ref{prop:2} holds. Choose $\tau_l^2=l^{-5}$ for $l\ge 1$.
Suppose $\log{p}=n^k$ for $k\in(0,4/5)$, $s_n\propto1$, $\theta_n\propto1$, $l_n\propto1$, and $a=4$.
Let $m_1=\zeta n^{1/5}+c_{1n}$ and $m_2=\zeta n^{1/5}+c_{2n}$, where $\zeta>0$ is constant,
$c_{1n}=o((\log{n})^r)$, $c_{2n}=o((\log{n})^r)$, $c_{1n}\le c_{2n}$, and $r>0$ is a constant. Let $h_m=(\log{m})^r$.
Choose $\log(\underline{\phi}_n)=n^{k_1}$ with $k_1>k$. It can be shown in this special situation that
Assumption \ref{B3'} holds. Furthermore, the condition $s_n^2m_2^2\log{p}=o(n)$ (see Proposition \ref{prop:2})
implies $k<3/5$. So the growth rate of $p$ is again not exceeding $\exp(O(n^{3/5}))$.

\begin{Theorem}\label{main:thm2}
Suppose $0<t_n<s_n$ and Assumptions \ref{B1}--\ref{B3'} are satisfied.
\begin{enumerate}[(i).]
\item\label{thm2:i} As $n\rightarrow\infty$,
$\max\limits_{m\in[m_1,m_2]}
\sup\limits_{\underline{\phi}_n\le c_1,\ldots, c_p\le \bar{\phi}_n}
\frac{\max\limits_{\boldsymbol{\gamma}\in T_1(t_n)\cup T_2(t_n)}p(\boldsymbol{\gamma}|\textbf{D}_n)}
{\max\limits_{\boldsymbol{\gamma}\in T_0(t_n)}p(\boldsymbol{\gamma}|\textbf{D}_n)}\rightarrow0, \textrm{in probability}.
$

\item\label{thm2:ii}
Furthermore, suppose Assumption \ref{A3'} (\ref{A3':special}) is satisfied, and
there is $\boldsymbol{\gamma}\in T_0(t_n)\backslash\{\emptyset\}$ and a constant $b_0>0$ such that for all $m\in[m_1,m_2]$,
\begin{equation}\label{inform:model:f0}
\sum_{j\in\boldsymbol{\gamma}^0\backslash\boldsymbol{\gamma}}\|f_j^0\|_j^2\le b_0
\sum_{j\in\boldsymbol{\gamma}}\|f_j^0\|_j^2.
\end{equation}
Then, as $n\rightarrow\infty$,
$\max\limits_{m\in[m_1,m_2]}
\sup\limits_{\underline{\phi}_n\le c_1,\ldots, c_p\le \bar{\phi}_n}
\frac{p(\emptyset|\textbf{D}_n)}
{p(\boldsymbol{\gamma}|\textbf{D}_n)}\rightarrow0, \textrm{in probability}$.
\end{enumerate}
\end{Theorem}

When the hyperparameter $t_n$ is incorrectly specified as being smaller than the size of the true model,
the selected model $\widehat{\boldsymbol{\gamma}}$
cannot be the true model since necessarily $|\widehat{\boldsymbol{\gamma}}|<s_n$.
Theorem \ref{main:thm2} (\ref{thm2:i}) shows that in this false setting,
$\widehat{\boldsymbol{\gamma}}$ can be asymptotically nested to the true model with probability approaching one.
This means, as $n$ approaches infinity, all the selected components are the significant ones
which ought to be included in the model.
Here, the result holds uniformly for $m$ and $c_j$'
s within certain ranges,
showing insensitivity of the choice of these hyperparameters.
To the best of our knowledge, Theorem \ref{main:thm2} is the first theoretical examination of
the function selection approach when the model space is misspecified.

We should mention that in Theorem \ref{main:thm2} (\ref{thm2:i}),
it is possible that $\widehat{\boldsymbol{\gamma}}=\emptyset$
since $\emptyset$ is a natural subset of $\boldsymbol{\gamma}^0$. When $\boldsymbol{\gamma}^0$ is nonull,
we expect $\widehat{\boldsymbol{\gamma}}$ to include some significant variables.
Theorem \ref{main:thm2} (\ref{thm2:ii}) says that this is possible if there exists
a nonnull model that can be separated from the null model.
Explicitly,
the condition (\ref{inform:model:f0})
says that the functions $\{f_j^0, j\in\boldsymbol{\gamma}\}$ dominate
the functions $\{f_j^0, j\in\boldsymbol{\gamma}^0\backslash\boldsymbol{\gamma}\}$,
in terms of the corresponding norms $\|\cdot\|_j$'s.
This can be interpreted as that the model $\boldsymbol{\gamma}$
includes a larger amount of the information from the true model than its completion
$\boldsymbol{\gamma}^0\backslash\boldsymbol{\gamma}$.
Theorem \ref{main:thm2} (\ref{thm2:ii}) says that under this condition, with probability approaching one,
$\boldsymbol{\gamma}$ is more preferred than the null. Therefore,
$\widehat{\boldsymbol{\gamma}}$ is asymptotically nonnull.

\subsection{Basis Functions}

The proposed approach relies on a proper set of orthonormal basis functions $\{\varphi_{jl},l\ge1\}$ in $\mathcal{H}_j$
under the inner product $\langle\cdot,\cdot\rangle_j$.
In this section we briefly describe how to empirically construct such functions.

Suppose for each $j=1,\ldots,p$, $\{B_{jl}, l\ge0\}$ form a set of basis functions in $L^2[0,1]$.
Without loss of generality,
assume $B_{j0}$ to be the constant function.
For example, in empirical study we can choose the trigonometric polynomial basis, i.e.,
$B_{j0}=1$, $B_{jl}(x)=\sqrt{2}\cos(2\pi kx)$ if $l=2k-1$, and $B_{jl}(x)=\sqrt{2}\sin(2\pi kx)$ if $l=2k$,
for integer $k\ge1$. Other choices such as Legendre's polynomial basis can also be used; see \cite{CH53}.
We may choose a sufficiently large integer $M$ with $M<n$.
For $j=1,\ldots,p$ and $1\le l\le M$, define $\tilde{B}_{jl}$ to be a real-valued function whose value at $X_{ji}$
is $\tilde{B}_{jl}(X_{ji})=B_{jl}(X_{ji})-\frac{1}{n}\sum_{i=1}^n B_{jl}(X_{ji})$.
Define $\textbf{W}_{ji}=(\tilde{B}_{j1}(X_{ji}),\ldots,\tilde{B}_{jM}(X_{ji}))^T$,
and $\hat{\boldsymbol{\Sigma}}_j=\frac{1}{n}\sum_{i=1}^n \textbf{W}_{ji}\textbf{W}_{ji}^T$.
Let $\textbf{A}_j$ be an $M$ by $M$ invertible matrix such that
$\textbf{A}_j^T\hat{\boldsymbol{\Sigma}}_j\textbf{A}_j=\textbf{I}_M$.
Write $\textbf{A}_j=(\textbf{a}_{j1},\ldots,\textbf{a}_{jM})$, where $\textbf{a}_{jl}$ is the $l$-th
column, an $M$-vector. Then define $\varphi_{jl}$ as a real-valued function whose value at $X_{ji}$
is $\varphi_{jl}(X_{ji})=\textbf{a}_{jl}^T\textbf{W}_{ji}$, for $j=1,\ldots,p$ and $l=1,\ldots,M$.
In the simplest situation where $X_{ji}$'s are \textit{iid} uniform in [0,1], for $j=1,\ldots,p$,
it can be seen that $\hat{\boldsymbol{\Sigma}}_j\approx \textbf{I}_M$, for which we can choose
$\textbf{A}_j=\textbf{I}_M$, leading to $\varphi_{jl}=\tilde{B}_{jl}$ for $l=1,\ldots,M$.

Next we heuristically show that the functions $\varphi_{jl}$'s approximately form an orthonormal basis.
By the law of large numbers,
$E\{\varphi_{jl}(X_j)\}\approx\frac{1}{n}\sum_{i=1}^n\varphi_{jl}(X_{ji})=\frac{1}{n}\textbf{a}_{jl}^T\sum_{i=1}^n\textbf{W}_{ji}=0$,
and $E\{\varphi_{jl}(X_j)\varphi_{jl'}(X_j)\}\approx
\frac{1}{n}\sum_{i=1}^n\varphi_{jl}(X_{ji})\varphi_{jl'}(X_{ji})=\textbf{a}_{jl}^T\hat{\boldsymbol{\Sigma}}\textbf{a}_{jl'}
=\delta_{ll'}$, for $l,l'=1,\ldots,M$,
where $\delta_{ll'}=1$ if $l=l'$, and zero otherwise. Thus, $\{\varphi_{jl},l=1,\ldots,M\}$ approximately form
an orthonormal system.
Furthermore, any $f_j\in\mathcal{H}_j$ admits the approximate expansion
$f_j(X_{ji})\approx\sum_{l=0}^M\tilde{\beta}_{jl}B_{jl}(X_{ji})$ for some real sequence $\tilde{\beta}_{jl}$.
So $0=E\{f_j(X_j)\}\approx\frac{1}{n}\sum_{i=1}^n f_j(X_{ji})\approx\frac{1}{n}\sum_{i=1}^n\sum_{l=0}^M\tilde{\beta}_{jl}B_{jl}(X_{ji})$.
Therefore, we get that
$f_j(X_{ji})\approx\sum_{l=0}^M\tilde{\beta}_{jl}B_{jl}(X_{ji})-
\frac{1}{n}\sum_{i=1}^n\sum_{l=0}^M\tilde{\beta}_{jl}B_{jl}(X_{ji})=\sum_{l=1}^M\tilde{\beta}_{jl}\tilde{B}_{jl}(X_{ji})
=\tilde{\boldsymbol{\beta}}_j^T\textbf{W}_{ji}=
(\textbf{A}_j^{-1}\tilde{\boldsymbol{\beta}}_j)^T(\varphi_{j1}(X_{ji}),\ldots,\varphi_{jM}(X_{ji}))^T$,
where $\tilde{\boldsymbol{\beta}}_j=(\tilde{\beta}_{j1},\ldots,\tilde{\beta}_{jM})^T$.
This means that the function $f_j$ can be approximately represented by the $\varphi_{jl}$'s for $l=1,\ldots,M$.
Consequently, $\{\varphi_{jl},l=1,\ldots,M\}$ approximately form an orthonormal basis in $\mathcal{H}_j$
given that $M$ is large enough.

\subsection{Mixtures of $g$-prior}\label{g:prior}

The results in Sections \ref{sec:wstms} and \ref{sec:mtms} can also be extended to the $g$-prior setting.
Suppose $c_j=c$ for $j=1,\ldots,p$. We assume $c$ to have prior density $g(c)$, a function of positive values
over $(0,\infty)$ satisfying $\int_0^\infty g(c)dc=1$, i.e., $g$ is a proper prior.
Then (\ref{p:gamma}) is actually $p(\boldsymbol{\gamma}|c,\textbf{D}_n)$.
The posterior distribution of $\boldsymbol{\gamma}$ is
therefore $p_g(\boldsymbol{\gamma}|\textbf{D}_n)=\int_0^\infty p(\boldsymbol{\gamma}|c,\textbf{D}_n)g(c)dc$,
with the subscript $g$ emphasizing the $g$-prior situation. Then we have the following results
parallel to Theorems \ref{main:thm1} and \ref{main:thm2}.
The interpretations are similar to those for Theorems \ref{main:thm1} and \ref{main:thm2}.
Their proofs are similar to those in \cite{SL13}, and thus are omitted.

\begin{Theorem}\label{main:gp:thm1}
Suppose Assumptions \ref{A1}--\ref{A3'} are satisfied. Furthermore, $g$ is proper and,
as $n\rightarrow\infty$, $\int_0^{\underline{\phi}_n}g(c)dc=o(1)$
and $\int_{\bar{\phi}_n}^\infty g(c)dc=o(1)$. Then as $n\rightarrow\infty$,
$\min_{m\in[m_1,m_2]} p_g(\boldsymbol{\gamma}^0|\textbf{D}_n)\rightarrow 1$, in probability.
\end{Theorem}

\begin{Theorem}\label{main:gp:thm2}
Suppose $0<t_n<s_n$.
Let Assumptions \ref{B1}--\ref{B3'} be satisfied, and $g$ be proper and supported in $[\underline{\phi}_n,\bar{\phi}_n]$, i.e.,
$g(c)=0$ if $c\notin[\underline{\phi}_n,\bar{\phi}_n]$.
\begin{enumerate}[(i).]
\item As $n\rightarrow\infty$,
$\max_{m\in[m_1,m_2]}\frac{\max_{\boldsymbol{\gamma}\in T_1(t_n)\cup T_2(t_n)}p_g(\boldsymbol{\gamma}|\textbf{D}_n)}
{\max_{\boldsymbol{\gamma}\in T_0(t_n)}p_g(\boldsymbol{\gamma}|\textbf{D}_n)}\rightarrow0$, in probability.

\item
If, in addition, Assumption \ref{A3'} (\ref{A3':special}) holds,
and there exist a $\boldsymbol{\gamma}\in T_0(t_n)\backslash\{\emptyset\}$ and a constant $b_0>0$ such that for all $m\in[m_1,m_2]$,
$\sum_{j\in\boldsymbol{\gamma}^0\backslash\boldsymbol{\gamma}}\|f_j^0\|_j^2\le b_0
\sum_{j\in\boldsymbol{\gamma}}\|f_j^0\|_j^2$. Then as $n\rightarrow\infty$,
$\max_{m\in[m_1,m_2]}\frac{p_g(\emptyset|\textbf{D}_n)}{p_g(\boldsymbol{\gamma}|\textbf{D}_n)}\rightarrow0$, in probability.
\end{enumerate}
\end{Theorem}

We propose two types of $g$-priors that generalize the Zellner-Siow prior by \cite{ZS80} and
generalize the hyper-$g$ prior by \cite{LPMCB08}. We name them as
the \emph{generalized Zellner-Siow} (GZS) prior and the \emph{generalized hyper-$g$} (GHG)
prior respectively. Let $b,\mu>0$ be fixed hyperparameters.
The GZS prior is defined to have the form
\begin{equation}\label{gzs}
g(c)=\frac{p^{b}}{\Gamma(b)}
c^{-b-1}\exp(-p^\mu/c),
\end{equation}
and the GHG prior is defined to have the form
\begin{equation}\label{ghg}
g(c)=\frac{\Gamma(p^\mu+1+b)}{\Gamma(p^\mu+1)\Gamma(b)}\cdot\frac{c^{p^\mu}}{(1+c)^{p^\mu+1+b}}.
\end{equation}

We conclude that both GZS and GHG priors can yield posterior consistency. To see this, since we assume $p\gg n$, we have
$p^{\mu/\sqrt{\log{n}}}\rightarrow\infty$ as $n\rightarrow\infty$.
Let $\underline{\phi}_n=p^{\mu/\sqrt{\log{n}}}$ and $\bar{\phi}_n=p^{\mu(\log{n})^2}$.
It can be directly examined that, as $n\rightarrow\infty$,
the GZS prior satisfies $\int_0^{\underline{\phi}_n} g(c)dc=(\Gamma(b))^{-1}\int_{p^{\mu}\underline{\phi}_n^{-1}}^\infty c^{a-1}\exp(-c)dc=o(1)$,
and $\int_{\bar{\phi}_n}^\infty g(c)dc=(\Gamma(b))^{-1}\int_0^{p^{\mu}\bar{\phi}_n^{-1}}c^{a-1}\exp(-c)dc=o(1)$;
the GHG prior satisfies
$\int_0^{\underline{\phi}_n} g(c)dc=O((p^\mu+1)^{b-1}\exp(-(p^\mu+1)/(1+\underline{\phi}_n)))=o(1)$
and $\int_{\bar{\phi}_n}^\infty g(c)dc=O((p^\mu+1)^b/(1+\bar{\phi}_n))=o(1)$.
Furthermore, suppose $\tau_l^2=l^{-5}$, $a=4$, $\log{p}=n^k$ with $k\in (0,3/5)$,
$\theta_n\propto1$, $s_n\propto1$, $l_n\propto1$, $t_n\propto1$ with $t_n\ge s_n$,
$h_m=(\log{n})^r$,
and $m_1=\zeta n^{1/5}+o((\log{n})^r)$, $m_2=\zeta n^{1/5}+o((\log{n})^r)$,
where $\zeta$ and $r$ are positive constants and $r\in (0,1/2)$.
It can be examined directly that the above
$\underline{\phi}_n$ and $\bar{\phi}_n$ satisfy the assumptions of Theorem \ref{main:gp:thm1},
implying posterior consistency of the $g$-prior methods.
Clearly, the modes of the GZS and GHG priors
are both $p^\mu/(b+1)$ which converges to infinity as $n$ inflates, yielding
consistent selection results.

\section{Computational Details}\label{sec:MCMC}

In this section, sampling details will be provided. Instead of fixing $t_n$ and $m$,
we may place priors over them to make the procedure more flexible.
Let $c_j=c$ for $j=1,\ldots,p$. Assume a $g$-prior (either GZS or GHG)
$g(c)$ for $c$, and denote the priors for $t_n$ and $m$ by $p(t_n)$ and $p(m)$ respectively.
For convenience, we consider the flat priors $p(\boldsymbol{\gamma}|t_n)=I(|\boldsymbol{\gamma}|\le t_n)$,
$p(t_n)=I(1\le t_n\le T_n)$ for some prefixed positive integer $T_n$,
and $p(m)=I(m_1\le m\le m_2)$ for some fixed positive integers $m_1$ and $m_2$. For other choices of $p(t_n)$ and $p(m)$,
the computational details in this section require corresponding modifications.

It follows by (\ref{post:dist})
that the joint posterior distribution of
$(\boldsymbol{\gamma},\boldsymbol{\beta},\sigma^2,c,m,t_n)$ is
\begin{eqnarray}\label{joint:post}
&&p(\boldsymbol{\gamma},\boldsymbol{\beta},\sigma^2,c,m,t_n|\textbf{D}_n)\nonumber\\
&\propto&
\sigma^{-(n+\nu+2)}
\exp\left(-\frac{\|\textbf{Y}-\textbf{Z}\boldsymbol{\beta}\|^2+1}{2\sigma^2}\right)(\sqrt{2\pi c}\sigma)^{-m|\boldsymbol{\gamma}|}
\det(\textbf{T}_m)^{-|\boldsymbol{\gamma}|/2}
\exp\left(-\frac{\sum_{j\in\boldsymbol{\gamma}}\boldsymbol{\beta}_j^T\textbf{T}_m^{-1}\boldsymbol{\beta}_j}
{2c\sigma^2}\right)\nonumber\\
&&\prod_{j\in-\boldsymbol{\gamma}}
\delta_{\textbf{0}}(\boldsymbol{\beta}_j)\cdot p(\boldsymbol{\gamma}|t_n)g(c)p(m)p(t_n),
\end{eqnarray}
where $\delta_{\textbf{0}}$ denotes the probability measure concentrating on the $m$-dimensional zero vector.
The MCMC sampling procedure is described as follows. For initial values, let $\boldsymbol{\gamma}^{(0)}=\emptyset$,
$\boldsymbol{\beta}^{(0)}=\textbf{0}$. Let $\sigma^2_{(0)}$ and $c^{(0)}$ be uniformly drawn from some compact subsets of $(0,\infty)$,
and $m^{(0)}$ and $t_n^{(0)}$ be drawn from $p(m)$ and $p(t_n)$ respectively. Suppose at the $q$-th iteration, we have obtained
samples $(\boldsymbol{\gamma}^{(q)},\boldsymbol{\beta}^{(q)},\sigma^2_{(q)},c^{(q)},m^{(q)},t_n^{(q)})$.

\textit{Sampling} $(\boldsymbol{\beta},m,\boldsymbol{\gamma})$. The sampling procedure proceeds in two steps.
First, one draws $m^{(q+1)}$ given $c^{(q)},\sigma^2_{(q)},t_n^{(q)}$.
Second, one draws $(\boldsymbol{\beta}^{(q+1)},\boldsymbol{\gamma}^{(q+1)})$ given
$m^{(q+1)},c^{(q)},\sigma^2_{(q)},t_n^{(q)}$.
To complete the first step, by integrating out
$\boldsymbol{\beta}$ in (\ref{joint:post}), the conditional distribution of $m$ given $c^{(q)},\sigma^2_{(q)},t_n^{(q)}$
is found by
\begin{eqnarray}\label{marginal:m}
&&p(m|c^{(q)},\sigma^2_{(q)},t_n^{(q)},\textbf{D}_n)\nonumber\\
&\propto& p(m)\sum_{|\boldsymbol{\gamma}|\le t_n^{(q)}}
(c^{(q)})^{-m|\boldsymbol{\gamma}|/2}\det(\textbf{S}^{(q)}_{\boldsymbol{\gamma},m})^{-1/2}
\exp\left(-\frac{\textbf{Y}^T(\textbf{I}_n-\textbf{Z}_{\boldsymbol{\gamma}}(\textbf{U}_{\boldsymbol{\gamma},m}^{(q)})^{-1}
\textbf{Z}_{\boldsymbol{\gamma}}^T)\textbf{Y}}{2\sigma^2_{(q)}}\right),
\end{eqnarray}
where $\textbf{U}_{\boldsymbol{\gamma},m}^{(q)}=
(c^{(q)})^{-1}\boldsymbol{\Lambda}_{\boldsymbol{\gamma},m}^{-1}
+\textbf{Z}_{\boldsymbol{\gamma}}^T\textbf{Z}_{\boldsymbol{\gamma}}$,
$\textbf{S}_{\boldsymbol{\gamma},m}^{(q)}=\boldsymbol{\Lambda}_{\boldsymbol{\gamma},m}
\textbf{U}_{\boldsymbol{\gamma},m}^{(q)}$, and
$\boldsymbol{\Lambda}_{\boldsymbol{\gamma},m}
=\textrm{diag}(\textbf{T}_m,\ldots,\textbf{T}_m)$
with $\textbf{T}_m$ therein appearing $|\boldsymbol{\gamma}|$ times.
In principle, one can
draw $m^{(q+1)}$ based on (\ref{marginal:m}). However, (\ref{marginal:m}) involves
a computationally expensive sum which is hard to handle in practice.
To overcome this difficulty, we propose an alternative (approximate) way of sampling $m$.
When $\boldsymbol{\gamma}^{(q)}=\emptyset$,
draw $m^{(q+1)}$ randomly from $[m_1,m_2]$.
When $\boldsymbol{\gamma}^{(q)}\neq\emptyset$,
from (\ref{joint:post}) the conditional distribution of $m$ given
$\boldsymbol{\gamma}^{(q)},c^{(q)},\sigma^2_{(q)},t_n^{(q)}$ is found by
\begin{eqnarray}\label{marginal:gamma}
&&p(m|\boldsymbol{\gamma}^{(q)},c^{(q)},\sigma^2_{(q)},t_n^{(q)},\textbf{D}_n)\nonumber\\
&\propto& p(m)(c^{(q)})^{-m|\boldsymbol{\gamma}^{(q)}|/2}\det(\textbf{S}^{(q)}_{\boldsymbol{\gamma}^{(q)},m})^{-1/2}
\exp\left(-\frac{\textbf{Y}^T(\textbf{I}_n-\textbf{Z}_{\boldsymbol{\gamma}^{(q)}}(\textbf{U}^{(q)}_{\boldsymbol{\gamma}^{(q)},m})^{-1}
\textbf{Z}_{\boldsymbol{\gamma}^{(q)}}^T)\textbf{Y}}{2\sigma^2_{(q)}}\right).
\end{eqnarray}
In practice, one can draw
$m$ from (\ref{marginal:gamma}) by an Metropolis-Hasting step given the current value $\boldsymbol{\gamma}^{(q)}$,
which avoids computing the expensive sum and thus is more efficient.
Explicitly, given the current value $m_{old}$, one draws
$m_{new}$ from some proposal distribution $Q(m_{new}| m_{old})$. Then accept $m_{new}$ with probability
$\frac{p(m_{new}|\boldsymbol{\gamma},c,\sigma^2,t_n,\textbf{D}_n)}{p(m_{old}|\boldsymbol{\gamma},c,\sigma^2,t_n,\textbf{D}_n)}
\cdot \frac{Q(m_{old}|m_{new})}{Q(m_{new}|m_{old})}$. The choice of the proposal distribution $Q(m'|m)$ is not unique,
but can be made very simple. For instance, when $m=m_1$ (or $m_2$), one draws $m'$ randomly from $\{m,m+1\}$ (or $\{m,m-1\}$);
when $m_1<m<m_2$, one draws $m'$ randomly from $\{m-1,m,m+1\}$.

To complete the second step, we apply a nontrivial variation of the conventional blockwise technique
(see \cite{GR98,WGN04}) to sample $\boldsymbol{\beta}_j$'s and $\gamma_j$'s, given an updated sample $m^{(q+1)}$.
Note that the sample $\boldsymbol{\beta}^{(q)}_j$'s from the previous $q$-th iteration have dimension $m^{(q)}$
which might be different from $m^{(q+1)}$. This phenomenon of different dimensions
makes the conventional blockwise sampling approach fail since there is an underlying conflict
between the current state of $m$ and the (conditioning) blocks from the previous iteration.
Motivated from the ``dimension-matching" technique in the
reversible jump MCMC approach (see \cite{P95}), we propose to modify the $\boldsymbol{\beta}^{(q)}_j$'s to be of dimension $m^{(q+1)}$
to match the current state of $m$.
Specifically, if $m^{(q+1)}<m^{(q)}$, define $\tilde{\boldsymbol{\beta}}^{(q)}_j$ to be an $m^{(q+1)}$-dimensional vector
which consists of the first $m^{(q+1)}$ elements of $\boldsymbol{\beta}^{(q)}_j$. If $m^{(q+1)}>m^{(q)}$,
then define $\tilde{\boldsymbol{\beta}}^{(q)}_j=((\boldsymbol{\beta}^{(q)}_j)^T,\textbf{0}_{m^{(q+1)}-m^{(q)}}^T)^T$,
where $\textbf{0}_h$ denotes the $h$-dimensional zero vector.
That is, $\tilde{\boldsymbol{\beta}}^{(q)}_j$ is $m^{(q+1)}$-dimensional with the first $m^{(q)}$ elements being exactly
the ones of $\boldsymbol{\beta}^{(q)}_j$, and the remaining $m^{(q+1)}-m^{(q)}$ elements being zero.
If $m^{(q+1)}=m^{(q)}$, then set $\tilde{\boldsymbol{\beta}}^{(q)}_j=\boldsymbol{\beta}^{(q)}_j$.
Repeating the above procedure for all $j=1,\ldots,p$, one gets
$\tilde{\boldsymbol{\beta}}_j^{(q)}$'s, a ``modified" set of samples from the previous stage.

Suppose we have updated samples
$(\boldsymbol{\beta}_1^{(q+1)},\gamma_1^{(q+1)}),\ldots,(\boldsymbol{\beta}_{j-1}^{(q+1)},\gamma_{j-1}^{(q+1)})$,
in which all $\boldsymbol{\beta}_{j'}^{(q+1)}$, for $j'=1,\ldots,j-1$, are $m^{(q+1)}$-dimensional.
Define $\textbf{b}_{j'}=(\boldsymbol{\beta}^{(q+1)}_{j'},\gamma^{(q+1)}_{j'})$
for $j'=1,\ldots,j-1$, $\textbf{b}_{j'}=(\tilde{\boldsymbol{\beta}}^{(q)}_{j'},\gamma^{(q)}_{j'})$ for $j'=j+1,\ldots,p$,
and $\textbf{b}_j=(\boldsymbol{\beta}_j,\gamma_j)$, where $\boldsymbol{\beta}_j$ is an $m^{(q+1)}$-dimensional
nominal vector and both $\boldsymbol{\beta}_j$ and $\gamma_j$ will be updated.
For convenience, define $\textbf{b}_{-j}=\{\textbf{b}_{j'},j'=1,\ldots,p,j'\neq j\}$ to be the conditioning blocks.
The full conditional of $\textbf{b}_j$
given $\textbf{b}_{-j}$ and other variables highly depends on the size of $\boldsymbol{\gamma}_{-j}
=(\gamma_1^{(q+1)},\ldots,\gamma_{j-1}^{(q+1)},\gamma_{j+1}^{(q)},\ldots,\gamma_{p}^{(q)})$.
Specifically, for effective sampling,
$|\boldsymbol{\gamma}_{-j}|$ cannot exceed $t_n^{(q)}$ since otherwise the block $\textbf{b}_j$
will have zero posterior probability.
When $|\boldsymbol{\gamma}_{-j}|=t_n^{(q)}$,
$\gamma_j^{(q+1)}$ has to be zero since otherwise the conditional probability becomes zero.
In this case, one simply sets $\boldsymbol{\beta}^{(q+1)}_j=\textbf{0}_{m^{(q+1)}}$.

Next we suppose $|\boldsymbol{\gamma}_{-j}|<t_n^{(q)}$. For $j',j=1,\ldots,p$, define
$\textbf{Z}^{(q+1)}_{j'}=(\boldsymbol{\Phi}_{j'1},\ldots,\boldsymbol{\Phi}_{j'm^{(q+1)}})$, an $n$ by $m^{(q+1)}$ matrix,
and define $\textbf{Z}^{(q+1)}_{-j}=(\textbf{Z}^{(q+1)}_{j'},j'\neq j)$,
an $n$ by $m^{(q+1)}(p-1)$ matrix.
Similarly, define $\tilde{\boldsymbol{\beta}}_{-j}^{(q)}$ to be the $m^{(q+1)}(p-1)$-vector
formed by $\boldsymbol{\beta}^{(q+1)}_1,\ldots,\boldsymbol{\beta}^{(q+1)}_{j-1}$,
$\tilde{\boldsymbol{\beta}}^{(q)}_{j+1},\ldots,\tilde{\boldsymbol{\beta}}^{(q)}_p$.
Let $\textbf{u}_j=\textbf{Y}-\textbf{Z}^{(q+1)}_{-j}\tilde{\boldsymbol{\beta}}_{-j}^{(q)}$.
Note $p(\gamma_j,\boldsymbol{\gamma}_{-j})=1$.
Then we have from (\ref{joint:post}) that
\begin{eqnarray}\label{block:conditional1}
&&p(\boldsymbol{\beta}_j,\gamma_j=1|\textbf{b}_{-j},\sigma_{(q)}^2,c^{(q)},t_n^{(q)},m^{(q+1)},\textbf{D}_n)\nonumber\\
&\propto&
\exp\left(-\frac{\|\textbf{u}_j-\textbf{Z}^{(q+1)}_j\boldsymbol{\beta}_j\|^2}{2\sigma_{(q)}^2}\right)
(\sqrt{2\pi c^{(q)}}\sigma_{(q)})^{-m^{(q+1)}}\det(\textbf{T}_{m^{(q+1)}})^{-1/2}
\exp\left(-\frac{\boldsymbol{\beta}_j^T\textbf{T}_{m^{(q+1)}}^{-1}\boldsymbol{\beta}_j}{2c^{(q)}\sigma_{(q)}^2}\right).\nonumber\\
\end{eqnarray}
Integrating out $\boldsymbol{\beta}_j$ in (\ref{block:conditional1}), one gets that
\begin{eqnarray}\label{gamma1:eq}
&&p(\gamma_j=1|\textbf{b}_{-j},\sigma_{(q)}^2,c^{(q)},t_n^{(q)},m^{(q+1)},\textbf{D}_n)\nonumber\\
&\propto&
\det(\textbf{Q}_j^{(q)}\textbf{T}_{m^{(q+1)}})^{-1/2}
(c^{(q)})^{-m^{(q+1)}/2}\exp\left(-\frac{\|\textbf{u}_j\|^2-\textbf{u}_j^T\textbf{Z}_j^{(q+1)}
(\textbf{Q}_j^{(q)})^{-1}(\textbf{Z}_j^{(q+1)})^T\textbf{u}_j}
{2\sigma^2_{(q)}}\right),
\end{eqnarray}
where $\textbf{Q}_j^{(q)}=(c^{(q)})^{-1}\textbf{T}_{m^{(q+1)}}^{-1}+
(\textbf{Z}_j^{(q+1)})^T\textbf{Z}_j^{(q+1)}$.
Similarly, one gets from (\ref{joint:post}) that
\begin{equation}\label{block:conditional0}
p(\boldsymbol{\beta}_j,\gamma_j=0|\textbf{b}_{-j},\sigma_{(q)}^2,
c^{(q)},t_n^{(q)},m^{(q+1)},\textbf{D}_n)\propto
\exp\left(-\frac{\|\textbf{u}_j-\textbf{Z}_j^{(q+1)}\boldsymbol{\beta}_j\|^2}{2\sigma^2_{(q)}}\right)\delta_{\textbf{0}}(\boldsymbol{\beta}_j).
\end{equation}
Integrating out $\boldsymbol{\beta}_j$ in (\ref{block:conditional0}) one gets that
\begin{equation}\label{gamma0:eq}
p(\gamma_j=0|\textbf{b}_{-j},\sigma_{(q)}^2,
c^{(q)},t_n^{(q)},m^{(q+1)},\textbf{D}_n)\propto
\exp\left(-\frac{\|\textbf{u}_j\|^2}{2\sigma^2_{(q)}}\right).
\end{equation}
Consequently, from (\ref{gamma1:eq}) and (\ref{gamma0:eq}) we draw $\gamma_j^{(q+1)}$ from
\begin{equation}\label{block1}
p(\gamma_j=1|\textbf{b}_{-j},\sigma^2_{(q)},c^{(q)},t_n^{(q)},m^{(q+1)},\textbf{D}_n)=\frac{1}{1+\theta_j},
\end{equation}
where $\theta_j=\det(\textbf{Q}_j^{(q)}\textbf{T}_{m^{(q+1)}})^{1/2}(c^{(q)})^{m^{(q+1)}/2}
\exp\left(-\frac{\textbf{u}_j^T\textbf{Z}_j^{(q+1)}
(\textbf{Q}_j^{(q)})^{-1}(\textbf{Z}_j^{(q+1)})^T\textbf{u}_j}{2\sigma^2_{(q)}}\right)$.
It can be shown from (\ref{block:conditional1}) and (\ref{block:conditional0}) that
\begin{eqnarray}\label{block2}
&&\boldsymbol{\beta}_j| \gamma_j^{(q+1)}=1,\textbf{b}_{-j},\sigma^2_{(q)},c^{(q)},t_n^{(q)},m^{(q+1)},\textbf{D}_n
\sim
N\left((\textbf{Q}_j^{(q)})^{-1}(\textbf{Z}_j^{(q+1)})^T\textbf{u}_j,\sigma^2_{(q)}(\textbf{Q}_j^{(q)})^{-1}\right),\nonumber\\
&&p(\boldsymbol{\beta}_j=\textbf{0}| \gamma_j^{(q+1)}=0,\textbf{b}_{-j},\sigma^2_{(q)},c^{(q)},t_n^{(q)},m^{(q+1)},\textbf{D}_n)=1,
\end{eqnarray}
from which $\boldsymbol{\beta}_j^{(q+1)}$ is drawn. In the above procedure,
finding the matrix product $\textbf{Z}^{(q+1)}_{-j}\tilde{\boldsymbol{\beta}}^{(q)}_{-j}$ is a time-consuming step.
It is possible to avoid computing this matrix product by iteratively using the following relation
\begin{equation}\label{a:rel}
\textbf{Z}^{(q+1)}_{-(j+1)}\tilde{\boldsymbol{\beta}}^{(q)}_{-(j+1)}
=\textbf{Z}^{(q+1)}_{-j}\tilde{\boldsymbol{\beta}}^{(q)}_{-j}+
\textbf{Z}_j^{(q+1)}\boldsymbol{\beta}_j^{(q+1)}-\textbf{Z}_{j+1}^{(q+1)}\tilde{\boldsymbol{\beta}}_{j+1}^{(q)}.
\end{equation}
In practice, one only needs to compute $\textbf{Z}^{(q+1)}_{-1}\tilde{\boldsymbol{\beta}}^{(q)}_{-1}$
since the subsequent products can be iteratively updated through (\ref{a:rel}).

The proposed blockwise sampling scheme (\ref{block1}) and (\ref{block2}) can be viewed as a generalization of \cite{GR98}
from $m=1$ (without group structure) to general $m$ (with group structure).
This generalization is nontrivial because we allow $m$, the dimension of $\boldsymbol{\beta}_j$,
to change across the consecutive iterations. When updating $\textbf{b}_j$ given the blocks from the previous iteration
whose dimensions might be different from the current value of $m$, we have to apply the ``dimension-matching" technique
to the previous samples of $\boldsymbol{\beta}_j$'s so that they have the same dimension as the current $m$.
By doing so, one can apply the conventional blockwise techniques to update the blocks consecutively.
Note that when $m$ does not change across the iterations, there is no need to use such ``dimension-matching" procedure.
Furthermore, the proposed blockwise technique can only be used for the
constrained situation, i.e., when $|\boldsymbol{\gamma}_{-j}|\le t_n^{(q)}$,
which is essentially a constrained version (with group structure) of the conventional blockwise sampling approaches.

\textit{Sampling} $\sigma^2$. From (\ref{joint:post}), it can be easily seen that the full conditional of $\sigma^2$ is
\begin{eqnarray*}
&&\sigma^2|\boldsymbol{\gamma}^{(q+1)},\boldsymbol{\beta}^{(q+1)},m^{(q+1)},c^{(q)},t_n^{(q)},\textbf{D}_n\\
&&\sim IG\left(\frac{n+\nu+m^{(q+1)}|\boldsymbol{\gamma}^{(q+1)}|}{2},
\frac{\|\textbf{Y}-\textbf{Z}^{(q+1)}\boldsymbol{\beta}^{(q+1)}\|^2
+(\boldsymbol{\beta}^{(q+1)}_{\boldsymbol{\gamma}^{(q+1)}})^T
\boldsymbol{\Lambda}_{\boldsymbol{\gamma}^{(q+1)},m^{(q+1)}}^{-1}\boldsymbol{\beta}^{(q+1)}_{\boldsymbol{\gamma}^{(q+1)}}
(c^{(q)})^{-1}+1}{2}\right),
\end{eqnarray*}
where $IG(a,b)$ denotes the inverse Gamma distribution. Denote $\sigma^2_{(q+1)}$ as the updated sample.

\textit{Sampling} $c$. When $g(c)$ is chosen to be the GZS prior specified as (\ref{gzs}),
we can use a Gibbs sampling step to draw $c^{(q+1)}$. Indeed, the full conditional of $c$ is found
to be
\[
IG\left(m^{(q+1)}|\boldsymbol{\gamma}^{(q+1)}|/2+b,p^\mu+
(\boldsymbol{\beta}^{(q+1)}_{\boldsymbol{\gamma}^{(q+1)}})^T
\boldsymbol{\Lambda}_{\boldsymbol{\gamma}^{(q+1)},m^{(q+1)}}^{-1}
\boldsymbol{\beta}^{(q+1)}_{\boldsymbol{\gamma}^{(q+1)}}/(2\sigma^2_{(q+1)})\right).
\]

When $g(c)$ is chosen to be the GHG prior specified as (\ref{ghg}),
we need an Metropolis-Hasting step. Explicitly,
the full conditional of $c$ is
\begin{eqnarray*}
&&p(c|\boldsymbol{\gamma}^{(q+1)},\boldsymbol{\beta}^{(q+1)},\sigma^2_{(q+1)},m^{(q+1)},t_n^{(q)},\textbf{D}_n)\\
&\propto&
c^{-m^{(q+1)}|\boldsymbol{\gamma}^{(q+1)}|/2}
\exp\left(-(\boldsymbol{\beta}^{(q+1)}_{\boldsymbol{\gamma}^{(q+1)}})^T
\boldsymbol{\Lambda}_{\boldsymbol{\gamma}^{(q+1)},m^{(q+1)}}^{-1}
\boldsymbol{\beta}^{(q+1)}_{\boldsymbol{\gamma}^{(q+1)}}/(2c\sigma^2_{(q+1)})\right)g(c).
\end{eqnarray*}
Write $c=\exp(\kappa)$, then the full conditional of $\kappa$ is
\begin{eqnarray*}
&&p(\kappa|\boldsymbol{\gamma}^{(q+1)},\boldsymbol{\beta}^{(q+1)},\sigma^2_{(q+1)},m^{(q+1)},t_n^{(q)},\textbf{D}_n)\\
&\propto&
\exp\left(-\left(m^{(q+1)}|\boldsymbol{\gamma}^{(q+1)}|/2-1\right)
\kappa-(\boldsymbol{\beta}^{(q+1)}_{\boldsymbol{\gamma}^{(q+1)}})^T
\boldsymbol{\Lambda}_{\boldsymbol{\gamma}^{(q+1)},m^{(q+1)}}^{-1}
\boldsymbol{\beta}^{(q+1)}_{\boldsymbol{\gamma}^{(q+1)}}/(2\exp(\kappa)\sigma^2_{(q+1)})\right)g(\exp(\kappa)).
\end{eqnarray*}
Given an old value $\kappa_{old}$, draw $\kappa_{new}\sim N(\kappa_{old},\sigma_\kappa^2)$ for some fixed $\sigma_\kappa^2$.
Then accept $\kappa_{new}$ with probability
$p(\kappa_{new}|\boldsymbol{\gamma}^{(q+1)},\boldsymbol{\beta}^{(q+1)},\sigma^2_{(q+1)},m^{(q+1)},t_n^{(q)},\textbf{D}_n)
/p(\kappa_{old}|\boldsymbol{\gamma}^{(q+1)},\boldsymbol{\beta}^{(q+1)},\sigma^2_{(q+1)},m^{(q+1)},t_n^{(q)},\textbf{D}_n)$.

\textit{Sampling} $t_n$. It is easy to see that the full conditional of $t_n$ is uniform over $[|\boldsymbol{\gamma}^{(q+1)}|,T_n]$,
from which we obtain $t_n^{(q+1)}$.

\section{Numerical Study}\label{numer:study}

In this section we demonstrate the performance of the proposed method through empirical studies.
Specifically, we compare our Bayesian method based on GZS and GHG priors, denoted as BGZS and BGHG respectively,
with the iterative nonparametric independence screening combined with penGAM, denoted as
INIS-penGAM, and its greedy modification, denoted as g-INIS-penGAM, both
proposed by \cite{FFS11}.
Other well-known approaches include the penalized method for additive model (penGAM) proposed by
\cite{MSB09}, and the iterative sure independence screening (ISIS) combined with SCAD
proposed by \cite{FL08,FSW09}; see \cite{FFS11} for numerical details.

We adopted two simulation settings considered by \cite{HHW10,FFS11}
in the following examples in which $p=1000$ and $n=400$.
We chose somewhat arbitrarily the hyperparameter $\nu=6$ in the prior (\ref{fmb:3}).
In both the GZS and GHG priors defined by (\ref{gzs}) and (\ref{ghg}), we chose $b=0$.
To see how sensitive the results are with respect to the choice of $\mu$,
we considered difference values of $\mu$. The test functions are defined by
\[
f_1(x)=x,\,\,\,\, f_2(x)=(2x-1)^2,\,\,\,\, f_3(x)=\frac{\sin(2\pi x)}{2-\sin(2\pi x)},\,\,\,\,\textrm{and}
\]
\[
f_4(x)=0.1\sin(2\pi x)+0.2\cos(2\pi x)+0.3\sin(2\pi x)^2+0.4\cos(2\pi x)^3+0.5\sin(2\pi x)^3.
\]

\begin{Example}\label{exam:1}
We adopted the simulation setting of Example 3 in \cite{FFS11}.
Specifically, the data were generated from the additive model
$Y=5f_1(X_1)+3f_2(X_2)+4f_3(X_3)+6f_4(X_4)+\sqrt{1.74}\epsilon$, where $\epsilon\sim N(0,1)$.
The covariates were simulated by $X_j=(W_j+\rho U)/(1+\rho)$, $j=1,\ldots,p$,
where $W_j$'s and $U$ are \textit{iid} draws from uniform distribution over $[0,1]$.
$\rho=0$ yields independent $X_j$'s and $\rho=1$ yields dependent covariates with pairwise correlation $0.5$.
\end{Example}

\begin{Example}\label{exam:2}
We adopted the simulation setting of Example 4 in \cite{FFS11}.
This example is more challenging in that it contains more true functions than Example \ref{exam:1}.
Specifically, the data were generated from the following model
\begin{eqnarray*}
Y&=&f_1(X_1)+f_2(X_2)+f_3(X_3)+f_4(X_4)\\
&&+1.5f_1(X_5)+1.5f_2(X_6)+1.5f_3(X_7)+1.5f_4(X_8)\\
&&+2f_1(X_9)+2f_2(X_{10})+2f_3(X_{11})+2f_4(X_{12})+\sqrt{0.5184}\epsilon,
\end{eqnarray*}
where $\epsilon\sim N(0,1)$. The covariates $X_j$'s were generated according to Example \ref{exam:1}.
\end{Example}

In Examples \ref{exam:1} and \ref{exam:2}, \cite{FFS11} used five spline basis functions to represent the nonparametric functions.
In the present paper we considered both Legendre polynomial basis and trigonometric polynomial basis. In both cases,
we chose $m_1=4$ and $m_2=6$ so that the number of basis functions $m$ is varying around $5$
to enhance flexibility. We used $\mu=0.5,0.6,0.8$ and $0.8,0.9,1.1$ for the above two bases, respectively,
to demonstrate the insensitivity of the results. The MCMC algorithm introduced in Section \ref{sec:MCMC}
was implemented for posterior sampling.
Results were based on 100 replicated data sets.
Based on each data, we generated Markov chains with length 4000 for each model parameter.
The prior for $t_n$ was chosen as uniform in $\{1,\ldots,T_n\}$.
Note in model (\ref{fbm:1}) there are at most $m|\boldsymbol{\gamma}|$ nonzero Fourier coefficients.
In the present setup, this quantity is upper bounded by $mT_n$. We chose $T_n=[n/(3m)]$
so that the maximum number of nonzero coefficients does not exceed $n/3$.
In \cite{DE03,LF09} it was shown that the number of nonzero coefficients cannot exceed $n/2$
for uniqueness of the solution in sparse recovery. Here we reduced the upper bound to $n/3$
to gain more sparse solutions. For GHG prior, we chose $\sigma_\kappa^2=0.2$ for the MH update of $\kappa$
in sampling $c$; see Section \ref{sec:MCMC} for detailed description.

Recall that the Fourier coefficient vector $\boldsymbol{\beta}_j$ may change dimension across iterations, i.e.,
the so-called trans-dimensional problem.
The resulting chains may include varying-dimension components. It is well known in the literature that
the classic approaches for convergence diagnostics may fail. Following \cite{GH09}, we used the chains of MSE,
a natural scalar statistics, to monitor MCMC convergence of the Fourier coefficients,
which successfully resolves the trans-dimensional problem. Although we are aware that such scalar statistics
cannot guarantee convergence of the full chains, its computational convenience is attractive. Moreover,
the scope of the current paper focuses more about the selection and estimation issues,
for which monitoring convergence of the MSE chains is believed to be a reasonable strategy.
In our study we used Gelman-Rubin's statistics (see \cite{GCSR03}) to monitor convergence of the chains relating to MSE
and the remaining parameters. Confirming chain convergence, we dropped the first half of the posterior samples as burnins
and only used the second half to conduct statistical procedures.

We reported the average number of true positives (TP), the average number of false positives (FP),
the prediction errors (PE) based on BGZS and BGHG, and compared them with INIS and g-INIS.
Marginal inclusion rule is adopted to select the model.
That is, the $j$th variable is selected if its posterior exclusion
probability $P_j=1-p(\gamma_j=1|\textbf{D}_n)\le \widehat{p}$
for some quantity $\widehat{p}\in (0,1)$. We chose $\widehat{p}=0.5$ to yield median probability models;
see \cite{BB04}.
The TP/FP is the number of true/false inclusions in the selected model.
The PE was calculated as $\sum_{q=1}^Q\|\textbf{Y}-\widehat{\textbf{Y}}^{(q)}\|^2/(nQ)$,
where $\widehat{\textbf{Y}}^{(q)}=\textbf{Z}^{(q)}\boldsymbol{\beta}^{(q)}$ is
the fitted response value obtained from the $q$th iteration. In other words, PE is the average value of the mean
square errors (MSE) along with the iterations.

Results on TP, FP and PE using BGZS and BGHG were summarized in Tables \ref{example:1:L}--\ref{example:2:L}
and Tables \ref{example:1:F}--\ref{example:2:F}, based on Legendre polynomial basis and trigonometric polynomial basis,
respectively.
Results on INIS and g-INIS were directly summarized from \cite{FFS11}.
In Example \ref{exam:1},
we observed that, for both bases, BGZS and BGHG perform equally well as INIS and g-INIS in terms of TP, but perform better
in terms of FP and PE.

In Example \ref{exam:2} where Legendre polynomial basis was used,
both Bayesian approaches perform better than INIS and g-INIS. Specifically,
when $\rho=1$ and $\mu=0.6$, both BGZS and BGHG yield larger TP, smaller PE, and comparable FP;
when $\mu=0.8$, both BGZS and BGHG yield smaller FP and PE, and comparable TP.

In Example \ref{exam:2} where trigonometric basis was used, the performance is not as good as using Legendre polynomial basis,
but is still satisfactory. Specifically,
when $\rho=1$ and $\mu=0.8$, both BGZS and BGHG yield slightly larger TP and FP than INIS and g-INIS
(implying less conservative selection results),
and when $\mu=1.1$, both methods yield slightly smaller TP and FP (implying more conservative selection results);
when $\rho=0$, $\mu=0.8$ or $0.9$, both BGZS and BGHG can
select all the significant variables though they yield slightly larger FP.
In all the cases, the proposed Bayesian methods yield smaller PE.

The above results are not sensitive to the choice of $\mu$, though certain $\mu$ may yield slightly better performance.
Due to the essentially different basis structures, the feasible ranges of $\mu$
should be slightly different. We found that, at least in the above examples, $\mu\in[0.5,0.8]$ and $\mu\in[0.8,1.1]$
are feasible ranges for Legendre polynomial basis and trigonometric polynomial basis.
Any choice of $\mu$ within these ranges can provide satisfactory results. Values outside the ranges
may slightly lower the level of accuracy.

\begin{table}[htp]
\begin{center}{\footnotesize
\begin{tabular}{cccccc}
$\rho$&\multicolumn{2}{c}{Method}&TP           & FP          & PE\\ \hline
     0&\multicolumn{2}{c}{INIS}  &4.00 (0.00) & 2.58 (2.24) & 3.02 (0.34)\\
      &\multicolumn{2}{c}{g-INIS}&4.00 (0.00) & 0.67 (0.75) & 2.92 (0.30)\\
      &BGZS  &$\mu=0.5$          &4.00 (0.00) & 0.03 (0.17) & 2.25 (0.20)\\
      &      &$\mu=0.6$          &4.00 (0.00) & 0.02 (0.14) & 2.25 (0.16)\\
      &      &$\mu=0.8$          &4.00 (0.00) & 0.03 (0.17) & 2.23 (0.17)\\
      &BGHG  &$\mu=0.5$          &4.00 (0.00) & 0.03 (0.17) & 2.25 (0.20)\\
      &      &$\mu=0.6$          &4.00 (0.00) & 0.02 (0.14) & 2.25 (0.16)\\
      &      &$\mu=0.8$          &4.00 (0.00) & 0.03 (0.17) & 2.24 (0.17)\\ \hline
     1&\multicolumn{2}{c}{INIS}  &3.98 (0.00) & 15.76 (6.72) & 2.97 (0.39)\\
      &\multicolumn{2}{c}{g-INIS}&4.00 (0.00) &  0.98 (1.49) & 2.61 (0.26)\\
      &BGZS  &$\mu=0.5$          &3.99 (0.10) &  0.06 (0.28) & 2.02 (0.16)\\
      &      &$\mu=0.6$          &3.99 (0.10) &  0.05 (0.22) & 2.00 (0.15)\\
      &      &$\mu=0.8$          &3.99 (0.10) &  0.05 (0.22) & 2.04 (0.15)\\
      &BGHG  &$\mu=0.5$          &3.98 (0.14) &  0.08 (0.30) & 2.02 (0.16)\\
      &      &$\mu=0.6$          &3.99 (0.10) &  0.06 (0.24) & 2.00 (0.15)\\
      &      &$\mu=0.8$          &3.99 (0.10) &  0.05 (0.22) & 2.04 (0.15)\\ \hline
\end{tabular}}
\end{center}
\caption{Simulation results of Example \ref{exam:1} using Legendre polynomial basis.}
\label{example:1:L}
\end{table}
\begin{table}[htp]
\begin{center}{\footnotesize
\begin{tabular}{cccccc}
$\rho$&\multicolumn{2}{c}{Method}&TP            & FP          & PE\\ \hline
     0&\multicolumn{2}{c}{INIS}  & 11.97 (0.00) & 3.22 (1.49) & 0.97 (0.11)\\
      &\multicolumn{2}{c}{g-INIS}& 12.00 (0.00) & 0.73 (0.75) & 0.91 (0.10)\\
      &BGZS  &$\mu=0.5$          & 11.98 (0.14) & 0.74 (1.00) & 0.60 (0.05)\\
      &      &$\mu=0.6$          & 11.98 (0.14) & 0.54 (0.86) & 0.59 (0.05)\\
      &      &$\mu=0.8$          & 11.98 (0.14) & 0.41 (0.65) & 0.60 (0.05)\\
      &BGHG  &$\mu=0.5$          & 11.98 (0.14) & 0.70 (0.93) & 0.60 (0.05)\\
      &      &$\mu=0.6$          & 11.98 (0.14) & 0.58 (0.90) & 0.59 (0.05)\\
      &      &$\mu=0.8$          & 11.98 (0.14) & 0.44 (0.67) & 0.60 (0.05)\\ \hline
     1&\multicolumn{2}{c}{INIS}  & 10.01 (1.49) & 15.56 (0.93) & 1.03 (0.13)\\
      &\multicolumn{2}{c}{g-INIS}& 10.78 (0.75) &  1.08 (1.49) & 0.87 (0.11)\\
      &BGZS  &$\mu=0.5$          & 10.75 (0.80) &  1.25 (1.30) & 0.54 (0.05)\\
      &      &$\mu=0.6$          & 10.92 (0.69) &  1.08 (1.29) & 0.54 (0.05)\\
      &      &$\mu=0.8$          & 10.76 (0.79) &  0.88 (1.27) & 0.54 (0.05)\\
      &BGHG  &$\mu=0.5$          & 10.74 (0.75) &  1.13 (1.20) & 0.54 (0.05)\\
      &      &$\mu=0.6$          & 10.86 (0.72) &  1.10 (1.18) & 0.54 (0.05)\\
      &      &$\mu=0.8$          & 10.72 (0.80) &  0.82 (1.13) & 0.54 (0.05)\\ \hline
\end{tabular}}
\end{center}
\caption{Simulation results of Example \ref{exam:2} using Legendre polynomial basis.}
\label{example:2:L}
\end{table}

\begin{table}[htp]
\begin{center}{\footnotesize
\begin{tabular}{cccccc}
$\rho$&\multicolumn{2}{c}{Method}&TP           & FP          & PE\\ \hline
     0&\multicolumn{2}{c}{INIS}  &4.00 (0.00) & 2.58 (2.24) & 3.02 (0.34)\\
      &\multicolumn{2}{c}{g-INIS}&4.00 (0.00) & 0.67 (0.75) & 2.92 (0.30)\\
      &BGZS  &$\mu=0.8$          &4.00 (0.00) & 0.04 (0.19) & 2.07 (0.14)\\
      &      &$\mu=0.9$          &4.00 (0.00) & 0.06 (0.24) & 2.07 (0.15)\\
      &      &$\mu=1.1$          &4.00 (0.00) & 0.02 (0.14) & 2.09 (0.17)\\
      &BGHG  &$\mu=0.8$          &4.00 (0.00) & 0.04 (0.19) & 2.07 (0.14)\\
      &      &$\mu=0.9$          &4.00 (0.00) & 0.06 (0.24) & 2.07 (0.15)\\
      &      &$\mu=1.1$          &4.00 (0.00) & 0.02 (0.14) & 2.09 (0.17)\\ \hline
     1&\multicolumn{2}{c}{INIS}  &3.98 (0.00) & 15.76 (6.72) & 2.97 (0.39)\\
      &\multicolumn{2}{c}{g-INIS}&4.00 (0.00) &  0.98 (1.49) & 2.61 (0.26)\\
      &BGZS  &$\mu=0.8$          &4.00 (0.00) &  0.08 (0.44) & 1.76 (0.15)\\
      &      &$\mu=0.9$          &4.00 (0.00) &  0.04 (0.20) & 1.78 (0.12)\\
      &      &$\mu=1.1$          &4.00 (0.00) &  0.00 (0.00) & 1.76 (0.13)\\
      &BGHG  &$\mu=0.8$          &4.00 (0.00) &  0.10 (0.46) & 1.76 (0.15)\\
      &      &$\mu=0.9$          &4.00 (0.00) &  0.04 (0.20) & 1.78 (0.12)\\
      &      &$\mu=1.1$          &4.00 (0.00) &  0.00 (0.00) & 1.76 (0.13)\\ \hline
\end{tabular}}
\end{center}
\caption{Simulation results of Example \ref{exam:1} using trigonometric polynomial basis.}
\label{example:1:F}
\end{table}

\begin{table}[htp]
\begin{center}{\footnotesize
\begin{tabular}{cccccc}
$\rho$&\multicolumn{2}{c}{Method}&TP           & FP          & PE\\ \hline
     0&\multicolumn{2}{c}{INIS}  &11.97 (0.00) & 3.22 (1.49) & 0.97 (0.11)\\
      &\multicolumn{2}{c}{g-INIS}&12.00 (0.00) & 0.73 (0.75) & 0.91 (0.10)\\
      &BGZS  &$\mu=0.8$          &12.00 (0.00) & 1.22 (1.34) & 0.54 (0.05)\\
      &      &$\mu=0.9$          &12.00 (0.00) & 1.24 (1.27) & 0.54 (0.06)\\
      &      &$\mu=1.1$          &11.88 (0.32) & 0.34 (0.77) & 0.58 (0.05)\\
      &BGHG  &$\mu=0.8$          &12.00 (0.00) & 1.16 (1.40) & 0.54 (0.05)\\
      &      &$\mu=0.9$          &12.00 (0.00) & 1.10 (1.01) & 0.54 (0.05)\\
      &      &$\mu=1.1$          &11.88 (0.33) & 0.30 (0.68) & 0.58 (0.05)\\ \hline
     1&\multicolumn{2}{c}{INIS}  &10.01 (1.49) & 15.56 (0.93) & 1.03 (0.13)\\
      &\multicolumn{2}{c}{g-INIS}&10.78 (0.75) &  1.08 (1.49) & 0.87 (0.11)\\
      &BGZS  &$\mu=0.8$          &10.86 (0.67) &  2.18 (1.81) & 0.44 (0.05)\\
      &      &$\mu=0.9$          &10.76 (0.82) &  1.34 (1.56) & 0.47 (0.05)\\
      &      &$\mu=1.1$          &10.46 (0.86) &  0.50 (0.81) & 0.53 (0.05)\\
      &BGHG  &$\mu=0.8$          &10.88 (0.69) &  2.06 (1.81) & 0.44 (0.05)\\
      &      &$\mu=0.9$          &10.68 (0.82) &  1.58 (1.75) & 0.47 (0.05)\\
      &      &$\mu=1.1$          &10.44 (0.84) &  0.48 (0.76) & 0.53 (0.05)\\ \hline
\end{tabular}}
\end{center}
\caption{Simulation results of Example \ref{exam:2} using trigonometric polynomial basis.}
\label{example:2:F}
\end{table}

\section{Conclusions}\label{sec:conc}

A fully Bayesian approach is proposed to handle the ultrahigh-dimensional nonparametric additive models,
and the theoretical properties are carefully studied.
The numerical results demonstrate satisfactory performance of the method,
in terms of selection and estimation accuracy.
The method can achieve high level accuracy in both Legendre polynomial
basis and trigonometric polynomial basis.
Therefore, basis selection is not a critically important issue for the proposed approach,
though, to make the approach highly accurate, the choice of the hyperparameter $\mu$ in the proposed $g$-priors
should be slightly different in using different bases.
The numerical findings suggest us to use $\mu\in[0.5,0.8]$
and $\mu\in[0.8,1.1]$ for Legendre polynomial
basis and trigonometric polynomial basis, respectively.
The values outside these ranges are found
to merely slightly lower the accuracy within an acceptable range.

{\bf Acknowledge:}
Zuofeng Shang was a postdoctorate researcher supported by NSF-DMS 0808864, NSF-EAGER 1249316, a gift from Microsoft, a gift from Google, and the PI's salary recovery account.   Ping Li is partially supported by
ONR-N000141310261 and NSF-BigData 1249316.

\section{Appendix: Proofs}
To prove Theorem \ref{main:thm1}, we need the following preliminary lemma.
The proof is similar to that of Lemma 1 in \cite{SL13} and thus is omitted.

\begin{lemma}\label{lemma1} Suppose $\boldsymbol{\epsilon}\sim N(\textbf{0},\sigma_0^2  \textbf{I}_n)$ is independent of $\textbf{Z}_j$'s. Furthermore, $m_2\le n=o(p)$.
\begin{enumerate}[(i).]

\item\label{lemma1:i} Let $\boldsymbol{\nu}_{\boldsymbol{\gamma},m}$ be an $n$-dimensional vector indexed by $\boldsymbol{\gamma}\in \mathcal{S}$, a subset of the model space, and integer $1\le m\le m_2$.
Adopt the convention that $\boldsymbol{\nu}_{\boldsymbol{\gamma},m}^T\boldsymbol{\epsilon}/\|\boldsymbol{\nu}_{\boldsymbol{\gamma},m}\|=0$ when $\boldsymbol{\nu}_{\boldsymbol{\gamma},m}=0$.
Let $\#\mathcal{S}$ denote the cardinality of $\mathcal{S}$ with $\#\mathcal{S}\ge 2$. Then
\begin{equation}\label{lemma1:eq1}
\max\limits_{1\le m\le m_2}
\max\limits_{\boldsymbol{\gamma}\in \mathcal{S}}
\frac{|\boldsymbol{\nu}_{\boldsymbol{\gamma},m}^T\boldsymbol{\epsilon}|}{\|\boldsymbol{\nu}_{\boldsymbol{\gamma},m}\|}=O_P\left(\sqrt{\log(m_2\#\mathcal{S})}\right).
\end{equation}
In particular, let $\boldsymbol{\nu}_{\boldsymbol{\gamma},m}=(\textbf{I}_n-\textbf{P}_{\boldsymbol{\gamma}})
\textbf{Z}_{\boldsymbol{\gamma}^0\backslash\boldsymbol{\gamma}}\boldsymbol{\beta}_{\boldsymbol{\gamma}^0\backslash\boldsymbol{\gamma}}^0$ for $\boldsymbol{\gamma} \in S_2(t_n)$, we have
\begin{equation}\label{lemma1:eq2}
\max\limits_{1\le m\le m_2}\max\limits_{\boldsymbol{\gamma}\in S_2(t_n)}
\frac{|\boldsymbol{\nu}_{\boldsymbol{\gamma},m}^T\boldsymbol{\epsilon}|}{\|\boldsymbol{\nu}_{\boldsymbol{\gamma},m}\|}=O_P(\sqrt{\log{m_2}+t_n\log{p}})=O_P(\sqrt{t_n\log{p}}).
\end{equation}

\item\label{lemma1:ii} For any fixed $\alpha>4$,
\[
\lim\limits_{n\rightarrow\infty}P\left(\max\limits_{1\le m\le m_2}\max\limits_{\boldsymbol{\gamma}\in S_1(t_n)}\boldsymbol{\epsilon}^T
(\textbf{P}_{\boldsymbol{\gamma}}-\textbf{P}_{\boldsymbol{\gamma}^0})\boldsymbol{\epsilon}/(|\boldsymbol{\gamma}|-s_n) \le \alpha\sigma_0^2\log{p}\right)=1.
\]

\item\label{lemma1:iii} Adopt the convention that $\boldsymbol{\epsilon}^T \textbf{P}_{\boldsymbol{\gamma}}\boldsymbol{\epsilon}/|\gamma|=0$ when $\gamma$ is
null. Then for any fixed $\alpha>4$,
\[
\lim\limits_{n\rightarrow\infty}P\left(\max\limits_{1\le m\le m_2}\max\limits_{\boldsymbol{\gamma}\in S_2(t_n)}\boldsymbol{\epsilon}^T \textbf{P}_{\boldsymbol{\gamma}}\boldsymbol{\epsilon}/|\boldsymbol{\gamma}| \le \alpha\sigma_0^2\log{p}\right)=1.
\]

\end{enumerate}
\end{lemma}

\subsection*{Proof of Proposition \ref{prop:1}}

Let $C_\varphi=\max_{1\le j\le p}\sup_{l\ge1}\|\varphi_{jl}\|_{\sup}$.
We first show that (\ref{prop1:eq}) holds with $\frac{1}{n}\textbf{Z}_{\boldsymbol{\gamma}}^T\textbf{Z}_{\boldsymbol{\gamma}}$
therein replaced with $E\{\frac{1}{n}\textbf{Z}_{\boldsymbol{\gamma}}^T\textbf{Z}_{\boldsymbol{\gamma}}\}$.
Then we show (\ref{prop1:eq}) by using concentration inequalities which establish sharp approximations
between $\frac{1}{n}\textbf{Z}_{\boldsymbol{\gamma}}^T\textbf{Z}_{\boldsymbol{\gamma}}$
and $E\{\frac{1}{n}\textbf{Z}_{\boldsymbol{\gamma}}^T\textbf{Z}_{\boldsymbol{\gamma}}\}$.

For any $\textbf{a}_j=(a_{j1},\ldots,a_{jm})^T$, $j=1,\ldots,p$, note
$\textbf{Z}_j\textbf{a}_j=\sum\limits_{l=1}^m a_{jl}\boldsymbol{\Phi}_{jl}$. Define
$\textbf{a}_{\boldsymbol{\gamma}}$ to be the $m|\boldsymbol{\gamma}|$-vector
formed by $\textbf{a}_j$'s with $j\in\boldsymbol{\gamma}$. Therefore, we get that
\[
\textbf{a}_{\boldsymbol{\gamma}}^T E\{\textbf{Z}_{\boldsymbol{\gamma}}^T\textbf{Z}_{\boldsymbol{\gamma}}\}\textbf{a}_{\boldsymbol{\gamma}}
=
E\left\{\left(\sum_{j\in\boldsymbol{\gamma}}\textbf{Z}_j\textbf{a}_j\right)^T
\left(\sum_{j\in\boldsymbol{\gamma}}\textbf{Z}_j\textbf{a}_j\right)\right\}
=
\sum_{j\in\boldsymbol{\gamma}} E\{\textbf{a}_j^T\textbf{Z}_j^T\textbf{Z}_j\textbf{a}_j\}+
\sum_{\substack{j,j'\in\boldsymbol{\gamma} \\ j\neq j'}}
E\{\textbf{a}_j^T\textbf{Z}_j^T\textbf{Z}_{j'}\textbf{a}_{j'}\}.
\]
Since $\varphi_{jl}$'s are orthonormal in $\mathcal{H}_j$,
$E\{\textbf{a}_j^T\textbf{Z}_j^T\textbf{Z}_j\textbf{a}_j\}=
n E\left\{\left(\sum_{l=1}^m a_{jl}\varphi_{jl}(X_{ji})\right)^2\right\}=n\sum_{l=1}^m a_{jl}^2$.
On the other hand, for any $j,j'\in\boldsymbol{\gamma}$, $j\neq j'$,
$|E\{\textbf{a}_j^T\textbf{Z}_j^T\textbf{Z}_{j'}\textbf{a}_{j'}\}|=n |E\{\sum_{l=1}^m a_{jl}\varphi_{jl}(X_{ji})
\sum_{l=1}^m a_{j'l}\varphi_{j'l}(X_{j'i})\}|\le n\rho(|j-j'|) \sqrt{\sum_{l=1}^m a_{jl}^2}\sqrt{\sum_{l=1}^m a_{j'l}^2}$.
Therefore, by Cauchy's inequality
\begin{eqnarray*}
|\sum_{\substack{j,j'\in\boldsymbol{\gamma} \\ j\neq j'}}
E\{\textbf{a}_j^T\textbf{Z}_j^T\textbf{Z}_{j'}\textbf{a}_{j'}\}|
&\le&n\sum_{\substack{j,j'\in\boldsymbol{\gamma} \\ j\neq j'}}\rho(|j-j'|) \sqrt{\sum_{l=1}^m a_{jl}^2}\sqrt{\sum_{l=1}^m a_{j'l}^2}\\
&=&n\sum_{r=1}^\infty\rho(r)\sum_{j\in\boldsymbol{\gamma}}\sqrt{\sum_{l=1}^m a_{jl}^2}\sum_{j'\in\boldsymbol{\gamma},|j-j'|=r}
\sqrt{\sum_{l=1}^m a_{j'l}^2}\\
&\le&n\sum_{r=1}^\infty\rho(r)\sqrt{\sum_{j\in\boldsymbol{\gamma}}\sum_{l=1}^m a_{jl}^2}
\sqrt{\sum\limits_{j\in\boldsymbol{\gamma}}\left(\sum_{j'\in\boldsymbol{\gamma},|j-j'|=r}
\sqrt{\sum_{l=1}^m a_{j'l}^2}\right)^2}\\
&\le&n\sum_{r=1}^\infty\rho(r)\sqrt{\sum_{j\in\boldsymbol{\gamma}}\sum_{l=1}^m a_{jl}^2}
\sqrt{2\sum\limits_{j\in\boldsymbol{\gamma}}\sum_{j'\in\boldsymbol{\gamma},|j-j'|=r}\sum_{l=1}^m a_{j'l}^2}\\
&\le&2n\sum_{r=1}^\infty\rho(r)\sum_{j\in\boldsymbol{\gamma}}\sum_{l=1}^m a_{jl}^2.
\end{eqnarray*}
Therefore, for any $m\in[m_1,m_2]$ and $\boldsymbol{\gamma}\neq\emptyset$,
\begin{equation}\label{prop1:eq1}
1-2\sum_{r=1}^\infty \rho(r)\le
\lambda_-\left(E\{\frac{1}{n}\textbf{Z}_{\boldsymbol{\gamma}}^T\textbf{Z}_{\boldsymbol{\gamma}}\}\right)
\le\lambda_+\left(E\{\frac{1}{n}\textbf{Z}_{\boldsymbol{\gamma}}^T\textbf{Z}_{\boldsymbol{\gamma}}\}\right)
\le 1+2\sum_{r=1}^\infty \rho(r).
\end{equation}

Next we look at the difference
$\boldsymbol{\Delta}=\frac{1}{n}(\textbf{Z}_{\boldsymbol{\gamma}}^T\textbf{Z}_{\boldsymbol{\gamma}}
-E\{\textbf{Z}_{\boldsymbol{\gamma}}^T\textbf{Z}_{\boldsymbol{\gamma}}\})$.
The representative entry is
\[
\frac{1}{n}\sum_{i=1}^n [\varphi_{jl}(X_{ji})\varphi_{j'l'}(X_{j'i})
-E\{\varphi_{jl}(X_{ji})\varphi_{j'l'}(X_{j'i})\}],
\]
for $j,j'\in\boldsymbol{\gamma}$, and $l,l'=1,\ldots,m$.
Since $\varphi_{jl}$'s are uniformly bounded by $C_\varphi$, fixing $C>0$ such that
$C^2>8C_\varphi^4$, by Hoeffding's inequality,
\begin{eqnarray*}
&&P\left(\max_{\substack{j,j'=1,\ldots,p \\ l,l'=1,\ldots,m_2}}
\bigg|\sum_{i=1}^n [\varphi_{jl}(X_{ji})\varphi_{j'l'}(X_{j'i})
-E\{\varphi_{jl}(X_{ji})\varphi_{j'l'}(X_{j'i})\}]\bigg|\ge C\sqrt{n\log{p}}\right)\\
&\le&2\sum_{j,j=1}^p\sum_{l,l'=1}^{m_2} 2\exp\left(-\frac{2C^2 n\log{p}}{4nC_\varphi^4}\right)
\le2p^{4-C^2/(2C_\varphi^4)}\rightarrow0,\,\,\textrm{as $n\rightarrow\infty$.}
\end{eqnarray*}
Therefore, $\max_{\substack{j,j'=1,\ldots,p \\ l,l'=1,\ldots,m_2}}
|\sum_{i=1}^n [\varphi_{jl}(X_{ji})\varphi_{j'l'}(X_{j'i})
-E\{\varphi_{jl}(X_{ji})\varphi_{j'l'}(X_{j'i})\}]|=O_P(\sqrt{n\log{p}})$.
Denote $\Delta_{j,l;j',l'}$ to be the $(j,l;j',l')$-th entry of $\boldsymbol{\Delta}$.
By \cite{GVL89}, with probability approaching one, for any $\boldsymbol{\gamma}$ with $|\boldsymbol{\gamma}|\le 2t_n$,
and $m\in[m_1,m_2]$, the spectral norm of $\boldsymbol{\Delta}$
is upper bounded by $\|\boldsymbol{\Delta}\|_{\textrm{spectral}}\le
\max_{j',l'}\sum_{j\in\boldsymbol{\gamma},1\le l\le m} |\Delta_{j,l;j',l'}|\le
C'\frac{t_n^2 m_2^2\log{p}}{n}$, for some fixed large $C'>0$.
That is, when $n,p\rightarrow\infty$,
\[
\max_{|\boldsymbol{\gamma}|\le 2t_n}\max_{m\in[m_1,m_2]}
\|\boldsymbol{\Delta}\|_{\textrm{spectral}}\le
C'\frac{t_n^2 m_2^2\log{p}}{n}=o(1).
\]
By Weyl's inequality on eigenvalues (see \cite{GVL89}) and by (\ref{prop1:eq1}),
one can properly choose a small $c_0>0$
to satisfy (\ref{prop1:eq}), which completes the proof.
Using similar proofs of Proposition 2.1 in \cite{SC11}, it can be shown that (\ref{prop1:eq}) implies Assumption \ref{A1}.
The details are straightforward and thus are omitted.

\subsection*{Proof of Theorem \ref{main:thm1}}
\renewcommand{\theAssumption}{A.\arabic{Assumption}}
\setcounter{Assumption}{3}
Denote $\boldsymbol{\beta}_j^0=(\beta_{j1}^0,\ldots,\beta_{jm}^0)^T$ for $j=1,\ldots,p$.
Define $k_n=\sum_{j\in\boldsymbol{\gamma}^0}\|\boldsymbol{\beta}_j^0\|^2$
and $\psi_n=\min_{j\in\boldsymbol{\gamma}^0}\|\boldsymbol{\beta}_j^0\|$.
Before giving the proof of Theorem \ref{main:thm1}, we should mention that
Assumption \ref{A3'} is actually equivalent to
the following Assumption \ref{A3} which assumes the growing rates on terms involving the Fourier coefficients
of the partial Fourier series, i.e., $k_n$ and $\psi_n$.
The difference between Assumptions \ref{A3'} and \ref{A3}
is that $l_n$ and $\theta_n$ in the former are replaced with $k_n$ and $\psi_n$
in the latter, respectively.
This modified assumption is easier to use in technical proofs.

\begin{Assumption}\label{A3} There exists a positive sequence $\{h_m, m\ge1\}$
such that, as $m,m_1,m_2\rightarrow\infty$, $h_m\rightarrow\infty$,
$m^{-a}h_m$ decreasingly converges to zero, $mh_m$ increasingly converges to $\infty$,
and $\sum_{m_1\le m\le m_2}1/h_m=o(1)$.
Furthermore,
the sequences $m_1,m_2,h_m,s_n,t_n,\psi_n,k_n,\underline{\phi}_n,\bar{\phi}_n$ satisfy
\begin{enumerate}[(1).]
\item\label{A3:1} $m_2h_{m_2}s_n=o(n\min\{1,\psi_n^2\})$ and $m_1^{-a}h_{m_1}s_n^2=o(\min\{1,n^{-1}m_1\log(\underline{\phi}_n),\psi_n^2,\psi_n^4\})$;
\item\label{A3:2} $t_n\ge s_n$ and $t_n\log{p}=o(n\log(1+\min\{1,\psi_n^2\}))$;
\item\label{A3:3} $k_n=O(\underline{\phi}_n\tau_{m_2}^2)$ and $\log{p}=o(m_1\log{(n\underline{\phi}_n\tau_{m_2}^2)})$;
\item\label{A3:4} $m_2 s_n\log(1+n\bar{\phi}_n)=o(n\log(1+\min\{1,\psi_n^2\}))$.
\end{enumerate}
\end{Assumption}
To see the equivalence, it can be directly shown by (\ref{TFS:error:rate}) that uniformly for $m\in[m_1,m_2]$
\begin{equation}\label{theta:and:psi:1}
l_n-k_n=\sum_{j\in\boldsymbol{\gamma}^0}\sum_{l\ge m+1}|\beta_{jl}^0|^2\le C_\beta s_n m_1^{-a}.
\end{equation}
On the other hand, for any $j\in\boldsymbol{\gamma}^0$ and any $m\in [m_1,m_2]$, we have
$\|f_j^0\|_j^2=\sum_{l=1}^m |\beta_{jl}^0|^2+\sum_{l=m+1}^\infty|\beta_{jl}^0|^2\le \sum_{l=1}^m |\beta_{jl}^0|^2+
C_\beta m_1^{-a}$
and, obviously, $\|f_j^0\|_j^2\ge \sum_{l=1}^m |\beta_{jl}^0|^2$, which lead to
$\psi_n^2\le\theta_n^2\le\psi_n^2+C_\beta m_1^{-a}$.
Therefore,
\begin{equation}\label{theta:and:psi:2}
0\le \theta_n^2-\psi_n^2\le C_\beta m_1^{-a}.
\end{equation}
By (\ref{theta:and:psi:1}) and (\ref{theta:and:psi:2}) and
direct examinations, it can be verified that Assumption \ref{A3}
is equivalent to Assumption \ref{A3'}. We will prove the desired theorem
based on the equivalent Assumptions \ref{A1}, \ref{A2} and \ref{A3}.

Throughout the entire section of proof, we use ``w.p.a.1" to mean ``with probability approaching one".
Using the trivial fact
$p(\boldsymbol{\gamma}^0|\textbf{D}_n)=\frac{1}{1+\sum_{\boldsymbol{\gamma}\neq\boldsymbol{\gamma}^0}\frac{p(\boldsymbol{\gamma}|\textbf{D}_n)}{p(\boldsymbol{\gamma}^0|\textbf{D}_n)}}$,
to get the desired result it is sufficient to show $\sum_{\boldsymbol{\gamma}\neq\boldsymbol{\gamma}^0}\frac{p(\boldsymbol{\gamma}|\textbf{D}_n)}{p(\boldsymbol{\gamma}^0|\textbf{D}_n)}$
approaches zero in probability.
For any $\boldsymbol{\gamma}$ with $|\boldsymbol{\gamma}|\le t_n$, consider the following decomposition
\begin{eqnarray*}
&&-\log\left(\frac{p(\boldsymbol{\gamma}|\textbf{D}_n)}{p(\boldsymbol{\gamma}^0|\textbf{D}_n)}\right)\\
&=&\log\left(\frac{p(\boldsymbol{\gamma})}{p(\boldsymbol{\gamma}^0)}\right)
+\frac{1}{2}\log\left(\frac{\det(\textbf{W}_{\boldsymbol{\gamma}})}{\det(\textbf{W}_{\boldsymbol{\gamma}^0})}\right)
+\frac{n+\nu}{2}
\log\left(\frac{1+\textbf{Y}^T(\textbf{I}_n-\textbf{Z}_{\boldsymbol{\gamma}}
\textbf{U}_{\boldsymbol{\gamma}}^{-1}\textbf{Z}_{\boldsymbol{\gamma}}^T)
\textbf{Y}}{1+\textbf{Y}^T(\textbf{I}_n-\textbf{P}_{\boldsymbol{\gamma}})}\right)\\
&&-\frac{n+\nu}{2}
\log\left(\frac{1+\textbf{Y}^T(\textbf{I}_n-\textbf{Z}_{\boldsymbol{\gamma}^0}
\textbf{U}_{\boldsymbol{\gamma}^0}^{-1}\textbf{Z}_{\boldsymbol{\gamma}^0}^T)
\textbf{Y}}{1+\textbf{Y}^T(\textbf{I}_n-\textbf{P}_{\boldsymbol{\gamma}^0})\textbf{Y}}\right)
+\frac{n+\nu}{2}\log\left(\frac{1+\textbf{Y}^T(\textbf{I}_n-\textbf{P}_{\boldsymbol{\gamma}})\textbf{Y}}
{1+\textbf{Y}^T(\textbf{I}_n-\textbf{P}_{\boldsymbol{\gamma}^0})\textbf{Y}}\right).
\end{eqnarray*}
Denote the five terms by $J_1,J_2,J_3,J_4,J_5$. It follows by Assumption \ref{A2} that $J_1$
is bounded below uniformly for $\boldsymbol{\gamma}\in S(t_n)$.
It is also easy to see that $J_3\ge 0$ almost surely.
To prove $J_4$ is lower bounded, by Sherman-Morrison-Woodbury (see \cite{SL03}) ,
\[
(\textbf{Z}_{\boldsymbol{\gamma}^0}^T\textbf{Z}_{\boldsymbol{\gamma}^0}+\boldsymbol{\Sigma}_{\boldsymbol{\gamma}^0}^{-1})^{-1}
=(\textbf{Z}_{\boldsymbol{\gamma}^0}^T\textbf{Z}_{\boldsymbol{\gamma}^0})^{-1}-
(\textbf{Z}_{\boldsymbol{\gamma}^0}^T\textbf{Z}_{\boldsymbol{\gamma}^0})^{-1}
(\boldsymbol{\Sigma}_{\boldsymbol{\gamma}^0}+(\textbf{Z}_{\boldsymbol{\gamma}^0}^T\textbf{Z}_{\boldsymbol{\gamma}^0})^{-1})^{-1}
(\textbf{Z}_{\boldsymbol{\gamma}^0}^T\textbf{Z}_{\boldsymbol{\gamma}^0})^{-1},
\]
and by $\boldsymbol{\Sigma}_{\boldsymbol{\gamma}^0}\ge \underline{\phi}_n\tau_m^2 \textbf{I}_{ms_n}$
and similar calculations in the proof of Theorem 2.2 in \cite{SC11},
it can be shown that
\[
\frac{1+\textbf{Y}^T(\textbf{I}_n-\textbf{Z}_{\boldsymbol{\gamma}^0}
\textbf{U}_{\boldsymbol{\gamma}^0}^{-1}\textbf{Z}_{\boldsymbol{\gamma}^0}^T)\textbf{Y}}
{1+\textbf{Y}^T(\textbf{I}_n-\textbf{P}_{\boldsymbol{\gamma}^0})\textbf{Y}}
\le 1+\underline{\phi}_n^{-1}\tau_m^{-2}\frac{\textbf{Y}^T\textbf{Z}_{\boldsymbol{\gamma}^0}
(\textbf{Z}_{\boldsymbol{\gamma}^0}^T\textbf{Z}_{\boldsymbol{\gamma}^0})^{-2}\textbf{Z}_{\boldsymbol{\gamma}^0}^T\textbf{Y}}
{1+\textbf{Y}^T(\textbf{I}_n-\textbf{P}_{\boldsymbol{\gamma}^0})\textbf{Y}}.
\]
Note $\textbf{Y}=\textbf{Z}_{\boldsymbol{\gamma}^0}\boldsymbol{\beta}_{\boldsymbol{\gamma}^0}^0+\tilde{\boldsymbol{\eta}}$,
where $\tilde{\boldsymbol{\eta}}=\boldsymbol{\eta}+\boldsymbol{\epsilon}$,
$\boldsymbol{\eta}=\sum_{j\in\boldsymbol{\gamma}^0}\sum_{l=m+1}^\infty\beta_{jl}^0\boldsymbol{\Phi}_{jl}$,
$\boldsymbol{\Phi}_{jl}=(\varphi_{jl}(X_{j1}),\ldots,\varphi_{jl}(X_{jn}))^T$,
and $\boldsymbol{\epsilon}=(\epsilon_1,\ldots,\epsilon_n)^T$.
Since for any $m$,
\[
E\{\boldsymbol{\epsilon}^T\textbf{P}_{\boldsymbol{\gamma}^0}\boldsymbol{\epsilon}\}=ms_n\sigma_0^2,\,\,\textrm{and}
\]
\begin{eqnarray*}
E\{\|\boldsymbol{\eta}\|^2\}&=&n E\{(\sum_{j\in\boldsymbol{\gamma}^0}\sum_{l=m+1}^\infty\beta_{jl}^0\varphi_{jl}(X_{ji}))^2\}\\
&\le& ns_n\sum_{j\in\boldsymbol{\gamma}^0}E\{(\sum_{l=m+1}^\infty\beta_{jl}^0\varphi_{jl}(X_{ji}))^2\}\\
&=&ns_n\sum_{j\in\boldsymbol{\gamma}^0}\sum_{l=m+1}^\infty|\beta_{jl}^0|^2\le C_\beta ns_n^2m^{-a},
\end{eqnarray*}
where the last inequality follows by assumption (\ref{TFS:error:rate}),
it can be shown by Bonferroni inequality that as $n\rightarrow\infty$,
\begin{equation}\label{imp:result}
P\left(\max_{m_1\le m\le m_2}
m^{-1}h_m^{-1}\boldsymbol{\epsilon}^T\textbf{P}_{\boldsymbol{\gamma}^0}\boldsymbol{\epsilon}\le s_n \sigma_0^2\right)
\rightarrow1,\,\,
\textrm{and}\,\,P\left(\max_{m_1\le m\le m_2}
m^{a}h_m^{-1}\|\boldsymbol{\eta}\|^2\le C_\beta n s_n^2\right)\rightarrow1.
\end{equation}
(\ref{imp:result}) will be frequently used in the proof of the main results in this paper.
Since $\boldsymbol{\eta}^T
\textbf{P}_{\boldsymbol{\gamma}^0}\boldsymbol{\eta}\le\|\boldsymbol{\eta}\|^2$,
we have, w.p.a.1, for $m\in[m_1,m_2]$,
\begin{eqnarray*}
\textbf{Y}^T\textbf{Z}_{\boldsymbol{\gamma}^0}
(\textbf{Z}_{\boldsymbol{\gamma}^0}^T\textbf{Z}_{\boldsymbol{\gamma}^0})^{-2}\textbf{Z}_{\boldsymbol{\gamma}^0}^T\textbf{Y}
&\le& 2\left(\|\boldsymbol{\beta}_{\boldsymbol{\gamma}^0}^0\|^2+\tilde{\boldsymbol{\eta}}^T
\textbf{Z}_{\boldsymbol{\gamma}^0}
(\textbf{Z}_{\boldsymbol{\gamma}^0}^T\textbf{Z}_{\boldsymbol{\gamma}^0})^{-2}\textbf{Z}_{\boldsymbol{\gamma}^0}^T\tilde{\boldsymbol{\eta}}\right)\\
&\le&2\left(\|\boldsymbol{\beta}_{\boldsymbol{\gamma}^0}^0\|^2+c_0n^{-1}\tilde{\boldsymbol{\eta}}^T
\textbf{Z}_{\boldsymbol{\gamma}^0}
(\textbf{Z}_{\boldsymbol{\gamma}^0}^T\textbf{Z}_{\boldsymbol{\gamma}^0})^{-1}\textbf{Z}_{\boldsymbol{\gamma}^0}^T\tilde{\boldsymbol{\eta}}\right)\\
&\le&2\left(\|\boldsymbol{\beta}_{\boldsymbol{\gamma}^0}^0\|^2+2c_0n^{-1}\boldsymbol{\eta}^T
\textbf{P}_{\boldsymbol{\gamma}^0}\boldsymbol{\eta}+2c_0n^{-1}\boldsymbol{\epsilon}^T
\textbf{P}_{\boldsymbol{\gamma}^0}\boldsymbol{\epsilon}\right)\\
&\le&2\left(\|\boldsymbol{\beta}_{\boldsymbol{\gamma}^0}^0\|^2+2c_0C_\beta s_n^2m^{-a}h_m+2c_0\sigma_0^2n^{-1}mh_ms_n\right)\\
&\le&2\left(\|\boldsymbol{\beta}_{\boldsymbol{\gamma}^0}^0\|^2+2c_0C_\beta s_n^2m_1^{-a}h_{m_1}+2c_0\sigma_0^2n^{-1}m_2h_{m_2}s_n\right).
\end{eqnarray*}
Since $k_n\ge s_n\psi_n^2\gg s_n^2m_1^{-a}h_{m_1}+n^{-1}m_2h_{m_2}s_n$, w.p.a.1, for $m\in[m_1,m_2]$,
$\textbf{Y}^T\textbf{Z}_{\boldsymbol{\gamma}^0}
(\textbf{Z}_{\boldsymbol{\gamma}^0}^T\textbf{Z}_{\boldsymbol{\gamma}^0})^{-2}\textbf{Z}_{\boldsymbol{\gamma}^0}^T\textbf{Y}
\le 2k_n(1+o(1))$. On the other hand, w.p.a.1, for $m\in[m_1,m_2]$,
\begin{eqnarray}\label{thm1:eq1}
\textbf{Y}^T(\textbf{I}_n-\textbf{P}_{\boldsymbol{\gamma}^0})\textbf{Y}
&=&\tilde{\boldsymbol{\eta}}^T(\textbf{I}_n-\textbf{P}_{\boldsymbol{\gamma}^0})\tilde{\boldsymbol{\eta}}
=\boldsymbol{\eta}^T(\textbf{I}_n-\textbf{P}_{\boldsymbol{\gamma}^0})\boldsymbol{\eta}+2\boldsymbol{\eta}^T(\textbf{I}_n-\textbf{P}_{\boldsymbol{\gamma}^0})\boldsymbol{\epsilon}
-\boldsymbol{\epsilon}^T\textbf{P}_{\boldsymbol{\gamma}^0}\boldsymbol{\epsilon}+\boldsymbol{\epsilon}^T\boldsymbol{\epsilon}\nonumber\\
&=&O\left(n s_n^2m_1^{-a}h_{m_1}+n\sqrt{s_n^2m_1^{-a}h_{m_1}}+m_2h_{m_2}s_n\right)+
\boldsymbol{\epsilon}^T\boldsymbol{\epsilon}\nonumber\\
&=&\boldsymbol{\epsilon}^T\boldsymbol{\epsilon}+O\left(n\sqrt{s_n^2m_1^{-a}h_{m_1}}+m_2h_{m_2}s_n\right).
\end{eqnarray}
By (\ref{A3:1}) in Assumption \ref{A3}, (\ref{thm1:eq1}) implies $\textbf{Y}^T(\textbf{I}_n-\textbf{P}_{\boldsymbol{\gamma}^0})\textbf{Y}=n\sigma_0^2(1+o_P(1))$.
Therefore, w.p.a.1., for $m\in[m_1,m_2]$,
\begin{eqnarray*}
-J_4\le\frac{n+\nu}{2}\log\left(1+\frac{2k_n(1+o(1))}{n\underline{\phi}_n\tau_{m_2}^2\sigma_0^2}\right)=O(1),
\end{eqnarray*}
where the last upper bound follows by $k_n=O(\underline{\phi}_n\tau_{m_2}^2)$,
i.e., Assumption \ref{A3} (\ref{A3:3}). This shows that, w.p.a.1, $J_4$ is lower bounded uniformly for $m\in[m_1,m_2]$ and $c_j$'s $\in[\underline{\phi}_n,\bar{\phi}_n]$.

Next we approximate $J_5$ in two situations. First, for $\boldsymbol{\gamma}\in S_2(t_n)$,
a direct calculation leads to
\[
\textbf{Y}^T(\textbf{I}_n-\textbf{P}_{\boldsymbol{\gamma}})\textbf{Y}
=\|\boldsymbol{\nu}_{\boldsymbol{\gamma},m}\|^2+2\boldsymbol{\nu}_{\boldsymbol{\gamma},m}^T\tilde{\boldsymbol{\eta}}+\tilde{\boldsymbol{\eta}}^T(\textbf{I}_n-\textbf{P}_{\boldsymbol{\gamma}})\tilde{\boldsymbol{\eta}},
\]
where $\boldsymbol{\nu}_{\boldsymbol{\gamma},m}=
(\textbf{I}_n-\textbf{P}_{\boldsymbol{\gamma}})\textbf{Z}_{\boldsymbol{\gamma}^0
\backslash\boldsymbol{\gamma}}\boldsymbol{\beta}_{\boldsymbol{\gamma}^0\backslash\boldsymbol{\gamma}}$.
Since w.p.a.1., for $m\in[m_1,m_2]$, $\boldsymbol{\eta}^T(\textbf{I}_n-\textbf{P}_{\boldsymbol{\gamma}})
\boldsymbol{\eta}\le\|\boldsymbol{\eta}\|^2\le C_\beta n s_n^2 m_1^{-a}h_{m_1}$,
and $\boldsymbol{\epsilon}^T(\textbf{I}_n-\textbf{P}_{\boldsymbol{\gamma}})\boldsymbol{\epsilon}\le \boldsymbol{\epsilon}^T\boldsymbol{\epsilon}\le 2n\sigma_0^2$,
by Lemma \ref{lemma1} (\ref{lemma1:iii}), for a prefixed $\alpha>4$
\[
\tilde{\boldsymbol{\eta}}^T(\textbf{I}_n-\textbf{P}_{\boldsymbol{\gamma}})\tilde{\boldsymbol{\eta}}
\ge \boldsymbol{\epsilon}^T\boldsymbol{\epsilon}-\alpha\sigma_0^2t_n\log{p}-\sqrt{2C_\beta\sigma_0^2n^2s_n^2m_1^{-a}h_{m_1}}.
\]
Meanwhile, by Lemma \ref{lemma1} (\ref{lemma1:i}), for some large constant $C'>0$ and w.p.a.1., uniformly for $m\in[m_1,m_2]$,
$|\boldsymbol{\nu}_{\boldsymbol{\gamma},m}^T\boldsymbol{\epsilon}|\le C'\sqrt{t_n\log{p}}\|\boldsymbol{\nu}_{\boldsymbol{\gamma},m}\|$
and $|\boldsymbol{\nu}_{\boldsymbol{\gamma},m}^T\boldsymbol{\eta}|\le \sqrt{C_\beta n s_n^2 m_1^{-a}h_{m_1}}\|\boldsymbol{\nu}_{\boldsymbol{\gamma},m}\|$.
By Assumption \ref{A1}, $\|\boldsymbol{\nu}_{\boldsymbol{\gamma},m}\|^2\ge c_0^{-1}n\psi_n^2$, therefore we get that
\begin{eqnarray*}
&&\textbf{Y}^T(\textbf{I}_n-\textbf{P}_{\boldsymbol{\gamma}})\textbf{Y}\\&\ge&
c_0^{-1}n\psi_n^2\left(1+O_P\left(\sqrt{\frac{t_n\log{p}}{n\psi_n^2}}+\sqrt{\frac{n s_n^2 m_1^{-a}h_{m_1}}{n\psi_n^2}}\right)
+O_P\left(\frac{t_n\log{p}+n s_n^2 m_1^{-a}h_{m_1}}{n\psi_n^2}\right)\right)+\boldsymbol{\epsilon}^T\boldsymbol{\epsilon}\\
&=&c_0^{-1}n\psi_n^2(1+o_P(1))+\boldsymbol{\epsilon}^T\boldsymbol{\epsilon}.
\end{eqnarray*}
Note Assumption \ref{A3} (\ref{A3:1}) leads to $n\sqrt{s_n^2m_1^{-a}h_{m_1}}+m_2h_{m_2}s_n=o(n\psi_n^2)$
and $n\sqrt{s_n^2m_1^{-a}h_{m_1}}+m_2h_{m_2}s_n=o(n)$.
By (\ref{thm1:eq1}), we have, w.p.a.1., uniformly for $m\in[m_1,m_2]$,
\[
J_5\ge \frac{n+\nu}{2}\log\left(1+\frac{c_0^{-1}\psi_n^2(1+o(1))}{\sigma_0^2}\right)\ge \frac{n+\nu}{2}\log\left(1+C'\psi_n^2\right),
\]
for some large constant $C'>0$.

Next we consider $\boldsymbol{\gamma}\in S_1(t_n)$.
It can be checked by (\ref{thm1:eq1}), Lemma \ref{lemma1} and straightforward calculations
that for a fixed $\alpha>4$, w.p.a.1., uniformly for $m\in[m_1,m_2]$,
\begin{eqnarray*}
J_5&=&\frac{n+\nu}{2}\log\left(1-\frac{\tilde{\boldsymbol{\eta}}^T(\textbf{P}_{\boldsymbol{\gamma}}-\textbf{P}_{\boldsymbol{\gamma}^0})\tilde{\boldsymbol{\eta}}}
{1+\tilde{\boldsymbol{\eta}}^T(\textbf{I}_n-\textbf{P}_{\boldsymbol{\gamma}^0})\tilde{\boldsymbol{\eta}}}\right)\\
&\ge&\frac{n+\nu}{2}\log\left(1-\frac{2\|\boldsymbol{\eta}\|^2+2\boldsymbol{\epsilon}^T(\textbf{P}_{\boldsymbol{\gamma}}-\textbf{P}_{\boldsymbol{\gamma}^0})\boldsymbol{\epsilon}}
{1+\tilde{\boldsymbol{\eta}}^T(\textbf{I}_n-\textbf{P}_{\boldsymbol{\gamma}^0})\tilde{\boldsymbol{\eta}}}\right)\\
&\ge&\frac{n+\nu}{2}\log\left(1-\frac{2C_\beta n s_n^2 m_1^{-a}h_{m_1}+2(|\boldsymbol{\gamma}|-s_n)\alpha\sigma_0^2\log{p}}
{1+\boldsymbol{\epsilon}^T\boldsymbol{\epsilon}+O\left(n\sqrt{s_n^2m_1^{-a}h_{m_1}}+m_2h_{m_2}s_n\right)}\right)\\
&\ge&\frac{n+\nu}{2}\log\left(1-\frac{2C_\beta n s_n^2 m_1^{-a}h_{m_1}+2(|\boldsymbol{\gamma}|-s_n)\alpha\sigma_0^2\log{p}}{n\sigma_0^2(1+o(1))}\right)\\
&\ge&-(3C_\beta\sigma_0^{-2} n s_n^2 m_1^{-a}h_{m_1}+2(|\boldsymbol{\gamma}|-s_n)\alpha_0\log{p}),
\end{eqnarray*}
where the last inequality follows by $t_n\log{p}=o(n)$, i.e., Assumption \ref{A3} (\ref{A3:2}),
the inequality that $\log(1-x)\ge -2x$ when $x\in (0,1/2)$,
and a suitably fixed $\alpha_0\in (4,\alpha)$.

In the end we analyze the term $J_2$. Using the proof of Lemma A.2 in \cite{SC11},
it can be shown that for any $c_j$'s $\in[\underline{\phi}_n,\bar{\phi}_n]$ and $m\in[m_1,m_2]$,
\begin{eqnarray}\label{thm1:eq2}
&& J_2\ge \frac{1}{2}m_1(|\boldsymbol{\gamma}|-s_n)
\log\left(1+c_0^{-1}n\underline{\phi}_n\tau_{m_2}^2\right)\,\,\textrm{for any $\boldsymbol{\gamma}\in S_1(t_n)$, and}\nonumber\\
&& J_2\ge -\frac{m_2s_n}{2}\log\left(1+c_0n\bar{\phi}_n\tau_1^2\right)\,\,\textrm{for any $\boldsymbol{\gamma}\in S_2(t_n)$}.
\end{eqnarray}
To make the proofs more readable, we give the brief proof of (\ref{thm1:eq2}).
When $\boldsymbol{\gamma}\in S_1(t_n)$, by Sylvester's determinant formula (see \cite{SL03}), Assumption \ref{A1} and straightforward calculations we have
\begin{eqnarray*}
\det(\textbf{U}_{\boldsymbol{\gamma}})&=&\det(\textbf{U}_{\boldsymbol{\gamma}^0})
\det\left(\boldsymbol{\Sigma}_{\boldsymbol{\gamma}\backslash\boldsymbol{\gamma}^0}^{-1}+
\textbf{Z}_{\boldsymbol{\gamma}\backslash\boldsymbol{\gamma}^0}^T(\textbf{I}_n-\textbf{Z}_{\boldsymbol{\gamma}^0}\textbf{U}_{\boldsymbol{\gamma}^0}^{-1}\textbf{Z}_{\boldsymbol{\gamma}^0}^T)
\textbf{Z}_{\boldsymbol{\gamma}\backslash\boldsymbol{\gamma}^0}\right)\\
&\ge&\det(\textbf{U}_{\boldsymbol{\gamma}^0})
\det\left(\boldsymbol{\Sigma}_{\boldsymbol{\gamma}\backslash\boldsymbol{\gamma}^0}^{-1}+
\textbf{Z}_{\boldsymbol{\gamma}\backslash\boldsymbol{\gamma}^0}^T(\textbf{I}_n-\textbf{P}_{\boldsymbol{\gamma}^0})
\textbf{Z}_{\boldsymbol{\gamma}\backslash\boldsymbol{\gamma}^0}\right)\\
&\ge&\det(\textbf{U}_{\boldsymbol{\gamma}^0})\det\left(\boldsymbol{\Sigma}_{\boldsymbol{\gamma}\backslash\boldsymbol{\gamma}^0}^{-1}+c_0^{-1}n\textbf{I}_{m|\boldsymbol{\gamma}\backslash\boldsymbol{\gamma}^0|}\right).
\end{eqnarray*}
Therefore,
\[
\frac{\det(\textbf{W}_{\boldsymbol{\gamma}})}{\det(\textbf{W}_{\boldsymbol{\gamma}^0})}
=\frac{\det(\boldsymbol{\Sigma}_{\boldsymbol{\gamma}})}{\det(\boldsymbol{\Sigma}_{\boldsymbol{\gamma}^0})}\cdot\frac{\det(\textbf{U}_{\boldsymbol{\gamma}})}{\det(\textbf{U}_{\boldsymbol{\gamma}^0})}
\ge \left(1+c_0^{-1}n\underline{\phi}_n\tau_m^2\right)^{m(|\boldsymbol{\gamma}|-s_n)}\ge \left(1+c_0^{-1}n\underline{\phi}_n\tau_{m_2}^2\right)^{m_1(|\boldsymbol{\gamma}|-s_n)}.
\]
Taking logarithm on both sides, we get the first inequality in (\ref{thm1:eq2}). When $\boldsymbol{\gamma}\in S_2(t_n)$,
since $\det(\textbf{W}_{\boldsymbol{\gamma}})\ge1$, the second inequality in (\ref{thm1:eq2}) follows by
\begin{eqnarray*}
J_2&\ge& -\frac{1}{2}\log(\det(\textbf{W}_{\boldsymbol{\gamma}^0}))=
-\frac{1}{2}\log\left(\det\left(\textbf{I}_{ms_n}+\boldsymbol{\Sigma}_{\boldsymbol{\gamma}^0}^{1/2}\textbf{Z}_{\boldsymbol{\gamma}^0}^T
\textbf{Z}_{\boldsymbol{\gamma}^0}\boldsymbol{\Sigma}_{\boldsymbol{\gamma}^0}^{1/2}\right)\right)\\
&\ge&-\frac{ms_n}{2}\log\left(1+c_0n\bar{\phi}_n\tau_1^2\right)\ge -\frac{m_2s_n}{2}\log\left(1+c_0n\bar{\phi}_n\tau_1^2\right).
\end{eqnarray*}

To the end of the proof, we notice that based on the above approximations of $J_1$ to $J_5$,
there exist some large positive constants $\tilde{C}$ and $N$ such that when $n\ge N$,
w.p.a.1., for any $c_j$'s $\in[\underline{\phi}_n,\bar{\phi}_n]$ and $m\in[m_1,m_2]$,
\begin{eqnarray*}
&&\sum\limits_{\boldsymbol{\gamma}\in S_1(t_n)}\frac{p(\boldsymbol{\gamma}|\textbf{D}_n)}{p(\boldsymbol{\gamma}^0|\textbf{D}_n)}\\
&\le&
\tilde{C}\sum\limits_{\boldsymbol{\gamma}\in S_1(t_n)}
\exp\left(3C_\beta\sigma_0^{-2} n s_n^2 m_1^{-a}h_{m_1}
+2\alpha_0
(|\boldsymbol{\gamma}|-s_n)\log{p}-\frac{m_1(|\boldsymbol{\gamma}|-s_n)}{2}\log(1+c_0^{-1}n\underline{\phi}_n\tau_{m_2}^2)\right)\\
&=&\tilde{C}\sum\limits_{v=s_n+1}^{t_n}{p-s_n\choose v-s_n}\left(\frac{p^{2\alpha_0}
\exp(3C_\beta\sigma_0^{-2}n s_n^2 m_1^{-a}h_{m_1})}{(1+c_0^{-1}n\underline{\phi}_n\tau_{m_2}^2)^{m_1/2}}\right)^{v-s_n}\\
&=&\tilde{C}\sum\limits_{v=1}^{t_n-s_n}{p-s_n\choose v}\left(\frac{p^{2\alpha_0}
\exp(3C_\beta\sigma_0^{-2}n s_n^2 m_1^{-a}h_{m_1})}{(1+c_0^{-1}n\underline{\phi}_n\tau_{m_2}^2)^{m_1/2}}\right)^{v}\\
&\le&\tilde{C}\sum\limits_{v=1}^{t_n-s_n}\frac{p^v}{v!}\left(\frac{p^{2\alpha_0}
\exp(3C_\beta\sigma_0^{-2}n s_n^2 m_1^{-a}h_{m_1})}{(1+c_0^{-1}n\underline{\phi}_n\tau_{m_2}^2)^{m_1/2}}\right)^{v}\\
&\le&\tilde{C}\left(\exp\left(\frac{p^{2\alpha_0+1}
\exp(3C_\beta\sigma_0^{-2}n s_n^2 m_1^{-a}h_{m_1})}
{(1+c_0^{-1}n\underline{\phi}_n\tau_{m_2}^2)^{m_1/2}}\right)-1\right)\rightarrow0,\,\,\textrm{as $n\rightarrow\infty$},
\end{eqnarray*}
where the last limit follows by Assumption \ref{A3} (\ref{A3:1})\&(\ref{A3:3}),
and by Assumption \ref{A3} (\ref{A3:4}) we can make $N$ large enough so that $m_2s_n\log(1+c_0n\bar{\phi}_n\tau_1^2)\le \frac{n+\nu}{2}\log(1+C'\psi_n^2)$ for $n\ge N$, which leads to
\begin{eqnarray*}
\sum\limits_{\boldsymbol{\gamma}\in S_2(t_n)}\frac{p(\boldsymbol{\gamma}|\textbf{D}_n)}{p(\boldsymbol{\gamma}^0|\textbf{D}_n)}&\le&
\tilde{C}\sum\limits_{\boldsymbol{\gamma}\in S_2(t_n)}\exp\left(\frac{1}{2}m_2s_n\log(1+c_0n\bar{\phi}_n\tau_1^2)-\frac{n+\nu}{2}\log(1+C'\psi_n^2)\right)\\
&\le&
\tilde{C}\sum\limits_{\boldsymbol{\gamma}\in S_2(t_n)}\exp\left(-\frac{n+\nu}{4}\log(1+C'\psi_n^2)\right)\\
&\le&\tilde{C}\cdot\#S_2(t_n)\cdot (1+C'\psi_n^2)^{-(n+\nu)/4}\\
&\le&\tilde{C}\cdot p^{t_n}\cdot (1+C'\psi_n^2)^{-(n+\nu)/4}\rightarrow0,\,\,\textrm{as $n\rightarrow\infty$},
\end{eqnarray*}
where the last limit follows by Assumption \ref{A3} (\ref{A3:2}). This completes the proof of Theorem \ref{main:thm1}.

Before proving Theorem \ref{main:thm2}, we need the following lemma.
The proof is similar to that of Lemma 2 in \cite{SL13} and thus is omitted.

\begin{lemma}\label{lemma2} Suppose $\boldsymbol{\epsilon}\sim N(\textbf{0},\sigma_0^2  \textbf{I}_n)$. Adopt the convention that $\boldsymbol{\nu}_{\boldsymbol{\gamma}}^T\boldsymbol{\epsilon}/\|\boldsymbol{\nu}_{\boldsymbol{\gamma}}\|=0$ when $\boldsymbol{\nu}_{\boldsymbol{\gamma}}=0$,
and $\boldsymbol{\epsilon}^T \textbf{P}_{\boldsymbol{\gamma}}\boldsymbol{\epsilon}/|\boldsymbol{\gamma}|=0$ when $\boldsymbol{\gamma}$ is
null. Furthermore, $m_2\le n=o(p)$.

\begin{enumerate}[(i).]

\item\label{lemma2:i} For $\boldsymbol{\gamma}\in T_0(t_n)$, define $\boldsymbol{\nu}_{\boldsymbol{\gamma}}=(\textbf{I}_n-\textbf{P}_{\boldsymbol{\gamma}})
    \textbf{Z}_{\boldsymbol{\gamma}^0\backslash\boldsymbol{\gamma}}
    \boldsymbol{\beta}_{\boldsymbol{\gamma}^0\backslash\boldsymbol{\gamma}}^0$.
Then $\max\limits_{1\le m\le m_2}\max\limits_{\boldsymbol{\gamma}\in T_0(t_n)}
\frac{|\boldsymbol{\nu}_{\boldsymbol{\gamma}}^T\boldsymbol{\epsilon}|}{\|\boldsymbol{\nu}_{\boldsymbol{\gamma}}\|}
= O_P(\sqrt{s_n+\log{m_2}})$.


\item\label{lemma2:ii} For $\boldsymbol{\gamma}\in T_1(t_n)$,
denote $\boldsymbol{\gamma}^*=\boldsymbol{\gamma}\cap\boldsymbol{\gamma}^0$ which is nonnull. For any fixed $\alpha>6$,
\[
\lim\limits_{n\rightarrow\infty}P\left(\max\limits_{1\le m\le m_2}\max\limits_{\boldsymbol{\gamma}\in T_1(t_n)}
\frac{\boldsymbol{\epsilon}^T(\textbf{P}_{\boldsymbol{\gamma}}-\textbf{P}_{\boldsymbol{\gamma}^*})\boldsymbol{\epsilon}}{|\boldsymbol{\gamma}|-|\boldsymbol{\gamma}^*|}\le \alpha\sigma_0^2s_n\log{p}\right)=1.
\]

\item\label{lemma2:iii} Then for any fixed $\alpha>4$,
\[
\lim\limits_{n\rightarrow\infty}P\left(\max\limits_{1\le m\le m_2}
\max\limits_{\boldsymbol{\gamma}\in T_2(t_n)}\boldsymbol{\epsilon}^T \textbf{P}_{\boldsymbol{\gamma}}\boldsymbol{\epsilon}/|\boldsymbol{\gamma}| \le \alpha\sigma_0^2\log{p}\right)=1.
\]


\end{enumerate}
\end{lemma}

\subsection*{Proof of Proposition \ref{prop:2}}

Let $C_\varphi=\max_{1\le j\le p}\sup_{l\ge1}\|\varphi_{jl}\|_{\sup}$.
By Proposition \ref{prop:1}, we get that (\ref{B1:eq}) holds.
Next we show that (\ref{B1:eq2}) holds with $\rho_n\propto m_2s_n^2\log{p}$.
Define $\boldsymbol{\Delta}=
\textbf{Z}_{\boldsymbol{\gamma}^0\backslash\boldsymbol{\gamma}}^T\textbf{P}_{\boldsymbol{\gamma}}
\textbf{Z}_{\boldsymbol{\gamma}^0\backslash\boldsymbol{\gamma}}$.
The diagonal entry of $\boldsymbol{\Delta}$
is $\boldsymbol{\Delta}_{j,l}=\boldsymbol{\Phi}_{jl}^T \textbf{P}_{\boldsymbol{\gamma}}
\boldsymbol{\Phi}_{jl}$ for $j\in\boldsymbol{\gamma}^0\backslash\boldsymbol{\gamma}$, and $l=1,\ldots,m$.
By \cite{BK00}, any random variable $\xi$ almost surely bounded by a number $b>0$
satisfies $E\{\exp(a\xi)\}\le \exp(a^2b^2/2)$, i.e., $\xi$ is sub-Gaussian.
Since $\varphi_{jl}(X_{ji})$, $i=1,\ldots,n$,
are independent and uniformly bounded by $C_\varphi$,
for any $n$-vector $\textbf{a}=(a_1,\ldots,a_n)^T$,
$E\{\exp(\textbf{a}^T\boldsymbol{\Phi}_{jl})\}=\prod_{i=1}^n E\{\exp(a_i\varphi_{jl}(X_{ji}))\}\le
\prod_{i=1}^n \exp(a_i^2 C_\varphi^2/2)=\exp(\|\textbf{a}\|^2 C_\varphi^2/2)$,
that is, $\boldsymbol{\Phi}_{jl}$ is sub-Gaussian.
By Theorem 2.1 of \cite{HKZ12}, for some $C>2$ which implies
$5CC_\varphi^2|\boldsymbol{\gamma}|\log{p}>C_\varphi^2(|\boldsymbol{\gamma}|
+2\sqrt{|\boldsymbol{\gamma}|t}+2t)$ with $t=C|\boldsymbol{\gamma}|\log{p}$,
we have
\begin{eqnarray*}
&&P\left(\max_{m\in[m_1,m_2]}\max_{0<|\boldsymbol{\gamma}|<s_n}
\max_{\substack{j\in\boldsymbol{\gamma}^0\backslash\boldsymbol{\gamma}\\ l=1,\ldots,m}}
\boldsymbol{\Phi}_{jl}^T\textbf{P}_{\boldsymbol{\gamma}}\boldsymbol{\Phi}_{jl}/|\boldsymbol{\gamma}|
\ge CC_\varphi^2\log{p}\right)\\
&\le&
\sum_{1\le m\le m_2}\sum_{0<|\boldsymbol{\gamma}|<s_n}
\sum_{\substack{j\in\boldsymbol{\gamma}^0\backslash\boldsymbol{\gamma}\\ l=1,\ldots,m}}
P\left(\boldsymbol{\Phi}_{jl}^T\textbf{P}_{\boldsymbol{\gamma}}\boldsymbol{\Phi}_{jl}\ge CC_\varphi^2|\boldsymbol{\gamma}|\log{p}\right)\\
&\le&\sum_{1\le m\le m_2}\sum_{0<|\boldsymbol{\gamma}|<s_n}
\sum_{\substack{j\in\boldsymbol{\gamma}^0\backslash\boldsymbol{\gamma}\\ l=1,\ldots,m}}
E\left\{P\left(\boldsymbol{\Phi}_{jl}^T\textbf{P}_{\boldsymbol{\gamma}}\boldsymbol{\Phi}_{jl}\ge
CC_\varphi^2|\boldsymbol{\gamma}|\log{p}\bigg| \textbf{P}_{\boldsymbol{\gamma}}\right)\right\}\\
&\le&\sum_{1\le m\le m_2}\sum_{0<|\boldsymbol{\gamma}|<s_n}
\sum_{\substack{j\in\boldsymbol{\gamma}^0\backslash\boldsymbol{\gamma}\\ l=1,\ldots,m}}
\exp(-C|\boldsymbol{\gamma}|\log{p})\\
&\le&m_2^2s_n\sum_{r=1}^{s_n-1}{p\choose r} p^{-Cr}
\le m_2^2s_n\sum_{r=1}^{s_n-1}\frac{p^r}{r!} p^{-Cr}
\le m_2^2s_n(\exp(p^{1-C})-1)=O(m_2^2s_n/p)=o(1),
\end{eqnarray*}
therefore, $\max_{m\in[m_1,m_2]}\max_{0<|\boldsymbol{\gamma}|<s_n}
\max_{\substack{j\in\boldsymbol{\gamma}^0\backslash\boldsymbol{\gamma}\\ l=1,\ldots,m}}
\boldsymbol{\Phi}_{jl}^T\textbf{P}_{\boldsymbol{\gamma}}\boldsymbol{\Phi}_{jl}/|\boldsymbol{\gamma}|
=O_P(\log{p})$. So with probability approaching one,
for any $m\in[m_1,m_2]$ and $\boldsymbol{\gamma}\in T(s_n-1)\backslash\{\emptyset\}$,
$\lambda_+\left(\textbf{Z}_{\boldsymbol{\gamma}}^T\textbf{P}_{\boldsymbol{\gamma}}\textbf{Z}_{\boldsymbol{\gamma}}\right)
\le \textrm{trace}\left(\textbf{Z}_{\boldsymbol{\gamma}}^T\textbf{P}_{\boldsymbol{\gamma}}\textbf{Z}_{\boldsymbol{\gamma}}\right)
\le C' m_2s_n^2\log{p}$, for some large constant $C'>0$. This completes the proof.

\subsection*{Proof of Theorem \ref{main:thm2} (\ref{thm2:i})}
\renewcommand{\theAssumption}{B.\arabic{Assumption}}
\setcounter{Assumption}{3}
Like in Assumption \ref{A3}, one can replace $\theta_n$ and $l_n$ in Assumption \ref{B3'} by $\psi_n$ and $k_n$
while preserving an equivalent condition. Specifically, by the statements in the beginning of Theorem \ref{main:thm1},
it can be shown that the following assumption is an equivalent version of Assumption \ref{B3'}.
\begin{Assumption}\label{B3} There exists a positive sequence $\{h_m,m\ge1\}$
such that, as $m,m_1,m_2\rightarrow\infty$, $h_m\rightarrow\infty$,
$m^{-a}h_m$ decreasingly converges to zero, $mh_m$ increasingly converges to $\infty$,
and $\sum_{m_1\le m\le m_2}1/h_m=o(1)$. Furthermore,
the sequences $m_1,m_2,h_m,s_n,\psi_n,k_n,\underline{\phi}_n$ satisfy
\begin{enumerate}[(1).]
\item\label{B3:1} $m_2h_{m_2}s_n=o(n\min\{1,\psi_n^2\})$ and $m_1^{-a}h_{m_1}s_n^2
=o(\min\{1,n^{-1}m_1\log(\underline{\phi}_n),\psi_n^2\})$;
\item\label{B3:2} $k_n=O(\underline{\phi}_n\tau_{m_2}^2)$;
\item\label{B3:3} $\max\{\rho_n,s_n^2\log{p}\}=o(\min\{n,m_1\log(n\underline{\phi}_n\tau_{m_2}^2)\})$.
\end{enumerate}
\end{Assumption}
Next we will prove the theorem based on Assumptions \ref{B1}, \ref{B2} and \ref{B3}.
We first show that w.p.a.1, for $m\in[m_1,m_2]$,
$\max\limits_{\boldsymbol{\gamma}\in T_1(t_n)}p(\boldsymbol{\gamma}|\textbf{D}_n)/p\boldsymbol{(\gamma}\cap\boldsymbol{\gamma}^0|\textbf{D}_n)$
converges to zero. Since the denominator is bounded by $\max\limits_{\boldsymbol{\gamma}\in
T_0(t_n)}p(\boldsymbol{\gamma}|\textbf{D}_n)$,
it follows that $\max\limits_{\boldsymbol{\gamma}\in T_1(t_n)}p(\boldsymbol{\gamma}|\textbf{D}_n)/\max\limits_{\boldsymbol{\gamma}\in
T_0(t_n)}p(\boldsymbol{\gamma}|\textbf{D}_n)\rightarrow0$ in probability.
Second, we show, w.p.a.1, for $m\in[m_1,m_2]$, $\max\limits_{\boldsymbol{\gamma}\in
T_2(t_n)}p(\boldsymbol{\gamma}|\textbf{D}_n)/p(\emptyset|\textbf{D}_n)\rightarrow0$. This will complete the proof. Next we proceed in two steps.

\textbf{Step 1}: Consider the following decomposition for $\boldsymbol{\gamma}\in T_1(t_n)$,
\begin{eqnarray*}
&&-\log\left(\frac{p(\boldsymbol{\gamma}|\textbf{D}_n)}{p(\boldsymbol{\gamma}^*|\textbf{D}_n)}\right)\\
&=&-\log\left(\frac{p(\boldsymbol{\gamma})}{p(\boldsymbol{\gamma}^*)}\right)
+\frac{1}{2}\log\left(\frac{\det(\textbf{W}_{\boldsymbol{\gamma}})}{\det(\textbf{W}_{\boldsymbol{\gamma}^*})}\right)
+\frac{n+\nu}{2}\log\left(\frac{1+\textbf{Y}^T(\textbf{I}_n-\textbf{Z}_{\boldsymbol{\gamma}} \textbf{U}_{\boldsymbol{\gamma}}^{-1}
\textbf{Z}_{\boldsymbol{\gamma}}^T)\textbf{Y}}{1+\textbf{Y}^T(\textbf{I}_n-\textbf{P}_{\boldsymbol{\gamma}})\textbf{Y}}\right)\\
&&-\frac{n+\nu}{2}\log\left(\frac{1+\textbf{Y}^T(\textbf{I}_n-\textbf{Z}_{\boldsymbol{\gamma}^*}\textbf{U}_{\boldsymbol{\gamma}^*}^{-1}
\textbf{Z}_{\boldsymbol{\gamma}^*}^T)\textbf{Y}}
{1+\textbf{Y}^T(\textbf{I}_n-\textbf{P}_{\boldsymbol{\gamma}^*})\textbf{Y}}\right)+
\frac{n+\nu}{2}\log\left(\frac{1+\textbf{Y}^T(\textbf{I}_n-\textbf{P}_{\boldsymbol{\gamma}})\textbf{Y}}
{1+\textbf{Y}^T(\textbf{I}_n-\textbf{P}_{\boldsymbol{\gamma}^*})\textbf{Y}}\right),
\end{eqnarray*}
where $\boldsymbol{\gamma}^*=\boldsymbol{\gamma}\cap\boldsymbol{\gamma}^0\neq\emptyset$.
Denote the five items by $J_1,J_2,J_3,J_4,J_5$. We use the methods in the proof of Theorem \ref{main:thm1} to analyze the five terms.
Note that $J_1$ is bounded below by Assumption \ref{B2}, and $J_3\ge0$ almost surely.
To handle $J_4$, using Sherman-Morrison-Woodbury matrix identity,
\begin{eqnarray*}
&&\frac{1+\textbf{Y}^T(\textbf{I}_n-\textbf{Z}_{\boldsymbol{\gamma}^*}\textbf{U}_{\boldsymbol{\gamma}^*}^{-1}
\textbf{Z}_{\boldsymbol{\gamma}^*}^T)\textbf{Y}}
{1+\textbf{Y}^T(\textbf{I}_n-\textbf{P}_{\boldsymbol{\gamma}^*})\textbf{Y}}\\
&=&1+\frac{\textbf{Y}^T\textbf{Z}_{\boldsymbol{\gamma}^*}((\textbf{Z}_{\boldsymbol{\gamma}^{*}}^T
\textbf{Z}_{\boldsymbol{\gamma}^*})^{-1}-\textbf{U}_{\boldsymbol{\gamma}^*}^{-1})
\textbf{Z}_{\boldsymbol{\gamma}^*}^T\textbf{Y}}
{1+\textbf{Y}^T(\textbf{I}_n-\textbf{P}_{\boldsymbol{\gamma}^*})\textbf{Y}}\\
&=&1+\frac{\textbf{Y}^T\textbf{Z}_{\boldsymbol{\gamma}^*}(\textbf{Z}_{\boldsymbol{\gamma}^{*}}^T\textbf{Z}_{\boldsymbol{\gamma}^*})^{-1}
(\boldsymbol{\Sigma}_{\boldsymbol{\gamma}^*}+(\textbf{Z}_{\boldsymbol{\gamma}^{*}}^T\textbf{Z}_{\boldsymbol{\gamma}^*})^{-1})^{-1}
(\textbf{Z}_{\boldsymbol{\gamma}^{*}}^T\textbf{Z}_{\boldsymbol{\gamma}^*})^{-1}\textbf{Z}_{\boldsymbol{\gamma}^*}^T\textbf{Y}}
{1+\textbf{Y}^T(\textbf{I}_n-\textbf{P}_{\boldsymbol{\gamma}^*})\textbf{Y}}\\
&\le&1+\underline{\phi}_n^{-1}\tau_m^{-2}\frac{\textbf{Y}^T\textbf{Z}_{\boldsymbol{\gamma}^*}(\textbf{Z}_{\boldsymbol{\gamma}^{*}}^T
\textbf{Z}_{\boldsymbol{\gamma}^*})^{-2}\textbf{Z}_{\boldsymbol{\gamma}^*}^T\textbf{Y}}
{1+\textbf{Y}^T(\textbf{I}_n-\textbf{P}_{\boldsymbol{\gamma}^*})\textbf{Y}}\\
&\le&1+2\underline{\phi}_n^{-1}\tau_m^{-2}\frac{(\boldsymbol{\beta}^0_{\boldsymbol{\gamma}^0})^T\textbf{Z}_{\boldsymbol{\gamma}^0}^T
\textbf{Z}_{\boldsymbol{\gamma}^*}(\textbf{Z}_{\boldsymbol{\gamma}^{*}}^T
\textbf{Z}_{\boldsymbol{\gamma}^*})^{-2}\textbf{Z}_{\boldsymbol{\gamma}^*}^T\textbf{Z}_{\boldsymbol{\gamma}^0}
\boldsymbol{\beta}^0_{\boldsymbol{\gamma}^0}+
\boldsymbol{\tilde{\boldsymbol{\eta}}}^T\textbf{Z}_{\boldsymbol{\gamma}^*}(\textbf{Z}_{\boldsymbol{\gamma}^{*}}^T
\textbf{Z}_{\boldsymbol{\gamma}^*})^{-2}\textbf{Z}_{\boldsymbol{\gamma}^*}^T\boldsymbol{\tilde{\boldsymbol{\eta}}}}
{1+\textbf{Y}^T(\textbf{I}_n-\textbf{P}_{\boldsymbol{\gamma}^*})\textbf{Y}}.
\end{eqnarray*}

Without loss of generality, assume
$\textbf{Z}_{\boldsymbol{\gamma}^0}=(\textbf{Z}_{\boldsymbol{\gamma}^*},\textbf{Z}_{\boldsymbol{\gamma}^0\backslash\boldsymbol{\gamma}^*})$
and
$\boldsymbol{\beta}^0_{\boldsymbol{\gamma}^0}=((\boldsymbol{\beta}^0_{\boldsymbol{\gamma}^*})^T,(\boldsymbol{\beta}^0_{\boldsymbol{\gamma}^0
\backslash\boldsymbol{\gamma}^*})^T)^T$.
By a direct calculation it can be examined that
\begin{eqnarray*}
\textbf{Z}_{\boldsymbol{\gamma}^0}^T\textbf{Z}_{\boldsymbol{\gamma}^*}(\textbf{Z}_{\boldsymbol{\gamma}^{*}}^T\textbf{Z}_{\boldsymbol{\gamma}^*})^{-2}
\textbf{Z}_{\boldsymbol{\gamma}^*}^T\textbf{Z}_{\boldsymbol{\gamma}^0}=
\left(\begin{array}{cc}\textbf{I}_{|\boldsymbol{\gamma}^*|} &
(\textbf{Z}_{\boldsymbol{\gamma}^*}^T\textbf{Z}_{\boldsymbol{\gamma}^*})^{-1}\textbf{Z}_{\boldsymbol{\gamma}^*}^T\textbf{Z}_{\boldsymbol{\gamma}^0
\backslash\boldsymbol{\gamma}^*}\\
\textbf{Z}_{\boldsymbol{\gamma}^0\backslash\boldsymbol{\gamma}^*}^T\textbf{Z}_{\boldsymbol{\gamma}^*}(\textbf{Z}_{\boldsymbol{\gamma}^*}^T
\textbf{Z}_{\boldsymbol{\gamma}^*})^{-1} &
\textbf{Z}_{\boldsymbol{\gamma}^0\backslash\boldsymbol{\gamma}^*}^T\textbf{Z}_{\boldsymbol{\gamma}^*}(\textbf{Z}_{\boldsymbol{\gamma}^*}^T
\textbf{Z}_{\boldsymbol{\gamma}^*})^{-2}
\textbf{Z}_{\boldsymbol{\gamma}^*}^T\textbf{Z}_{\boldsymbol{\gamma}^0\backslash\boldsymbol{\gamma}^*}
\end{array}\right).
\end{eqnarray*}
By Assumption \ref{B1}, w.p.a.1, for $\boldsymbol{\gamma}\in T_2(t_n)$ and $m\in[m_1,m_2]$,
\[
\lambda_+\left(\textbf{Z}_{\boldsymbol{\gamma}^0\backslash\boldsymbol{\gamma}^*}^T\textbf{Z}_{\boldsymbol{\gamma}^*}
(\textbf{Z}_{\boldsymbol{\gamma}^*}^T\textbf{Z}_{\boldsymbol{\gamma}^*})^{-2}
\textbf{Z}_{\boldsymbol{\gamma}^*}^T\textbf{Z}_{\boldsymbol{\gamma}^0\backslash\boldsymbol{\gamma}^*}\right)
\le \frac{d_0}{n}\lambda_+\left(\textbf{Z}_{\boldsymbol{\gamma}^0\backslash\boldsymbol{\gamma}^*}
\textbf{P}_{\boldsymbol{\gamma}^*}\textbf{Z}_{\boldsymbol{\gamma}^0\backslash\boldsymbol{\gamma}^*}\right)\le \frac{d_0\rho_n}{n},
\]
which implies, w.l.p.,
$\lambda_+\left(\textbf{Z}_{\boldsymbol{\gamma}^0}^T\textbf{Z}_{\boldsymbol{\gamma}^*}(\textbf{Z}_{\boldsymbol{\gamma}^{*}}^T
\textbf{Z}_{\boldsymbol{\gamma}^*})^{-2}\textbf{Z}_{\boldsymbol{\gamma}^*}^T
\textbf{Z}_{\boldsymbol{\gamma}^0}\right)\le 1+\frac{d_0\rho_n}{n}$. Therefore, it can be shown that
\[
(\boldsymbol{\beta}^0_{\boldsymbol{\gamma}^0})^T\textbf{Z}_{\boldsymbol{\gamma}^0}^T
\textbf{Z}_{\boldsymbol{\gamma}^*}(\textbf{Z}_{\boldsymbol{\gamma}^{*}}^T\textbf{Z}_{\boldsymbol{\gamma}^*})^{-2}
\textbf{Z}_{\boldsymbol{\gamma}^*}^T
\textbf{Z}_{\boldsymbol{\gamma}^0}\boldsymbol{\beta}^0_{\boldsymbol{\gamma}^0}\le (1+\frac{d_0\rho_n}{n})k_n.
\]
On the other hand,
by (\ref{imp:result}) in the proof of Theorem \ref{main:thm1}, it can be shown that,
w.p.a.1, for $m\in[m_1,m_2]$,
$\boldsymbol{\tilde{\boldsymbol{\eta}}}^T\textbf{Z}_{\boldsymbol{\gamma}^*}(\textbf{Z}_{\boldsymbol{\gamma}^{*}}^T
\textbf{Z}_{\boldsymbol{\gamma}^*})^{-2}\textbf{Z}_{\boldsymbol{\gamma}^*}^T\boldsymbol{\tilde{\boldsymbol{\eta}}}
\le\frac{2d_0}{n}\left(\|\boldsymbol{\eta}\|^2+\boldsymbol{\epsilon}^T \textbf{P}_{\boldsymbol{\gamma}^0}\boldsymbol{\epsilon}\right)
\le \frac{2d_0}{n}\left(\sigma_0^2 s_n m_2h_{m_2}+C_\beta m_1^{-a}h_{m_1}n s_n^2\right)$.
Meanwhile, by (\ref{thm1:eq1}),
$\textbf{Y}^T(\textbf{I}_n-\textbf{P}_{\boldsymbol{\gamma}^*})\textbf{Y}\ge
\textbf{Y}^T(\textbf{I}_n-\textbf{P}_{\boldsymbol{\gamma}^0})\textbf{Y}=
\boldsymbol{\epsilon}^T\boldsymbol{\epsilon}+O\left(n\sqrt{s_n^2m_1^{-a}h_{m_1}}+m_2h_{m_2}s_n\right)$.
So for $m\in[m_1,m_2]$, and $c_j$'s $\in[\underline{\phi}_n,\bar{\phi}_n]$, we have
$0\le-J_4\le \frac{n+\nu}{2}\log\left(1+\frac{2(1+d_0\rho_n/n)k_n(1+o_P(1))}{n\underline{\phi}_n\tau_{m_2}^2\sigma_0^2}\right)
=O_P(1)$ since $k_n=O(\underline{\phi}_n\tau_{m_2}^2)$ (see Assumption \ref{B3}).

To approximate $J_5$, without loss of generality, we may assume
$\textbf{Z}_{\boldsymbol{\gamma}^0}=(\textbf{Z}_{\boldsymbol{\gamma}^*},\textbf{Z}_{\boldsymbol{\gamma}^0\backslash\boldsymbol{\gamma}^*})$
and
$\boldsymbol{\beta}^0_{\boldsymbol{\gamma}^0}=((\boldsymbol{\beta}^0_{\boldsymbol{\gamma}^*})^T,(\boldsymbol{\beta}^0_{\boldsymbol{\gamma}^0
\backslash\boldsymbol{\gamma}^*})^T)^T$. It can be shown by Assumption \ref{B1}, \ref{B3} (1),
(\ref{imp:result}), and Lemma \ref{lemma2} (\ref{lemma2:ii}) that
\begin{eqnarray*}
&&\textbf{Y}^T(\textbf{P}_{\boldsymbol{\gamma}}-\textbf{P}_{\boldsymbol{\gamma}^*})\textbf{Y}\\
&\le&2(\boldsymbol{\beta}^0_{\boldsymbol{\gamma}^0\backslash\boldsymbol{\gamma}^*})^T
\textbf{Z}_{\boldsymbol{\gamma}^0\backslash\boldsymbol{\gamma}^*}^T
(\textbf{P}_{\boldsymbol{\gamma}}-\textbf{P}_{\boldsymbol{\gamma}^*})
\textbf{Z}_{\boldsymbol{\gamma}^0\backslash\boldsymbol{\gamma}^*}
\boldsymbol{\beta}^0_{\boldsymbol{\gamma}^0\backslash\boldsymbol{\gamma}^*}+
4\boldsymbol{\eta}^T(\textbf{P}_{\boldsymbol{\gamma}}-\textbf{P}_{\boldsymbol{\gamma}^*})\boldsymbol{\eta}
+4\boldsymbol{\epsilon}^T(\textbf{P}_{\boldsymbol{\gamma}}-\textbf{P}_{\boldsymbol{\gamma}^*})\boldsymbol{\epsilon}\\
&\le&2\rho_n\|\boldsymbol{\beta}^0_{\boldsymbol{\gamma}^0\backslash\boldsymbol{\gamma}^*}\|^2
+4(C_\beta n s_n^2 m_1^{-a}h_{m_1}+\alpha \sigma_0^2 s_n^2\log{p})\\
&\le& 2g_n(\|\boldsymbol{\beta}^0_{\boldsymbol{\gamma}^0\backslash\boldsymbol{\gamma}^*}\|^2+\alpha_1),
\end{eqnarray*}
where $g_n=\max\{\rho_n,ns_n^2 m_1^{-a}h_{m_1}, s_n^2\log{p}\}$, $\alpha>4$ and $\alpha_1$ are fixed positive constants.
On the other hand, define $\boldsymbol{\nu}_{\boldsymbol{\gamma}^*,m}=(\textbf{I}_n-\textbf{P}_{\boldsymbol{\gamma}^*})
\textbf{Z}_{\boldsymbol{\gamma}^0\backslash\boldsymbol{\gamma}^*}
\boldsymbol{\beta}_{\boldsymbol{\gamma}^0\backslash\boldsymbol{\gamma}^*}^0$. Then by Assumption \ref{B1},
$\|\boldsymbol{\nu}_{\boldsymbol{\gamma}^*,m}\|^2\ge (d_0^{-1}n-\rho_n)
\|\boldsymbol{\beta}_{\boldsymbol{\gamma}^0\backslash\boldsymbol{\gamma}^*}^0\|^2$. By Lemma \ref{lemma2} (\ref{lemma2:i}),
w.p.a.1, for any $m\in [m_1,m_2]$ and $\boldsymbol{\gamma}\in T_1(t_n)$,
\begin{eqnarray*}
&&\textbf{Y}^T(\textbf{I}_n-\textbf{P}_{\boldsymbol{\gamma}^*})\textbf{Y}\\
&=&\|\boldsymbol{\nu}_{\boldsymbol{\gamma}^*,m}\|^2
+2\boldsymbol{\nu}_{\boldsymbol{\gamma}^*,m}^T\tilde{\boldsymbol{\eta}}+\tilde{\boldsymbol{\eta}}^T
(\textbf{I}_n-\textbf{P}_{\boldsymbol{\gamma}^*})\tilde{\boldsymbol{\eta}}\\
&\ge&\|\boldsymbol{\nu}_{\boldsymbol{\gamma}^*,m}\|^2\left(1+O\left(\frac{\sqrt{s_n+\log{m_2}}
+\sqrt{n s_n^2 m_1^{-a}h_{m_1}}}{\sqrt{n\psi_n^2}}\right)\right)
+\boldsymbol{\epsilon}^T\boldsymbol{\epsilon}-\boldsymbol{\epsilon}^T \textbf{P}_{\boldsymbol{\gamma}^0}\boldsymbol{\epsilon}
-2\|\boldsymbol{\eta}\|\cdot \|\boldsymbol{\epsilon}\|\\
&=&\|\boldsymbol{\nu}_{\boldsymbol{\gamma}^*,m}\|^2\left(1+O\left(\frac{\sqrt{s_n+\log{m_2}}
+\sqrt{n s_n^2 m_1^{-a}h_{m_1}}}{\sqrt{n\psi_n^2}}\right)\right)
+\boldsymbol{\epsilon}^T\boldsymbol{\epsilon}-\sigma_0^2 s_n m_2h_{m_2}\\
&&-2C'\sqrt{C_\beta n^2 s_n^2 m_1^{-a}h_{m_1}}\\
&=&((d_0^{-1}n-\rho_n)
\|\boldsymbol{\beta}_{\boldsymbol{\gamma}^0\backslash\boldsymbol{\gamma}^*}^0\|^2
+n\sigma_0^2)(1+o(1)),
\end{eqnarray*}
for some constant $C'>0$. Therefore, for some large positive constant $C''$, w.p.a.1,
for any $m\in [m_1,m_2]$ and $\boldsymbol{\gamma}\in T_1(t_n)$,
\begin{eqnarray*}
J_5&=&\frac{n+\nu}{2}\log\left(1-\frac{\textbf{Y}^T(\textbf{P}_{\boldsymbol{\gamma}}-\textbf{P}_{\boldsymbol{\gamma}^*})\textbf{Y}}
{\textbf{Y}^T(\textbf{I}_n-\textbf{P}_{\boldsymbol{\gamma}^*})\textbf{Y}}\right)\\
&\ge&\frac{n+\nu}{2}\log\left(1-
\frac{2g_n(\|\boldsymbol{\beta}^0_{\boldsymbol{\gamma}^0\backslash\boldsymbol{\gamma}^*}\|^2+\alpha_1)}
{((d_0^{-1}n-\rho_n)
\|\boldsymbol{\beta}_{\boldsymbol{\gamma}^0\backslash\boldsymbol{\gamma}^*}^0\|^2+n\sigma_0^2)(1+o(1))}\right)\ge -C'' g_n.
\end{eqnarray*}

By similar arguments in the proof of Theorem \ref{main:thm1},
it can be shown that for any $m\in [m_1,m_2]$, $c_j$'s $\in[\underline{\phi}_n,\bar{\phi}_n]$,
and $\boldsymbol{\gamma}\in T_1(t_n)$,
$J_2\ge \frac{m_1}{2}\log(1+(d_0^{-1}n-\rho_n)\underline{\phi}_n\tau_{m_2}^2)$.
So, w.p.a.1,
for any $m\in [m_1,m_2]$ and $\boldsymbol{\gamma}\in T_1(t_n)$, for some constant $\widetilde{C}>0$
\[
\frac{p(\boldsymbol{\gamma}|\textbf{D}_n)}{p(\boldsymbol{\gamma}^*|\textbf{D}_n)}\le \widetilde{C}
\exp\left(-\frac{m_1}{2}\log\left(1+(nd_0^{-1}-\rho_n)\underline{\phi}_n\tau_{m_2}^2\right)+C'' g_n\right)\rightarrow 0.
\]
Thus, $\max_{m\in[m_1,m_2]}\max_{c_j\in[\underline{\phi}_n,\bar{\phi}_n]}
\frac{\max_{\boldsymbol{\gamma}\in T_1(t_n)}
p(\boldsymbol{\gamma}|\textbf{D}_n)}{\max_{\boldsymbol{\gamma}\in T_0(t_n)}p(\boldsymbol{\gamma}|\textbf{D}_n)}=o_P(1)$.

\textbf{Step 2:} Next we consider the following decomposition for $\boldsymbol{\gamma}\in T_2(t_n)$,
\begin{eqnarray*}
&&-\log\left(\frac{p(\boldsymbol{\gamma}|\textbf{D}_n)}{p(\emptyset|\textbf{D}_n)}\right)\\
&=&-\log\left(\frac{p(\boldsymbol{\gamma})}{p(\emptyset)}\right)+\frac{1}{2}
\log\left(\frac{\det(\textbf{W}_{\boldsymbol{\gamma}})}{\det(\textbf{W}_\emptyset)}\right)\\
&&+\frac{n+\nu}{2}\log\left(\frac{1+\textbf{Y}^T(\textbf{I}_n-\textbf{Z}_{\boldsymbol{\gamma}} \textbf{U}_{\boldsymbol{\gamma}}^{-1}\textbf{Z}_{\boldsymbol{\gamma}}^T)\textbf{Y}}
{1+\textbf{Y}^T(\textbf{I}_n-\textbf{P}_{\boldsymbol{\gamma}})\textbf{Y}}\right)
+\frac{n+\nu}{2}\log\left(\frac{1+\textbf{Y}^T(\textbf{I}_n-\textbf{P}_{\boldsymbol{\gamma}})
\textbf{Y}}{1+\textbf{Y}^T\textbf{Y}}\right).
\end{eqnarray*}
Denote the above four terms by $J_1,J_2,J_3,J_4$. It is clear that $J_1$ is lower bounded,
and $J_3\ge 0$. We approximate $J_4$. For $\boldsymbol{\gamma}\in T_2(t_n)$, let
$\boldsymbol{\nu}_{\boldsymbol{\gamma},m}=
\textbf{P}_{\boldsymbol{\gamma}}\textbf{Z}_{\boldsymbol{\gamma}^0}\boldsymbol{\beta}_{\boldsymbol{\gamma}^0}^0$.
By Assumption \ref{B1}, $\|\boldsymbol{\nu}_{\boldsymbol{\gamma},m}\|^2\le \rho_n k_n$.
Thus, by Lemma \ref{lemma2}, for some fixed $\alpha>4$, for any $m\in [m_1,m_2]$,
\begin{eqnarray*}
\textbf{Y}^T\textbf{P}_{\boldsymbol{\gamma}}\textbf{Y}
&\le&2\left(\|\boldsymbol{\nu}_{\boldsymbol{\gamma},m}\|^2
+2\boldsymbol{\eta}^T\textbf{P}_{\boldsymbol{\gamma}}\boldsymbol{\eta}
+2\boldsymbol{\epsilon}^T\textbf{P}_{\boldsymbol{\gamma}}\boldsymbol{\epsilon}\right)\\
 &\le&2(\rho_n k_n+2C_\beta ns_n^2 m_1^{-a}h_{m_1}+2\alpha\sigma_0^2 s_n\log{p})\le
2g_n(k_n+\alpha_2),
\end{eqnarray*}
where $g_n=\max\{\rho_n,ns_n^2 m_1^{-a}h_{m_1},s_n\log{p}\}$, and $\alpha_2$ is some fixed positive constant.
On the other hand, since $E\{|(\textbf{Z}_{\boldsymbol{\gamma}^0}\boldsymbol{\beta}^0_{\boldsymbol{\gamma}^0})^T
\boldsymbol{\epsilon}|^2/\|\textbf{Z}_{\boldsymbol{\gamma}^0}\boldsymbol{\beta}^0_{\boldsymbol{\gamma}^0}\|^2\}=\sigma_0^2$
we have $|(\textbf{Z}_{\boldsymbol{\gamma}^0}\boldsymbol{\beta}^0_{\boldsymbol{\gamma}^0})^T
\boldsymbol{\epsilon}|/\|\textbf{Z}_{\boldsymbol{\gamma}^0}\boldsymbol{\beta}^0_{\boldsymbol{\gamma}^0}\|=O_P(1)$.
Thus, $|(\textbf{Z}_{\boldsymbol{\gamma}^0}\boldsymbol{\beta}^0_{\boldsymbol{\gamma}^0})^T\tilde{\boldsymbol{\eta}}|
\le \|\textbf{Z}_{\boldsymbol{\gamma}^0}\boldsymbol{\beta}^0_{\boldsymbol{\gamma}^0}\|\cdot
\left(\|\boldsymbol{\eta}\|+O_P(1)\right)$. Since $\sqrt{s_n^2m_1^{-a}h_{m_1}}=o(\psi_n^2)=o(k_n)$, we have
\begin{eqnarray*}
\textbf{Y}^T\textbf{Y}&=&\|\textbf{Z}_{\boldsymbol{\gamma}^0}\boldsymbol{\beta}^0_{\boldsymbol{\gamma}^0}\|^2
+2(\textbf{Z}_{\boldsymbol{\gamma}^0}\boldsymbol{\beta}^0_{\boldsymbol{\gamma}^0})^T\tilde{\boldsymbol{\eta}}
+\tilde{\boldsymbol{\eta}}^T\tilde{\boldsymbol{\eta}}\\
&=&\|\textbf{X}_{\boldsymbol{\gamma}^0}\boldsymbol{\beta}^0_{\boldsymbol{\gamma}^0}\|^2
\left(1+O_P\left(\sqrt{\frac{1+ns_n^2 m_1^{-a}h_{m_1}}{n k_n}}\right)\right)+
n\sigma_0^2(1+o_P(1))+O_P\left(\sqrt{n^2s_n^2m_1^{-a}h_{m_1}}\right)\\
&=&\|\textbf{X}_{\boldsymbol{\gamma}^0}\boldsymbol{\beta}^0_{\boldsymbol{\gamma}^0}\|^2
\left(1+o_P(1)\right)+n\sigma_0^2 (1+o_P(1))\\
&\ge&(d_0^{-1} n k_n+n\sigma_0^2)\cdot(1+o_P(1)).
\end{eqnarray*}
Therefore, w.p.a.1, for $\boldsymbol{\gamma}\in T_2(t_n)$ and $m\in[m_1,m_2]$,
$J_4\ge \frac{n+\nu}{2}\log\left(1-\frac{2g_n(k_n+\alpha_2)}
{(d_0^{-1} n k_n+n\sigma_0^2)}\right)\ge -C' g_n$,
for some large constant $C'>0$.

Meanwhile, by similar proof in Step 1, it can be verified that for $\boldsymbol{\gamma}\in T_2(t_n)$ and $m\in[m_1,m_2]$,
$J_2\ge \frac{m_1}{2}\log(1+nd_0^{-1}\underline{\phi}_n\tau_{m_2}^2)$ which holds for $c_j$'s $\in[\underline{\phi}_n,\bar{\phi}_n]$.
Then  w.p.a.1, for $\boldsymbol{\gamma}\in T_2(t_n)$, $c_j$'s $\in[\underline{\phi}_n,\bar{\phi}_n]$ and $m\in[m_1,m_2]$,
\begin{eqnarray*}
\frac{p(\boldsymbol{\gamma}|\textbf{D}_n)}{p(\emptyset|\textbf{D}_n)}
\le \tilde{C}\exp\left(-\frac{m_1}{2}\log(1+nd_0^{-1}\underline{\phi}_n\tau_{m_2}^2)+C'g_n\right)=o_P(1),
\end{eqnarray*}
where $\tilde{C}$ is some large positive constant. This shows $\max_{m\in[m_1,m_2]}\max_{c_j\in [\underline{\phi}_n,\bar{\phi}_n]}
\frac{\max_{\boldsymbol{\gamma}\in T_2(t_n)}p(\boldsymbol{\gamma}|\textbf{D}_n)}{\max_{\boldsymbol{\gamma}\in T_0(t_n)}p(\boldsymbol{\gamma}|\textbf{D}_n)}=o_P(1)$.
This shows the desired result.

\subsection*{Proof of Theorem \ref{main:thm2} (\ref{thm2:ii})}

Under Assumption \ref{B3}, it can be shown using similar arguments in the beginning of the proof of Theorem \ref{main:thm1}
that Assumption \ref{A3'} (\ref{A3':special}) is equivalent to the following assumption, i.e., Assumption \ref{A3} (\ref{A3:4}),
\begin{equation}\label{A3':special:useful}
m_2s_n\log(1+n\bar{\phi}_n)=o(n\log(1+\min\{1,\psi_n^2\})).
\end{equation}
Similarly, (\ref{inform:model:f0}) can be shown to be equivalent to
\begin{equation}\label{inform:model}
\|\boldsymbol{\beta}^0_{\boldsymbol{\gamma}^0\backslash\boldsymbol{\gamma}}\|^2
\le b_0'\|\boldsymbol{\beta}^0_{\boldsymbol{\gamma}}\|^2,
\end{equation}
where $b_0'>0$ is constant.
To see this, using (\ref{TFS:error:rate}) and $\psi_n^2\gg m_1^{-a}$ (see Assumption \ref{B3} (\ref{B3:1})),
it can be shown that $\sum_{j\in\boldsymbol{\gamma}}\|f_j^0\|_j^2=\|\boldsymbol{\beta}_{\boldsymbol{\gamma}}^0\|^2(1+o(1))$
and $\sum_{j\in\boldsymbol{\gamma}^0\backslash\boldsymbol{\gamma}}\|f_j^0\|_j^2
=\|\boldsymbol{\beta}^0_{\boldsymbol{\gamma}^0\backslash\boldsymbol{\gamma}}\|^2(1+o(1))$,
uniformly for $m\in[m_1,m_2]$.
Then it can be seen that (\ref{inform:model}) is equivalent to (\ref{inform:model:f0}).
Next we will prove the theorem based on Assumptions \ref{B1}, \ref{B2}, \ref{B3}, (\ref{A3':special:useful}) and (\ref{inform:model}).

For the $\boldsymbol{\gamma}$ specified in the theorem, we consider the following decomposition
\begin{eqnarray*}
&&-\log\left(\frac{p(\emptyset|\textbf{Z})}{p(\boldsymbol{\gamma}|\textbf{Z})}\right)\\
&=&-\log\left(\frac{p(\emptyset)}{p(\boldsymbol{\gamma})}\right)+\frac{1}{2}\log\left(\frac{1}{\det(\textbf{W}_{\boldsymbol{\gamma}})}\right)
-\frac{n+\nu}{2}\log\left(\frac{1+\textbf{Y}^T(\textbf{I}_n-\textbf{Z}_{\boldsymbol{\gamma}} \textbf{U}_{\boldsymbol{\gamma}}^{-1} \textbf{Z}_{\boldsymbol{\gamma}}^T)\textbf{Y}}{1+\textbf{Y}^T(\textbf{I}_n-\textbf{P}_{\boldsymbol{\gamma}})\textbf{Y}}\right)\\
&&+\frac{n+\nu}{2}\log\left(\frac{1+\textbf{Y}^T\textbf{Y}}{1+\textbf{Y}^T(\textbf{I}_n-\textbf{P}_{\boldsymbol{\gamma}})\textbf{Y}}\right).
\end{eqnarray*}
Denote the above four terms by $J_1,J_2,J_3,J_4$. Again, $J_1$ has finite lower bound. By similar proof in Step 1 of Theorem \ref{main:thm2}, one can show that w.p.a.1, for $m\in[m_1,m_2]$ and $c_j$'s $\in[\underline{\phi}_n,\bar{\phi}_n]$, $0\le -J_3=O_P(1)$.

To analyze $J_4$, note $J_4=\frac{n+\nu}{2}\log\left(1+\frac{\textbf{Y}^T\textbf{P}_{\boldsymbol{\gamma}}\textbf{Y}}
{1+\textbf{Y}^T(\textbf{I}_n-\textbf{P}_{\boldsymbol{\gamma}})\textbf{Y}}\right)$.
Let $\boldsymbol{\nu}_{\boldsymbol{\gamma},m}=\textbf{P}_{\boldsymbol{\gamma}}
\textbf{Z}_{\boldsymbol{\gamma}^0}\boldsymbol{\beta}_{\boldsymbol{\gamma}^0}^0$.
It can be directly examined by property of $\textbf{P}_{\boldsymbol{\gamma}}$ that
$\boldsymbol{\nu}_{\boldsymbol{\gamma},m}=
\textbf{Z}_{\boldsymbol{\gamma}}\boldsymbol{\beta}_{\boldsymbol{\gamma}}^0
+\textbf{P}_{\boldsymbol{\gamma}}
\textbf{Z}_{\boldsymbol{\gamma}^0\backslash\boldsymbol{\gamma}}
\boldsymbol{\beta}_{\boldsymbol{\gamma}^0\backslash\boldsymbol{\gamma}}^0$.
By Assumption \ref{B1} and
$\|\boldsymbol{\beta}_{\boldsymbol{\gamma}^0\backslash\boldsymbol{\gamma}}^0\|^2\le b_0'
\|\boldsymbol{\beta}_{\boldsymbol{\gamma}}^0\|^2$,
i.e., (\ref{inform:model}), we have
$|(\boldsymbol{\beta}_{\boldsymbol{\gamma}}^0)^T\textbf{Z}_{\boldsymbol{\gamma}}^T
\textbf{P}_{\boldsymbol{\gamma}}
\textbf{Z}_{\boldsymbol{\gamma}^0\backslash\boldsymbol{\gamma}}
\boldsymbol{\beta}_{\boldsymbol{\gamma}^0\backslash\boldsymbol{\gamma}}^0|
\le \|\textbf{Z}_{\boldsymbol{\gamma}}\boldsymbol{\beta}_{\boldsymbol{\gamma}}^0\|\cdot
\sqrt{\rho_n b_0'}\|\boldsymbol{\beta}_{\boldsymbol{\gamma}}^0\|$.
Meanwhile, $\|\textbf{Z}_{\boldsymbol{\gamma}}
\boldsymbol{\beta}_{\boldsymbol{\gamma}}^0\|^2\ge nd_0^{-1}\|\boldsymbol{\beta}_{\boldsymbol{\gamma}}^0\|^2$.
Therefore, by $\rho_n=o(n)$, it can be shown that
$\|\boldsymbol{\nu}_{\boldsymbol{\gamma},m}\|^2=
\|\textbf{Z}_{\boldsymbol{\gamma}}\boldsymbol{\beta}^0_{\boldsymbol{\gamma}}\|^2
\left(1+\frac{2(\boldsymbol{\beta}^0_{\boldsymbol{\gamma}})^T\textbf{Z}_{\boldsymbol{\gamma}}^T \textbf{P}_{\boldsymbol{\gamma}}
\textbf{Z}_{\boldsymbol{\gamma}^0\backslash\boldsymbol{\gamma}}\boldsymbol{\beta}^0_{\boldsymbol{\gamma}^0\backslash\boldsymbol{\gamma}}}
{\|\textbf{Z}_{\boldsymbol{\gamma}}\boldsymbol{\beta}^0_{\boldsymbol{\gamma}}\|^2}+
\frac{\|\textbf{P}_{\boldsymbol{\gamma}}
\textbf{Z}_{\boldsymbol{\gamma}^0\backslash\boldsymbol{\gamma}}\boldsymbol{\beta}^0_{\boldsymbol{\gamma}^0\backslash\boldsymbol{\gamma}}\|^2}
{\|\textbf{Z}_{\boldsymbol{\gamma}}\boldsymbol{\beta}^0_{\boldsymbol{\gamma}}\|^2}\right)
=\|\textbf{Z}_{\boldsymbol{\gamma}}\boldsymbol{\beta}^0_{\boldsymbol{\gamma}}\|^2(1+o(1))$,
for all $m\in[m_1,m_2]$. Since for each $m\in[m_1,m_2]$,
$\boldsymbol{\nu}_{\boldsymbol{\gamma},m}^T\boldsymbol{\epsilon}/
\|\boldsymbol{\nu}_{\boldsymbol{\gamma},m}\|\sim N(0,\sigma_0^2)$,
we get $\max_{m\in [m_1,m_2]}|\boldsymbol{\nu}_{\boldsymbol{\gamma},m}^T\boldsymbol{\epsilon}|/
\|\boldsymbol{\nu}_{\boldsymbol{\gamma},m}\|=O_P(\sqrt{\log{m_2}})$.
Also note, w.p.a.1, for $m\in[m_1,m_2]$, $|\boldsymbol{\nu}_{\boldsymbol{\gamma},m}^T\boldsymbol{\eta}|\le
\|\boldsymbol{\nu}_{\boldsymbol{\gamma},m}\|\cdot\|\boldsymbol{\eta}\|\le
\sqrt{C_\beta ns_n^2m_1^{-a}h_{m_1}}\|\boldsymbol{\nu}_{\boldsymbol{\gamma},m}\|$, thus we get that
\begin{eqnarray*}
\textbf{Y}^T\textbf{P}_{\boldsymbol{\gamma}}\textbf{Y}&\ge& \|\boldsymbol{\nu}_{\boldsymbol{\gamma},m}\|^2
+2\boldsymbol{\nu}_{\boldsymbol{\gamma},m}^T\tilde{\boldsymbol{\eta}}\\
&=&\|\boldsymbol{\nu}_{\boldsymbol{\gamma},m}\|^2
\left(1+O_P\left(\sqrt{\frac{ns_n^2m_1^{-a}h_{m_1}+\log{m_2}}{n\psi_n^2}}\right)\right)
\ge nd_0^{-1}\|\boldsymbol{\beta}_{\boldsymbol{\gamma}}^0\|^2(1+o_P(1)).
\end{eqnarray*}

To approximate $\textbf{Y}^T(\textbf{I}_n-\textbf{P}_{\boldsymbol{\gamma}})\textbf{Y}$,
let $\tilde{\boldsymbol{\nu}}_{\boldsymbol{\gamma},m}=(\textbf{I}_n-\textbf{P}_{\boldsymbol{\gamma}})
\textbf{Z}_{\boldsymbol{\gamma}^0\backslash\boldsymbol{\gamma}}
\boldsymbol{\beta}_{\boldsymbol{\gamma}^0\backslash\boldsymbol{\gamma}}^0$.
It can be verified that $\max_{m\in [m_1,m_2]}
|\tilde{\boldsymbol{\nu}}_{\boldsymbol{\gamma},m}^T\boldsymbol{\epsilon}|/
\|\tilde{\boldsymbol{\nu}}_{\boldsymbol{\gamma},m}\|=O_P(\sqrt{\log{m_2}})$,
and, by Assumption \ref{B1}, we have
$\|\tilde{\boldsymbol{\nu}}_{\boldsymbol{\gamma},m}\|^2\ge (nd_0^{-1}-\rho_n)
\|\boldsymbol{\beta}_{\boldsymbol{\gamma}^0\backslash\boldsymbol{\gamma}}^0\|^2$,
and $\|\tilde{\boldsymbol{\nu}}_{\boldsymbol{\gamma},m}\|^2\le
nd_0\|\boldsymbol{\beta}_{\boldsymbol{\gamma}^0\backslash\boldsymbol{\gamma}}^0\|^2\le
nd_0b_0' \|\boldsymbol{\beta}_{\boldsymbol{\gamma}}^0\|^2$.
Therefore, it can be shown by direct calculation that w.p.a.1, for $m\in[m_1,m_2]$,
\[
\textbf{Y}^T(\textbf{I}_n-\textbf{P}_{\boldsymbol{\gamma}})\textbf{Y}\le
(\|\tilde{\boldsymbol{\nu}}_{\boldsymbol{\gamma},m}\|^2+n\sigma_0^2)(1+o_P(1))\le
(nd_0b_0' \|\boldsymbol{\beta}_{\boldsymbol{\gamma}}^0\|^2+n\sigma_0^2)(1+o_P(1)).
\]
Therefore, w.p.a.1, for $m\in[m_1,m_2]$,
\begin{eqnarray*}
J_4\ge \frac{n+\nu}{2}\log\left(1+\frac{nd_0^{-1}\|\boldsymbol{\beta}_{\boldsymbol{\gamma}}^0\|^2(1+o(1))}
{(nd_0b_0' \|\boldsymbol{\beta}_{\boldsymbol{\gamma}}^0\|^2+n\sigma_0^2)}\right)
\ge \frac{n+\nu}{2}\log\left(1+\frac{1+o(1)}{d_0^2 b_0'}\cdot\frac{\psi_n^2}{\psi_n^2+\zeta_0}\right),
\end{eqnarray*}
where $\zeta_0=\sigma_0^2/(d_0b_0')$ and the last inequality follows by $\|\boldsymbol{\beta}_{\boldsymbol{\gamma}}^0\|^2
\ge\psi_n^2$. Therefore, we can get that $J_4\ge \frac{n+\nu}{2}
\log\left(1+\frac{1+o(1)}{d_0^2 b_0'}\cdot\min\{1/2,\psi_n^2/(2\zeta_0)\}\right)$.

Finally, by the proof of (\ref{thm1:eq2}), it can be shown that w.p.a.1, for $m\in[m_1,m_2]$,
$J_2\ge -\frac{m_2s_n}{2}\log(1+d_0n\bar{\phi}_n\tau_1^2)$. So by (\ref{A3':special:useful}), w.p.a.1, for $m\in[m_1,m_2]$,
as $n\rightarrow\infty$,
\begin{eqnarray*}
&&\frac{p(\emptyset|\textbf{D}_n)}{p(\boldsymbol{\gamma}|\textbf{D}_n)}\\&
\le& \tilde{C}\exp\left(\frac{m_2s_n}{2}\log(1+d_0n\bar{\phi}_n\tau_1^2)
-\frac{n+\nu}{2}
\log\left(1+\frac{1+o(1)}{d_0^2 b_0'}\cdot\min\{1/2,\psi_n^2/(2\zeta_0)\}\right)\right)\rightarrow0,
\end{eqnarray*}
where $\tilde{C}$ is a large positive constant. This completes the proof.


\begin{thebibliography}{34}

\bibitem{BB04} Barbieri, M. M. and Berger, J. O. (2004).
Optimal predictive model selection.
\textit{Annals of Statistics} \textbf{32}, 870--897.

\bibitem{BP96} Berger, J. O. and Pericchi, L. (1996).
The intrinsic {B}ayes factor for model selection and prediction.
{\it Journal of the American Statistical Association}
{\bf 91}, 109--122.

\bibitem{BGM03} Berger, J. O., Ghosh, J. K. and Mukhopadhyay, N. (2003).
Approximations and consistency of Bayes factors as model dimension grows.
{\it Journal of Statistical Planning and Inference}
{\bf 112}, 241--258.

\bibitem{BG03} Belitser, E. and Ghosal, S. (2003).
Adaptive Bayesian inference on the mean of an infinite-dimensional normal distribution.
\textit{Annals of Statistics}, \textbf{31}, 536--559.

\bibitem{BK00} Buldygin, V. and Kozachenko, Y. (2000). \emph{Metric Characterization of Random Variables and Random Processes.}
Providence, RI: American Mathematical Society.

\bibitem{CH53} Courant, R. and Hilbert, D. (1953). \emph{Methods of Mathematical Physics}, Volume \textbf{1}.
New York: Interscience Publischer, Inc.

\bibitem{CGMM09} Casella, C., Gir{\'o}n, F. J., Mart\'{\i}nez, M.~L. and Moreno, E. (2009).
Consistency of {B}ayesian procedures for variable selection.
{\it Annals of Statistics}
{\bf 37}, 1207--1228.

\bibitem{CGM10} Chipman, H. George, E., and McCulloch, R. (2010).
BART: Bayesian adaptive regression trees.
\textit{Annals of Applied Statistics} \textbf{4}, 266--298.

\bibitem{CPV98} Clyde, M., Parmigiani, G. and Vidakovic, B. (1998).
Multiple shrinkage and subset selection in wavelets.
{\it Biometrika} {\bf 85}, 391--401.


\bibitem{DE03} Donoho, D. L. and Elad, M. (2003). Optimally sparse representation in general (nonorthogonal) dictionaries via
$\ell_1$ minimization.  \emph{Proc. Natl. Acad. Sci. U.S.A.}  \textbf{100}, 2197--2202.

\bibitem{FFS11} Fan, J., Feng, Y. and Song, R. (2011).
Nonparametric independence screening in sparse ultra-high dimensional additive models.
\emph{Journal of American Statistical Association} \textbf{116}, 544--557.

\bibitem{FL08} Fan, J. and Lv, J. (2008).
Sure independence screening for ultra-high dimensional feature space.
(with discussion) \emph{Journal of Royal Statistical Society B} \textbf{70}, 849--911.

\bibitem{FL10} Fan, J. and Lv, J. (2010).
A selective overview of variable selection in high dimensional feature space.
\textit{Statistica Sinica} \textbf{20}, 101--148.

\bibitem{FS10} Fan, J. and Song, R. (2010). Sure independence screening in generalized linear models with NP-dimensionality.
\textit{Annals of Statistics} \textbf{38}, 3567--3604.

\bibitem{FSW09} Fan, J., Samworth, R. and Wu, Y. (2009).
Ultrahigh dimensional variable selection: beyond the lienar model.
\emph{Journal of Machine Learning Research} \textbf{10}, 1829--1853.

\bibitem{FLS01} Fern{\'a}ndez, C.,  Ley, E. and Steel, M. F. J. (2001).
Benchmark priors for Bayesian model averaging.
{\it Journal of Econometrics}
{\bf 100}, 381--427.

\bibitem{GCSR03} Gelman, A., Carlin, J. B., Stern, H. S. and Rubin, D. B.
(2003). \textit{Bayesian Data Analysis} (2nd ed). Chapman $\&$ Hall/CRC.

\bibitem{P95} Green, P. J. (1995).
Reversible jump Markov chain Monte Carlo computation and Bayesian model determination.
\emph{Biometrika} \textbf{82}, 711--732.

\bibitem{GH09} Green, P. and Hastie, D. (2009). Reversible jump MCMC. Technical Report,
University of Bristol.

\bibitem{GMCM10} Gir\'{o}n, F. J., Moreno, E., Casella, G. and Mart\'{i}nez, M. L.
(2010). Consistency of objective Bayes factors for nonnested linear
models and increasing model dimension.
\textit{Revista de la Real Academia de Ciencias Exactas, Fisicas y Naturales. Serie A. Matematicas}
\textbf{104}, 57--67.

\bibitem{GR98} Godsill, J.~S. and Rayner, P. J. W. (1998).
Robust reconstruction and analysis of autoregressive signals in impulsive noise using the Gibbs sampler.
\textit{IEEE Trans. Speech Audio Process} \textbf{6}, 352--372.

\bibitem{GVL89} Golub, G. H. and Van Loan, C. F. (1989).
\emph{Matrix Computations}, 2nd ed. John Hopkins
Univ. Press, Baltimore.

\bibitem{HHM08} Huang, J., Horowitz, J. and Ma, S. (2008). Asymptotic properties of bridge estimators in sparse high-dimensional regression models.
\textit{Annals of Statistics} \textbf{36}, 587--613.


\bibitem{HHW10} Huang, J., Horowitz, J., and Wei, F. (2010).
Variable selection in nonparametric additive models. \textit{Annals of Statistics}
\textbf{38}, 2282--2313.

\bibitem{HT90} Hastie, T. J. and Tibshirani, R. J. (1990). \emph{Generalized Additive Models}.
Chapman \& Hall/CRC Monographs on Statistics \& Applied Probability.

\bibitem{HKZ12} Hsu, D., Kakade, S. M. and Zhang, T. (2012).
A tail inequality for quadratic forms of subgaussian random vectors.
\textit{Electronic Communication in Probability} \textbf{17}, 1--6.

\bibitem{JR12} Johnson, V. E. and Rossell, D. (2012). Bayesian model selection in high-dimensional settings.
\emph{Journal of the American Statistical Association} \textbf{107}, 649--660.

\bibitem{KY08} Koltchinskii, V. and Yuan, M. (2008). Sparse recovery in large ensembles of kernel machines.
21 st Annual Conference on Learning Theory-COLT 2008, Helsinki, Finland, July 9--12, 2008,
eds. R. A. Servedio and T. Zhang, Omnipress, pp. 229--238.

\bibitem{LF09} Lv, J. and Fan, Y. (2009).
A unified approach to model selection and sparse recovery using regularized least squares.
\emph{Annals of Statistics} \textbf{37} 3498--3528.

\bibitem{LPMCB08} Liang, F., Paulo, R., Molina, G., Clyde, M. and Berger, J. O. (2008).
Mixtures of $g$-priors for Bayesian variable selection.
\textit{Journal of the American Statistical Association} {\bf 103}, 410--423.

\bibitem{LSY13} Liang, F., Song, Q., and Yu, K. (2013).
Bayesian subset modeling for high dimensional generalized linear models.
{\it Journal of the American Statistical Association}, in press.

\bibitem{LZ06} Lin, Y. and Zhang, H. H. (2006).
Component Selection and Smoothing in Multivariate Nonparametric Regression.
\textit{Annals of Statistics} \textbf{34}, 2272--2297.

\bibitem{MB06} Meinshausen, N. and B\"{u}hlmann, P. (2006). High dimensional graphs and variable selection with the Lasso. \textit{Annals of Statistics} \textbf{34}, 1436--1462.

\bibitem{MSB09}
Meier, L., van de Geer, S. and Buehlmann, P. (2009). High-dimensional additive modeling.
\emph{Annals of Statistics} \textbf{37}, 3779--3821.

\bibitem{MY09} Meinshausen, N. and Yu, B. (2009). Lasso-type recovery of sparse representations for high-dimensional data. \textit{Annals of Statistics} \textbf{37}, 246--270.


\bibitem{RLLW09} Ravikumar, P., Lafferty, J., Liu, H. and Wasserman, L. (2009).
Sparse Additive Models. \textit{Journal of the Royal Statistical Society, Series B}
\textbf{71}, 1009--1030.


\bibitem{S85} Stone, C. (1985).
Additive regression and other nonparametric models.
\textit{Annals of Statistics} \textbf{13}, 689--705.


\bibitem{SC11} Shang, Z. and Clayton, M. K. (2011). Consistency of Bayesian model selection for linear models
with a growing number of parameters. {\it Journal of Statistical Planning and Inference}  \textbf{11}, 3463--3474.

\bibitem{SC12} Shang, Z. and Clayton, M. K. (2012). An application of Bayesian variable selection to spatial concurrent linear models.
\emph{Environmental and Ecological Statistics} \textbf{19}, 521--544.

\bibitem{SL13} Shang, Z. and Li, P. (2013).
Bayesian ultrahigh-dimensional screening via MCMC. Preprint.

\bibitem{SFK12} Scheipl, F., Fahrmeir, L., and Kneib, T. (2012).
Spike-and-slab priors for function selection in structured regression models.
\textit{Journal of the American Statistical Association} \textbf{107}, 1518--1532.


\bibitem{SHK11} Saban\'{e}s Bov\'{e}, D., Held, L., and Kauermann, G. (2011).
Mixtures of $g$-priors for generalised
ddditive model delection with penalised splines. Technical Report, University of Zurich.

\bibitem{SL03} Seber, G.~A.~F. and Lee, A. J. (2003).
{\it Linear Regression Analysis}, 2nd Ed.
Wiley-Interscience [John Wiley \& Sons], Hoboken, NJ.

\bibitem{SPZ12} Shen, X., Pan, W., Zhu, Y. (2012). Likelihood-based selection and sharp parameter estimation. \textit{Journal of American Statistical Association} \textbf{107}, 223-232.

\bibitem{WGN04} Wolfe, P. J., Godsill, S. J. and Ng, W.-J. (2004).
Bayesian variable selection and regularization for time-frequency surface estimation.
{\it Journal of the Royal Statistical Society, Series B}
{\bf 66}, 575--589.

\bibitem{VD08} van de Geer, S. A. (2008). High-dimensional generalized linear models and the Lasso. \textit{Annals of Statistics} \textbf{36}, 614--645.

\bibitem{XZ11} Xue, L. and Zou, H. (2011). Sure independence screening and compressed random sensing.
\textit{Biometrika}, \textbf{98}, 371--380.


\bibitem{YZ13} Yang, Y. and Zou, H. (2013). A cocktail algorithm for solving the elastic net penalized Cox's regression in high dimensions.
\textit{Statistics and Its Interface}, \textbf{6}, 167--173.

\bibitem{ZH08} Zhang, C.-H. and Huang, J. (2008). The sparsity and bias of the Lasso selection in high-dimensional linear regression.
\textit{Annals of Statistics} \textbf{36}, 1567--1594.


\bibitem{ZS80} Zellner, A. and Siow, A. (1980). Posterior odds ratios for selected regression hypotheses. In
\textit{Bayesian analysis in econometrics and statistics: the Zellner view and papers}, (ed. A. Zellner), 389--399.
Edward Elgar Publishing Limited.

\bibitem{ZY06} Zhao, P. and Yu, B. (2006). On model selection consistency of Lasso. \textit{Journal of Machine Learning Research} \textbf{7}, 2541--2567.


\end{thebibliography}
\end{document}